\documentclass[prx,aps,twocolumn,superscriptaddress]{revtex4-1}

\usepackage[colorlinks,urlcolor=blue,citecolor=blue,linkcolor=blue]{hyperref}

\usepackage[english]{babel} 
\usepackage[english]{layout}
\usepackage{amsmath,amsfonts,amssymb}
\usepackage{graphicx}
\usepackage{float}
\usepackage{physics}
\usepackage{color}
\usepackage{braket}
\usepackage{version}
\usepackage{array,multirow,makecell}
\usepackage{afterpage}
\usepackage[toc,page]{appendix} 
\usepackage{bbold}

\renewcommand{\k}{{\bf k}}

\newcommand{\q}{{\bf q}}
\newcommand{\eb}{\varepsilon_B}

\newcommand{\ob}[1]{{\color{black}#1}}

\begin{document}

\title{Bogoliubov excitations of a polariton condensate in dynamical equilibrium with an incoherent reservoir}

\author{M. Pieczarka}
\thanks{These two authors contributed equally to this work.}
\affiliation{ARC Centre of Excellence in Future Low-Energy Electronics Technologies and Department of Quantum Science and Technology, Research School of Physics, The Australian National University, Canberra, ACT 2601, Australia}
\affiliation{Department of Experimental Physics, Faculty of Fundamental Problems of Technology, Wroclaw University of Science and Technology, Wyb. Wyspia\'{n}skiego 27, 50-370 Wroc\l aw Poland}

\author{O. Bleu}
\thanks{These two authors contributed equally to this work.}
\affiliation{ARC Centre of Excellence in Future Low-Energy Electronics Technologies and School of Physics and Astronomy, Monash University, Victoria 3800, Australia}

\author{E.~Estrecho}
\affiliation{ARC Centre of Excellence in Future Low-Energy Electronics Technologies and Department of Quantum Science and Technology, Research School of Physics, The Australian National University, Canberra, ACT 2601, Australia}

\author{M.~Wurdack}
\affiliation{ARC Centre of Excellence in Future Low-Energy Electronics Technologies and Department of Quantum Science and Technology, Research School of Physics, The Australian National University, Canberra, ACT 2601, Australia}

\author{M.~Steger}
\thanks{Current address: National Renewable Energy Laboratory, Golden, CO 80401, USA}
\affiliation{Department of Physics and Astronomy, University of Pittsburgh, Pittsburgh, PA 15260, USA}%

\author{D.~W.~Snoke}
\affiliation{Department of Physics and Astronomy, University of Pittsburgh, Pittsburgh, PA 15260, USA}%

\author{K.~West}
\affiliation{Department of Electrical Engineering, Princeton University, Princeton, NJ 08544, USA}%

\author{L.~N.~Pfeiffer}
\affiliation{Department of Electrical Engineering, Princeton University, Princeton, NJ 08544, USA}

\author{A. G. Truscott}
\affiliation{Department of Quantum Science and Technology, Research School of Physics, The Australian National University, Canberra, ACT 2601, Australia}

\author{E. A. Ostrovskaya}
\affiliation{ARC Centre of Excellence in Future Low-Energy Electronics Technologies and Department of Quantum Science and Technology, Research School of Physics, The Australian National University, Canberra, ACT 2601, Australia}

\author{J. Levinsen}
\affiliation{ARC Centre of Excellence in Future Low-Energy Electronics Technologies and School of Physics and Astronomy, Monash University, Victoria 3800, Australia}

\author{M. M. Parish}
\affiliation{ARC Centre of Excellence in Future Low-Energy Electronics Technologies and School of Physics and Astronomy, Monash University, Victoria 3800, Australia}

\date{\today}

\begin{abstract}
   The classic Bogoliubov theory of weakly interacting Bose gases rests upon the assumption that nearly all the bosons condense into the lowest quantum state at sufficiently low temperatures. Here we develop a generalized version of Bogoliubov theory for the case of a driven-dissipative exciton-polariton condensate with a large incoherent uncondensed component, 
   or excitonic reservoir. We argue that such a reservoir can consist of both excitonic high-momentum polaritons and optically dark superpositions of excitons across different optically active layers, such as multiple quantum wells in a microcavity. 
   In particular, %our theory 
   we predict interconversion between the dark and bright (light-coupled) excitonic states that can lead to a dynamical equilibrium between the condensate and reservoir populations.
   We show that the presence of the reservoir 
   fundamentally modifies both the energy and the amplitudes of the Bogoliubov quasiparticle excitations %across all momenta
   due to the non-Galilean-invariant nature of polaritons. 
   Our theoretical findings are supported by our experiment, where we directly detect the Bogoliubov excitation branches of an optically trapped polariton condensate in the high-density regime. %Furthermore, 
   By analyzing the measured occupations of the excitation branches,
  we extract the Bogoliubov amplitudes across a range of momenta and show that they agree with our generalized theory. 
  %This thus constitutes the first measurement of the Bogoliubov amplitudes in a non-equilibrium condensate.
\end{abstract}
 
\maketitle

\section{Introduction}

When a dilute gas of bosons achieves quantum degeneracy, it forms a Bose-Einstein condensate (BEC), where a macroscopically large number of bosons occupies the lowest single-particle state. For an equilibrium system, this novel phase of matter is governed by the well-established Bogoliubov theory~\cite{bogoliubov1947theory}, which describes the emergent %quasiparticle 
collective excitations and the accompanying energy shifts due to boson-boson interactions. In particular, the Bogoliubov excitations underpin important hydrodynamic behavior such as superfluidity and the speed of sound, as well as determining the thermodynamic properties of the Bose gas~\cite{pitaevskii2016bose}. 
Following the first observation of a BEC with ultracold atomic gases~\cite{anderson1995observation,Bradley1995,Davis1995}, 
conventional Bogoliubov theory was cleanly demonstrated in pioneering cold-atom experiments which reported direct measurements of the Bogoliubov excitation spectrum and %the Bogoliubov 
quasiparticle amplitudes~\cite{Steinhauer2002,Vogels2002}. 

However, it is less clear how to apply Bogoliubov theory to highly non-equilibrium Bose systems, where drive and dissipation can lead to a condensate that coexists with an incoherent reservoir.  
This is particularly relevant to bosons in the solid state such as exciton polaritons, which are hybrid quasiparticles arising from the strong coupling between microcavity photons and semiconductor excitons (bound electron-hole pairs) \cite{DengRMP2010,RMP2013QFL,kavokin2017microcavities}. 
%On the one hand,
The photonic component endows polaritons with an exceptionally small effective mass---orders of magnitude smaller than that of the exciton---thus enabling condensation at elevated temperatures \cite{kasprzak2006bose,balili2007bose,Baumberg2008,Li2013,Plumhof2014,Su2017}.
%On the other hand, 
Since photons can escape from the microcavity, a polariton condensate requires continuous external pumping, either with a laser~\cite{DengRMP2010,RMP2013QFL} or by electric injection \cite{Schneider2013,Bhattacharya2013}, to sustain its density. Moreover, in order for a polariton BEC to arise spontaneously in this driven-dissipative system, one requires an off-resonant pump that excites 
electrons and holes which in turn relax into high-energy %optically dark 
excitons that eventually feed the condensate. 
%form an incoherent reservoir of excitons. 
%which can populate a polariton condensate. 
Thus, polariton condensation is necessarily accompanied by the formation of an incoherent uncondensed 
excitonic reservoir~\cite{DengRMP2010}. Such a reservoir can also manifest itself indirectly in experiments %as a  and 
via its repulsive interactions with polaritons~\cite{Ferrier2011,Schmidt2019,Brichkin2011,Klaas2017}, 
%and these interactions have been used to create 
allowing the creation of on-demand trapping potentials by spatially selective laser excitation~\cite{Askitopoulos2013,Cristofolini2013,Dall2014,Paschos2020,Orfanakis2021}.

Because of the inherent complexity of the nonequilibrium polariton-reservoir system, several simplified models have been introduced to describe condensation in the presence of incoherent drive and dissipation~\cite{Keeling2007}.
Such models predict that the %Bogoliubov 
collective excitation spectrum for polariton condensates is qualitatively modified at low momenta such that it becomes diffusive~\cite{Szymanska2006,Wouters2007,Hanai2018}, gapped~\cite{Byrnes2012} or remains conservative-like but damped~\cite{Solnyshkov2014}.
However, these predictions have yet to be confirmed experimentally, and indeed it has recently 
%recent experiments have 
been shown that any such modifications occur at momenta that are too low to be directly resolved in high-quality microcavities~\cite{Ballarini2020}. % and require more indirect probing \cite{Ballarini2009}. 
Furthermore, while the observed shape of the Bogoliubov spectrum in experiments has so far appeared to be consistent with conventional Bogoliubov theory~\cite{Utsunomiya2008,Kohnle2011,Pieczarka2015,Zajac2015,Horikiri2017,Stepanov2019,Pieczarka2020,Ballarini2020,bieganska2020collective,Steger2021}, there are notable unexplained observations 
such as the recently measured occupations of the excitation branches~\cite{Pieczarka2020}.

In this work, we devise a generalized Bogoliubov theory of a polariton BEC that accounts for the presence of an incoherent excitonic reservoir.  %Specifically, 
We consider the case of a high-quality microcavity with multiple semiconductor quantum wells (QWs), which can host a reservoir involving optically dark superpositions of otherwise optically active excitons in different QWs~\cite{Bleu2020}. In particular, our theory assumes that such a reservoir is in dynamical equilibrium with the polariton condensate due to interconversion between optically dark and bright exciton superpositions. %light-coupled excitonic states.
We show that the reservoir leads to changes in the Bogoliubov quasiparticles as a consequence of the intrinsic non-Galilean-invariant nature of polaritons.

Using an optically trapped condensate in the high-density
Thomas-Fermi regime, we experimentally probe the Bogoliubov spectrum directly and analyze it within the framework of our theory. This allows us to infer both the condensate and reservoir densities, and remarkably we find that they become locked to each other at large pump power, which %provides confirmation of the 
is consistent with interconversion  between polaritons and reservoir excitons. 
Furthermore, we extract the Bogoliubov amplitudes from the momentum-resolved occupations of the excitation branches and find a good agreement with the predictions of the generalised Bogoliubov theory for a wide range of momenta. 
To our knowledge, this is the first measurement of the Bogoliubov amplitudes in a non-equilibrium condensate.

\section{Theoretical description} \label{sec:Th}

\begin{figure}[tbp] % not "pt"
   \includegraphics[width=\columnwidth]{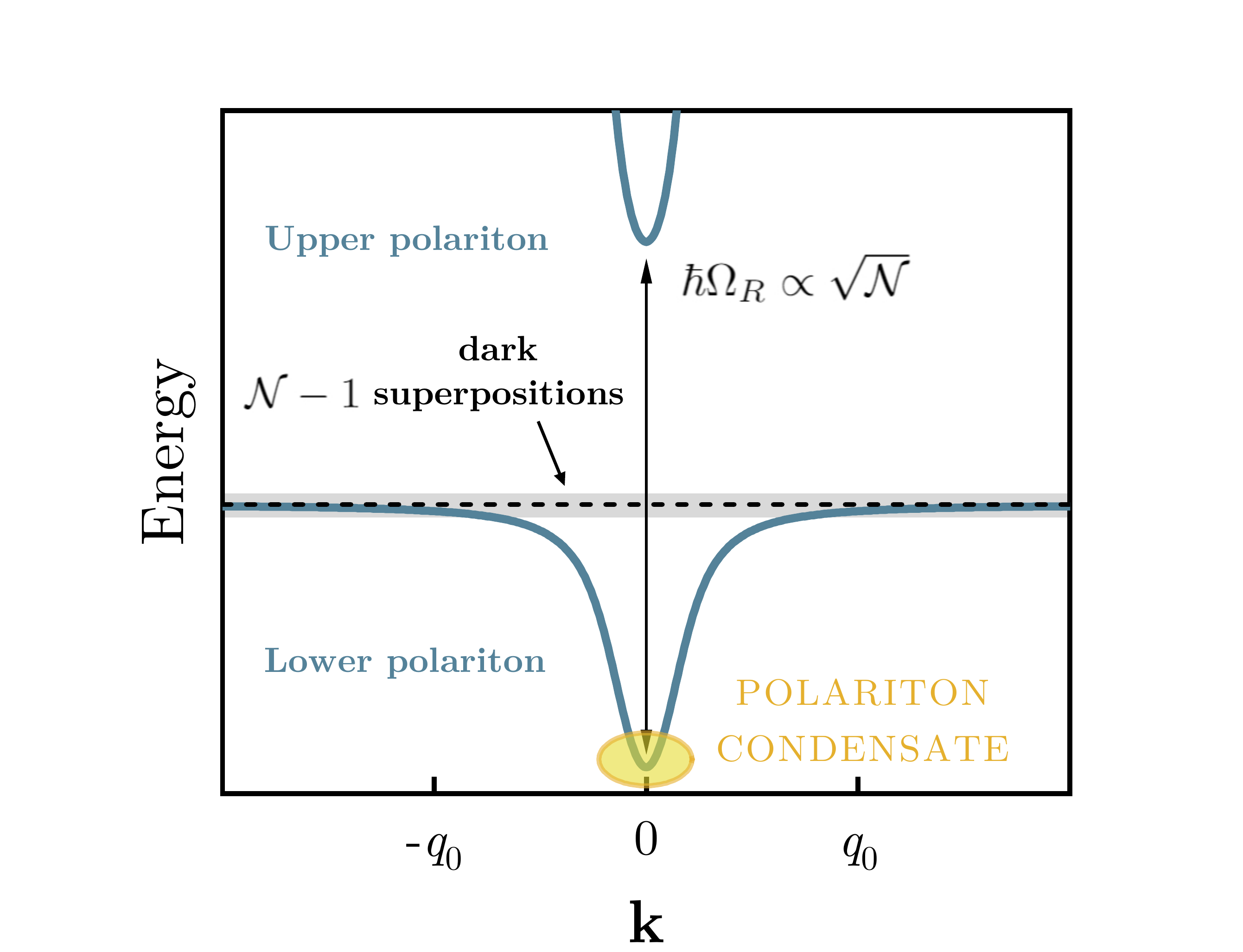}
\caption{Schematic representation of a polariton condensate (yellow ellipse) coexisting with an excitonic reservoir (shaded gray region) in a $\mathcal{N}$-layer microcavity. The solid-blue (dashed-black) lines represent the single-particle dispersion relations of polaritons  (excitons). In our description the reservoir corresponds to a population of the $\mathcal{N}-1$ degenerate exciton superpositions uncoupled to light and excitonic lower polaritons with wavevectors $|\k|>q_0$ (see text).}
\label{fig:fig1}
\end{figure}

In this section, we introduce a theoretical description of a polariton condensate in a microcavity with $\mathcal{N}$ quantum wells. %which will be used to analyse the experimental results.  
Our model accounts for the presence of dark exciton superpositions in the multilayer system, as well as the non-Galilean-invariant nature of polaritons, which results in a non-parabolic dispersion for the polaritons and a momentum-dependent exciton fraction (Hopfield coefficient). We furthermore consider the presence of an incoherent excitonic reservoir (depicted schematically in Fig.~\ref{fig:fig1}) and we assume that the condensate-reservoir system has achieved a dynamical equilibrium with spatially uniform and stationary densities. \ob{While photon loss from the cavity is required to achieve such a steady state, we assume that its energy scale $\hbar \gamma_C$ is much smaller than that set by the interactions and thus we do not include it explicitly in our model. This assumption is also justified in the present experiment which involves a high-quality sample with low losses and where only large $\k$-vectors are probed --- see Sec.~\ref{sec:ExpDet}}.

From our generalized Bogoliubov theory, we obtain analytical expressions for the condensate Bogoliubov excitation spectrum and quasiparticle amplitudes, and we relate them to experimentally accessible interaction-induced energy shifts.
In particular, our theory provides a self-consistent framework for analyzing the experiments and extracting the behavior of the polariton condensate and excitonic reservoir.

\subsection{Single-polariton Hamiltonian}

In a typical semiconductor multi-QW microcavity, such as the one used in our experiment, %described in Sec.~\ref{sec:ExpDet}, 
%the 
identical QWs are located in groups in the antinodes of the cavity photon field to maximize the light-matter coupling~\cite{kavokin2017microcavities}. Each group of QWs is grown with thin layers of material providing energy barriers, whose height ensures a negligible electronic coupling between neighboring QWs~\footnote{A significant tunneling between QWs would result in multiple polariton branches~\cite{Ouellet2015}, a feature that is absent in our sample.}. 
Hence, to model the single-particle physics in the $\mathcal{N}$-QW microcavity, we use the following exciton-photon Hamiltonian:
\begin{eqnarray} \nonumber
   \hat{H}_0&=&\sum_{\mathbf{k}} E_{\mathbf{k}}^C \hat{c}_{\mathbf{k}}^{\dagger} \hat{c}_{\mathbf{k}}+ \sum_{\mathbf{k}} \sum_{n=1}^{\mathcal{N}} E_{\mathbf{k}}^X \, \hat{x}_{\mathbf{k}, n}^{\dagger} \, \hat{x}_{\mathbf{k}, n}\\ \label{eq:ham0}
   & & + \frac{\hbar g_R}{2} \sum_{\mathbf{k}} \sum_{n=1}^{\mathcal{N}} \left(\hat{x}_{\mathbf{k}, n}^{\dagger} \, \hat{c}_{\mathbf{k}} +\hat{c}_{\mathbf{k}}^{\dagger} \hat{x}_{\mathbf{k}, n}\right) , 
\end{eqnarray}
where $g_R$ is the exciton-photon coupling strength. Here $\hat{c}_{\mathbf{k}}$ ($\hat{c}_{\mathbf{k}}^\dagger$) and $\hat{x}_{\mathbf{k},n}$ ($\hat{x}_{\mathbf{k},n}^\dagger$) are bosonic annihilation (creation) operators of cavity photons and quantum-well excitons, respectively, with in-plane momentum $\hbar\mathbf{k}$ and quantum-well index $n$. The kinetic energies at low momenta are $E_{\mathbf{k}}^C = \hbar^2k^2/2m_C +\delta$ and  $E_{\mathbf{k}}^X = \hbar^2k^2/2m_X$, where $k \equiv |\k|$ and
$m_C$ ($m_X$) is the photon (exciton) mass, while $\delta$ is the photon-exciton detuning. Here we measure energies with respect to the exciton energy at zero momentum, and we use a scalar theory that does not explicitly include the polarization of polaritons  
%describes polaritons with a given circular polarization 
since 
%the interactions between cross-circularly-polarized polaritons and other 
polarization effects are negligible in the present experiment \ob{(see Sec.~\ref{sec:ExpDet}).}

The light-matter coupling terms %coupling of light and matter 
in the second line of the Hamiltonian
\eqref{eq:ham0} immediately define a bright (symmetric) superposition of
the bare QW exciton operators
\begin{align}
    \hat{b}_{\mathbf{k}} =
\frac1{\sqrt{\mathcal{N}}}\sum_{n=1}^{\mathcal{N}}
\hat{x}_{\mathbf{k},n}.
\end{align}
Indeed, in the literature this is often the only excitonic mode that is considered. However, there are additionally $\mathcal{N}-1$ dark superposition states~\cite{RMP2013QFL}, corresponding to superpositions of the QW excitons that are orthogonal to the bright state~\footnote{For convenience, we denote these superpositions as bright or dark regardless of whether the momentum is within or outside the radiative region}. These may be conveniently represented by the discrete Fourier transforms
\begin{equation} \label{eq:Unitary}
  \hat{d}_{\mathbf{k},l}=\sum_{n=1}^{\mathcal{N}} u_{ln} \, \hat{x}_{\mathbf{k},n},
\end{equation}
with $l=1,\dots,\mathcal{N}$ and $u_{ln}=\frac{1}{\sqrt{\mathcal{N}}}e^{i2\pi n l/\mathcal{N}}$ \cite{Bleu2020}. Note that $\hat{d}_{\mathbf{k},\mathcal{N}}=\hat{b}_{\mathbf{k}}$. Importantly, as we now discuss, the existence of the
dark states allow excitons to exist in the low-momentum region, and (once we include interactions) dark excitons can be transformed into bright excitons.
%furthermore interactions allow the interconvertion between bright and dark excitons.

Since only the bright state is coupled to light, we can diagonalize the exciton-photon Hamiltonian in the usual manner:
\begin{equation}\label{eq:HamDiag}
 \hat{H}_0=\! \sum_{\mathbf{k}}\! \left[E_{\mathbf{k}}^L \hat{L}_{\mathbf{k}}^{\dagger} \hat{L}_{\mathbf{k}}+ E_{\mathbf{k}}^U \hat{U}_{\mathbf{k}}^{\dagger}\hat{U}_{\mathbf{k}}+\!\! \sum_{l=1}^{\mathcal{N}-1}  E_{\mathbf{k}}^X\hat{d}_{\mathbf{k},l}^{\dagger}\hat{d}_{\mathbf{k},l}\right] \! ,
\end{equation}
with $\hat{L}$ ($\hat{U}$) the lower (upper) polariton annihilation operators defined as superpositions of the bright exciton and the photon,
\begin{equation}
 \begin{pmatrix}  \hat{L}_{\mathbf{k}}\\ \hat{U}_{\mathbf{k}}  \end{pmatrix} = \begin{pmatrix}  X_{\mathbf{k}} && C_{\mathbf{k}}\\ -C_{\mathbf{k}} &&  X_{\mathbf{k}} \end{pmatrix}     \begin{pmatrix}  \hat{b}_{\mathbf{k}}\\ \hat{c}_{\mathbf{k}}  \end{pmatrix} .
\end{equation}
Here  $E_\mathbf{k}^{U,L}$ are the polariton eigenenergies,
\begin{align}
\label{eq:polDispersion}
E_\mathbf{k}^{U,L}=\frac{1}{2} \left(E_{\mathbf{k}}^X+E_{\mathbf{k}}^C \pm \sqrt{\left(E_{\mathbf{k}}^C-E_{\mathbf{k}}^X\right)^2+ \hbar^2\Omega^2 }\right) ,
\end{align}
with the enhanced Rabi splitting $\hbar \Omega=\hbar g_R \sqrt{\mathcal{N}}$, while
 $X_{\mathbf{k}},C_{\mathbf{k}}$ are the Hopfield coefficients, corresponding to the exciton and photon fractions of the  lower polariton:
\begin{align}
X_{\mathbf{k}}^2=\frac{1}{2} \left(1+ \frac{E_{\mathbf{k}}^C-E_{\mathbf{k}}^X}{E_{\mathbf{k}}^U-E_{\mathbf{k}}^L}\right) ,~~~ C_{\mathbf{k}}^2=1- X_{\mathbf{k}}^2 .
\end{align}
The single-particle spectrum of the lower and upper polaritons is shown in Fig.~\ref{fig:fig1}, including the energies of the dark exciton superpositions. This explicitly illustrates the existence of the bare exciton dispersion in multi-QW microcavities.

Physically, the $\mathcal{N}-1$ degenerate dark superpositions and the $\sqrt{\mathcal{N}}$-enhancement of the Rabi splitting 
%between the two polariton eigenmodes in Eq.~\eqref{eq:HamDiag} 
originate from the fact that $\mathcal{N}$ identical QWs are coupled to a single cavity mode. This is a known phenomenon which occurs in other models with $\mathcal{N}$ quantum emitters coupled to one photonic mode such as in the Tavis-Cummings model~\cite{Keeling2020}.  Thus, these dark superpositions should not be confused with spin-forbidden dark excitons which can exist in a single quantum well and which we do not explicitly consider here.

Note that %introducing 
%if there are also 
a strong coupling between QW excitons and additional nearby photonic modes %in \eqref{eq:ham0} 
(such as Bragg modes) can ``brighten'' some of the dark superpositions and create additional polariton lines \cite{Richard2005}. However, it is unlikely that all of the dark superpositions will become optically active, since this would require the QW excitons to strongly couple with $\mathcal{N}-1$ additional photonic modes.

\subsection{Two-body interactions and effective model} \label{sec:interactions}

Having introduced the single-particle eigenstates of the model, we now wish to include the effect of interactions between these. 
The interactions in the system originate from pairwise exciton-exciton interactions within a given quantum well $n$, described by the following interaction Hamiltonian
 \begin{equation}\label{eq:interL} 
\hat{V}=  \frac{ g_{}}{2\mathcal{A}} \sum_{n=1}^{\mathcal{N}} \sum_{\k,\k',\q} \hat{x}_{\mathbf{k+q},n}^{\dagger}\hat{x}_{\mathbf{k'-q},n}^{\dagger} \hat{x}_{\mathbf{k'},n}\hat{x}_{\mathbf{k},n} ,
\end{equation}
where $g$ is the bare exciton-exciton coupling constant and $\mathcal{A}$ is the system area.
%
%
%
%The interactions in a structure with $\mathcal{N}$ QWs are then described by $\hat{V}=\sum_n \hat{V}_n$.
When expressed in the basis of the bright and dark superpositions~\eqref{eq:Unitary}, %introduced above, 
it reads \cite{Bleu2020}:
 \begin{equation}\label{eq:interNB} 
\hat{V}= \frac{ g_{}}{2\mathcal{N}\mathcal{A}}\sum_{\{l_j\}}\delta_{\mathcal{M}} \! \! \sum_{\k,\k',\q}
 \hat{d}_{\mathbf{k+q},l_1}^{\dagger}\hat{d}_{\mathbf{k'-q},l_2}^{\dagger} \hat{d}_{\mathbf{k'},l_3}\hat{d}_{\mathbf{k},l_4} ,
\end{equation}
where  $\{l_j\}=\{l_1,l_2,l_3,l_4\}$. Here the Kronecker delta encodes a phase selection rule for pairwise scattering ($\delta_{\mathcal{M}}=1$ if $\mathcal{M}=0$, $\delta_{\mathcal{M}}=0$ otherwise, where $\mathcal{M}=\text{Mod}{[l_1+ l_2- l_3 -l_4,\mathcal{N}}]$), which is a discrete analog of momentum conservation. 
It is worth noting that the bare interaction coupling constant $g$ is 
reduced by the factor of $1/\mathcal{N}$ in the new bright-dark-superposition basis. Furthermore, written in this form, Eq.~\eqref{eq:interNB} involves a large number of terms ($\mathcal{N}^3$), which highlights the complexity of the scattering processes occurring in multi-QW structures in the strong-coupling regime. In particular, interconversion between different states (bright or dark) is allowed by the phase selection rule, which enables their respective populations to equilibrate at large momenta where the single-particle energies become degenerate. Furthermore, it allows dark excitons to be scattered into and out of a polariton condensate, 
which is important for achieving dynamical equilibrium. 

For the total Hamiltonian $\hat{H}_0+\hat{V}$, we thus have a complicated interacting many-body problem involving lower polaritons, upper polaritons and dark exciton superpositions. To make further progress, we neglect the upper polariton, since this is expected to have a negligible population once a condensate of lower polaritons is formed,
%in a polariton condensate,
and we assume that any excitonic particles are uncondensed and semiclassical due to the large exciton mass ($m_X \sim 10^{4} m_C$).  
%assume that the population of upper polaritons is negligible 
%and we assume that 
Thus, populations of dark superposition states and high-momentum polaritons with a large excitonic fraction %``bright'' states with large wavevectors $k>q_0$, 
%for which $E_\k^L\simeq E_\k^X$, 
will form an incoherent %and semiclassical 
reservoir (represented by the grey shading in Fig.~\ref{fig:fig1}). Within our model, the wavevector $q_0$ above which the lower polariton is excitonic can be roughly estimated by the condition $E_{\q_0}^L= E_0^X$, which gives $q_0=(\sqrt{m_C m_X} \Omega/\hbar)^{1/2}$ at zero photon-exciton detuning.

%Hence, 
We can describe such a semiclassical reservoir %of dark excitons 
within the Hartree-Fock approximation, allowing us to replace the operators by the total reservoir number  
\begin{align}\label{eq:Nr}
    N_R = \sum_{l=1}^{\mathcal{N}-1}\sum_{\k} N_{\k,l}+\sum_{\substack{\k\\|\k|>q_0}} N_{\k,\mathcal{N}},
\end{align}
with $N_{\k,l}\equiv \langle \hat{d}_{\k,l}^\dagger\hat{d}_{\k,l} \rangle$ the states' momentum occupations. 
Here we have assumed that the reservoir is spatially homogeneous, which is reasonable away from the pumped region in our experiment (see Sec.~\ref{sec:ExpDet}). 
Had we not accounted for the presence of dark states in the system, the first term in the right hand side of Eq.~\eqref{eq:Nr} would be absent. 
We also note that our model neglects other potential sources of a reservoir such as originating from spin-forbidden dark excitons with angular momentum $J=2$ or from populations of free carriers (electrons and holes).

As detailed in Appendix \ref{sec:Heff}, the Hartree-Fock approximation allows us to obtain an effective Hamiltonian for lower polaritons and the excitonic reservoir of the form:
  \begin{align} \label{eq:Heff}
\hat{H}_{\rm eff}&=E_{\text{res}}+\sum_{\k}\! \left[E_{\mathbf{k}}^L+2 \frac{ g_{pd}}{\mathcal{N}}%\mathcal{A}}
X_\k^2 %N_R
n_R \right] \hat{L}_{\mathbf{k}}^{\dagger} \hat{L}_{\mathbf{k}} \\ \nonumber
&+\frac{g_{pp}}{2\mathcal{N}\mathcal{A}} \sum_{\k,\k',\q} X_{\mathbf{k+q}}X_{\mathbf{k'-q}}X_{\mathbf{k'}}X_{\mathbf{k}} \hat{L}_{\mathbf{k+q}}^{\dagger}\hat{L}_{\mathbf{k'-q}}^{\dagger} \hat{L}_{\mathbf{k'}}\hat{L}_{\mathbf{k}},
  \end{align}
where $n_R=N_R/\mathcal{A}$ is the reservoir density.
Here, $E_{\text{res}}$ is the reservoir energy in the Hartree-Fock approximation; it contributes to the total energy of the system but it does not affect the lower polariton spectrum. 
%
%$g_{pp}X^4/\mathcal{N}$ is the effective polariton-polariton interaction strength and $g_{pd}X^2/\mathcal{N}$ the polariton-dark reservoir interaction strength.
%
\ob{We have introduced the polariton-polariton and polariton-reservoir effective interaction strengths $g_{pp}X^4/\mathcal{N}$ and $g_{pd}X^2/\mathcal{N}$,  respectively, which originate from the repeated two-body scattering processes --- for more details, see Appendix \ref{sec:Heff}.}
%$g_{pp}X^4/\mathcal{N}$ and $g_{pd}X^2/\mathcal{N}$ are the polariton-polariton and polariton-reservoir effective interaction strengths, respectively. 
These possess a momentum dependence through the Hopfield coefficients.
Both interactions are repulsive and contribute to the blueshift of the polariton line. The factor of 2 in the polariton-reservoir interactions originates from the fact that interconversion between polaritons and reservoir particles is possible within our model. %  this differs from the reservoir considered in ref. \cite{Grudinina2021}.
%Indeed, the Hamiltonian \eqref{eq:Heff} should also apply to any excitonic reservoir that contains spin-forbidden dark excitons 
The existence of such interconversion processes also allows the reservoir to achieve a dynamical equilibrium with the polariton condensate.

We emphasize that despite their two-dimensional nature, polaritons are able to interact pairwise efficiently at low momenta. This is in stark contrast to the standard low-energy scattering in two dimensions which vanishes logarithmically \cite{levinsen2015strongly}.
This difference can be viewed as a consequence of the strong light-matter coupling, as recently discussed in Ref.~\cite{Bleu2020}.
In the regime where the Rabi splitting exceeds the exciton binding energy $\hbar\Omega\gtrsim \eb$, as in the sample used in our experiments, we expect the Born approximation \cite{tassone1999exciton,Levinsen2019} to give a reasonable estimate of the polariton-polariton and polariton-reservoir interaction strengths \cite{li2021microscopic} as reported in experiments \cite{Estrecho2019}, and thus $g_{pd}\simeq g_{pp}$.

Finally, we note that an excitonic reservoir as described in Eq.~\eqref{eq:Heff} should not affect phenomena related to polariton-polariton interactions at the few-body level such as polariton antibunching~\cite{delteil2019,MunozMatutano2019}, since it only modifies the single-polariton energies. %rather than the two-body properties.

\subsection{Generalized Bogoliubov theory}
We now consider the situation where we have a polariton condensate, corresponding to a macroscopic occupation of the zero-momentum mode (Fig.~\ref{fig:fig1}). To this end, we introduce a generalized Bogoliubov theory of our model, Eq.~\eqref{eq:Heff}, that explicitly accounts for the effect of the incoherent reservoir on the Bogoliubov spectrum.

We start with the Heisenberg equation for the lower polariton which reads
    \begin{align}  \label{eq:Heis}
i\hbar \partial_t{\hat{L}_\k} &=[\hat{L}_\k,\hat{H}_{\rm eff}]
\\ \nonumber
&=\left(E_{\k}^L +2 \frac{ g_{pd}}{\mathcal{N}} X_\k^2 n_R\right)\hat{L}_\k
\\ \nonumber
&+\frac{g_{pp}}{\mathcal{N}\mathcal{A}}X_{\k} \sum_{\k',\q} X_{\k'+\q-\k}X_{\k'}X_{\q} \hat{L}_{\k'+\q-\k}^{\dagger} \hat{L}_{\k'}\hat{L}_{\q}.
  \end{align}  
%where $n_R=N_R/\mathcal{A}$ is the reservoir density.
Considering first the macroscopically occupied state at $\k=0$, we use the mean-field approximation which consists of defining $\langle \hat{L}_0\rangle \equiv \sqrt{\mathcal{A}}\psi_0$ and using the fact that  $N_0\equiv\mathcal{A}|\psi_0|^2\gg \sum_{\k\neq0} \langle \hat{L}_\k^{\dagger} \hat{L}_\k \rangle$. This allows us to obtain an equation for the time evolution of $\psi_0$
    \begin{align} \label{eq:psi0}
i\hbar \partial_t{\psi_0} &=\left[E_{0}^L +2 \frac{ g_{pd}}{\mathcal{N}} X_0^2 n_R+ \frac{ g_{pp}}{\mathcal{N}} X_0^4 |\psi_{0}|^2\right]\psi_{0}.
  \end{align}  
Its solution is of the form $\psi_0=\sqrt{n_0}e^{-i (E_{0}^L+\mu_T)t/\hbar}$, where $n_0=N_0/\mathcal{A}$ is the condensate density and
\begin{align} \label{eq:muT}
\mu_T &=2 \frac{ g_{pd}}{\mathcal{N}} X_0^2 n_R+ \frac{ g_{pp}}{\mathcal{N}} X_0^4 n_0
\end{align}  
%
%corresponds to the total interaction-induced blueshift of the polariton condensate, 
corresponds to the effective chemical potential %total interaction-induced energy shift 
of the polariton condensate, which contains both interactions with the homogeneous reservoir and interactions within the condensate.
%from the single-particle energy $E_0^L$,
%containing contributions from both the homogeneous reservoir and the condensate self-interaction.

\begin{figure}
    \centering
    \includegraphics[width=0.95\columnwidth]{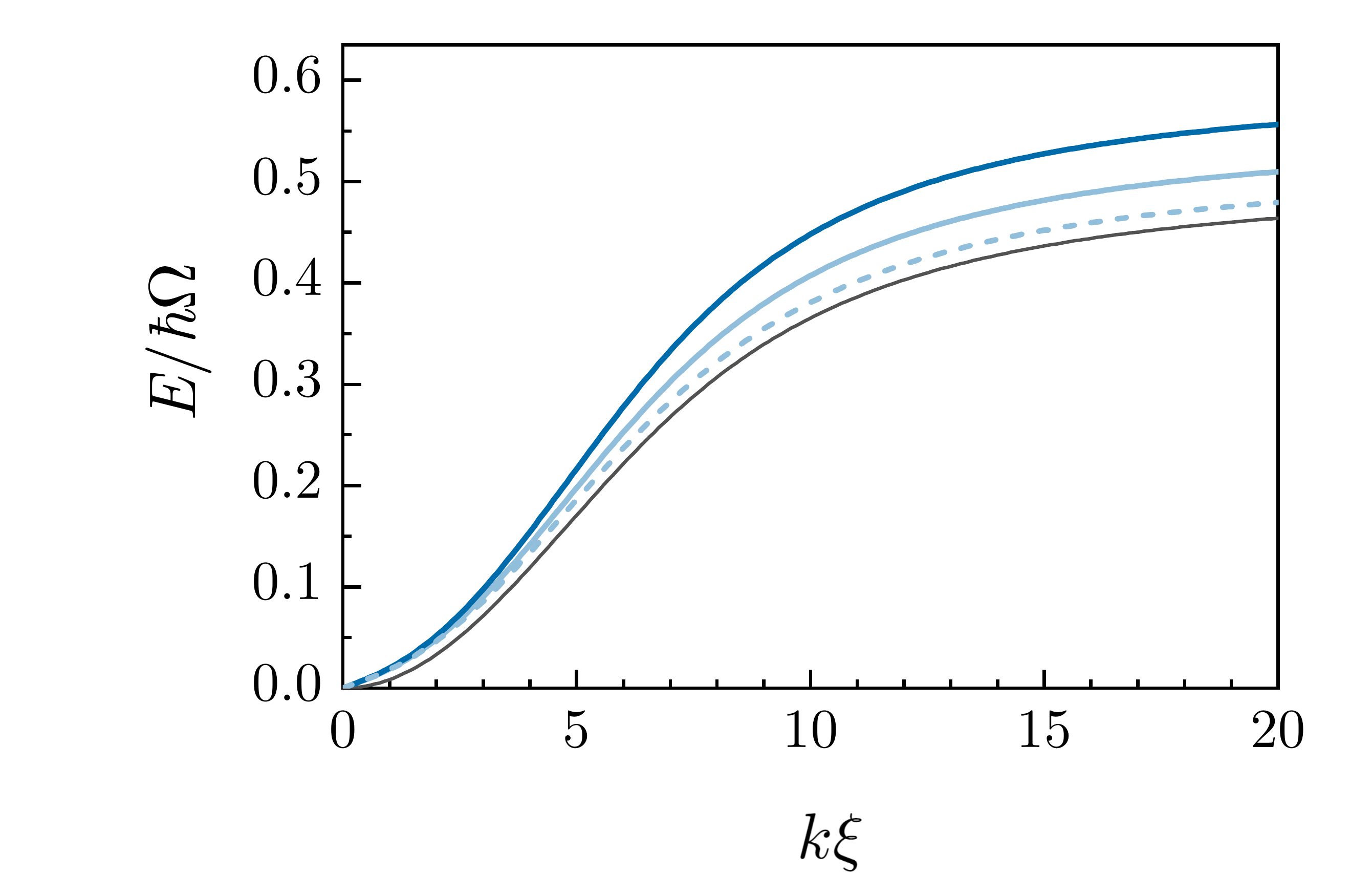}
    \caption{Polariton Bogoliubov excitation spectrum within various approximations.
The dark and light blue solid lines correspond to $\epsilon_{\bf{k}}$ obtained in Eq.~\eqref{eq:spectrum} in the presence ($\mu_T\neq\mu_C$) or absence of the reservoir ($\mu_T=\mu_C$), respectively.
For comparison, we also display the conventional Bogoliubov result $\epsilon_{\mathbf{k}}^{\rm Bg}$ with the dashed light blue line and the single polariton kinetic energy $T_\k^L$ with the thin black line.
 We have used experimentally relevant parameters: $m_C = 3.6\times10^{-5} m_0$,  $m_X=0.57m_0$~\cite{Yu2010}, $\hbar \Omega = 15.9~\text{meV}$, $\delta = 0$, $\mu_C = 250~\mu\text{eV}$; and we have taken $\mu_T = 1~\text{meV}$ for the dark blue line. The healing length is defined as $\xi=\hbar/\sqrt{m_L\mu_C}$.}
    \label{fig:fig2}
\end{figure}

For the fluctuation part at $\k\neq 0$, we make the replacement $\hat{L}_0 \rightarrow  \sqrt{\mathcal{A}}\psi_0$ in Eq.~\eqref{eq:Heis} and keep the leading order terms in $n_0$. This is equivalent to replacing  $\hat{H}_{\rm eff}$ by the mean-field Hamiltonian
\begin{align}\nonumber
\hat{H}_{\rm mf}&= %\left[E_{0}^L +2 \frac{ g_{pd}}{\mathcal{N}} X_0^2 n_R \right] \mathcal{A}|\psi_0|^2 + \frac{ g_{pp}}{2\mathcal{N}} \mathcal{A}X_0^4 |\psi_0|^4\\ \nonumber
\sum_{\k\neq 0}\! \left[E_{\mathbf{k}}^L+2 \frac{ g_{pd}}{\mathcal{N}} X_\k^2 n_R + \frac{2g_{pp}}{\mathcal{N}} X_\k^2 X_0^2 |\psi_0|^2 \right] \hat{L}_{\k}^{\dagger}\hat{L}_{\k}
\\ 
&+\frac{g_{pp}}{2\mathcal{N}} X_0^2 \sum_{\k\neq 0} X_\k^2 \left(\psi_0^2 \hat{L}_{\k}^{\dagger}\hat{L}_{-\k}^\dagger + (\psi_0^*)^2\hat{L}_{\k}\hat{L}_{-\k}\right), 
  \end{align}
which is time-dependent through the mean-field $\psi_0$. The time dependence can be absorbed into the polariton operators
via the transformation $\hat{L}_\k \rightarrow \hat{L}_\k e^{-i (E_{0}^L+\mu_T)t/\hbar}$, which effectively means that we measure energy with respect to that of the condensate. 
%
 %
%  The resulting Hamiltonian is time independent, and t
  The Heisenberg equation for the transformed operator then gives %$\hat{L}_\k e^{-i (E_{0}^L+\mu_T)t/\hbar}$ gives
     \begin{align} \label{eq:Heis1}
i\hbar \partial_t \hat{L}_{\k} = A_\k \hat{L}_{\k} + B_\k  \hat{L}_{-\k}^\dagger,
  \end{align}
%  while the one for $\hat{L}_{-\k}^{\dagger} e^{i (E_{0}^L+\mu_T)t/\hbar}$ gives
and likewise we have
       \begin{align} \label{eq:Heis2}
i\hbar \partial_t \hat{L}_{-\k}^{\dagger} =- B_\k \hat{L}_{\k}  -A_\k  \hat{L}_{-\k}^\dagger.
  \end{align}
In Eqs.~\eqref{eq:Heis1} and \eqref{eq:Heis2} we have introduced the functions
\begin{subequations}\label{eq:alpha&beta}
\begin{eqnarray}\label{eq:alpha}
 A_{\k}&=&T_\k^L+\left(\mu_T+\mu_C\right)\frac{X_\mathbf{k}^2}{X_0^2}-\mu_T,\\  \label{eq:beta}
   B_{\k}&=&\frac{\mu_C X_\mathbf{k}^2}{X_0^2},
\end{eqnarray}
\end{subequations}
where $T_\k^L=E_{\mathbf{k}}^L-E_{0}^L$ is the polariton kinetic energy and $\mu_C$ is defined as
\begin{eqnarray}\label{eq:muC}
\mu_C&=&\frac{g_{pp} X_0^4}{\mathcal{N}}n_0.
\end{eqnarray}
%Thus, $\mu_T$ corresponds to the total \textit{interaction-induced} blueshift of the polariton condensate, which accounts for both the %part related to the 
%condensate self-interaction and the interaction with the homogeneous reservoir, while 
In contrast to $\mu_T$, $\mu_C$ is the part related to the condensate self-interaction only. 
In an experiment, $\mu_T$ is measurable from the total interaction-induced energy shift of the polaritons at zero momentum (i.e., the blueshift), 
while $\mu_C$ can be extracted from the measured Bogoliubov spectrum $\epsilon_\k$ (see Fig.~\ref{fig:fig2}), 
as discussed below.

To extract the Bogoliubov spectrum, we perform the Bogoliubov transformation
\begin{subequations} \label{eq:bogtrans}
\begin{eqnarray}\label{eq:anni}
 \hat{L}_{\mathbf{k}}=u_{\k} \hat{\mathcal{L}}_\mathbf{k}-v_{\k} \hat{\mathcal{L}}_\mathbf{-k}^\dagger,   \label{eq:Lannihilate}
 \\
  \hat{L}_{\mathbf{k}}^\dagger=u_{\k} \hat{\mathcal{L}}_\mathbf{k}^\dagger-v_{\k} \hat{\mathcal{L}}_\mathbf{-k}.
\end{eqnarray}
\end{subequations}
Here, the coefficients take the form
    \begin{eqnarray}\label{eq:ukvks}
   u_{\k}^2=\frac{A_{\k}+\epsilon_{\k}}{2\epsilon_{\k}}, ~~       v_{\k}^2=\frac{A_{\k}-\epsilon_{\k}}{2\epsilon_{\k}},
 \end{eqnarray}
leading to the spectrum
  \begin{equation}\label{eq:spectrum}
   \epsilon_{\mathbf{k}}=\sqrt{A_{\k}^{2}-B_{\k}^2}.
\end{equation}
Using this transformation, one can rearrange Eqs.~\eqref{eq:Heis1}-\eqref{eq:Heis2}, to obtain the equation of motion for the Bogoliubov excitations, $i\hbar \partial_t  \hat{\mathcal{L}}_{\k} = \epsilon_\k  \hat{\mathcal{L}}_{\k}$.

%By contrast, 
If we neglect the momentum dependence of the Hopfield coefficients in Eqs.~\eqref{eq:alpha&beta} %is neglected 
(i.e., if we take $X_\k= X_0$), 
Eq.~\eqref{eq:spectrum} reduces to the conventional Bogoliubov result 
  \begin{equation}\label{eq:convBg}
   \epsilon_{\mathbf{k}}^{\rm Bg}=\sqrt{T_\k^L\left(T_\k^L+2\mu_C\right)},
\end{equation}
which is shown as a dashed line in Fig.~\ref{fig:fig2}. %Thus, 
In this approximation, the presence of a uniform excitonic reservoir does not affect the %condensate 
Bogoliubov excitation spectrum, nor the amplitudes $u_\k$ and $v_\k$ in Eq.~\eqref{eq:ukvks}. 
However, once we include the full momentum dependence of the Hopfield coefficients, 
%both the Bogoliubov amplitudes in Eq.~\eqref{eq:ukvks} and the spectrum in Eq.~\eqref{eq:spectrum} 
the pairwise interactions themselves inherit this dependence, % momentum dependent 
%such that
and the Bogoliubov excitations feature the total chemical potential %blueshift 
$\mu_T$ as well as $\mu_C$. %are affected.
In particular, in the long-wavelength limit where $\epsilon_{\mathbf{k}}\simeq \hbar c_s |\k|$, %and using $m_C/m_X\ll1$, 
the speed of sound $c_s$ is expressed as %has the modified expression %can be written as:
  \begin{equation}\label{eq:sound}
   c_s=\sqrt{1+\frac{2\mu_T}{\sqrt{\delta^2+\hbar^2\Omega^2}}}\sqrt{\frac{\mu_C}{m_L}} \equiv \sqrt{\frac{\mu_C}{m^*_L}},
\end{equation}
where we have used the fact that $m_C/m_X\ll1$ and have defined 
the lower-polariton %effective 
mass as $m_L=m_C/C_0^2$. 
%
%In Eq.~\eqref{eq:sound}, 
Since $\sqrt{\mu_C/m_L}$ is the conventional (Bogoliubov) speed of sound in a BEC composed of particles of mass $m_L$~\cite{pitaevskii2016bose}, 
%while the prefactor is a small correction. % originating from the momentum dependence of the Hopfield coefficient. 
%Thus, 
we see that Eq.~\eqref{eq:sound} corresponds to polaritons with a %modified and 
slightly smaller 
%the main effect in this limit is to slightly lower the polariton mass to the 
effective mass $m_L^*$, %slightly smaller than $m_L$, 
which includes many-body effects via $\mu_T$. 
Note that in the absence of the reservoir we have $\mu_T=\mu_C$ in which case Eq.~\eqref{eq:sound} matches the zero-temperature expression recently obtained in Ref.~\cite{Grudinina2021}. 

Similarly, the Bogoliubov amplitudes in the low-momentum limit become %are also corrected and read
  \begin{equation}\label{eq:lowkukvk}
  u_\k^2,v_\k^2\xrightarrow[k\rightarrow 0]{} %\left(1+\frac{2\mu_T}{\sqrt{\delta^2+\hbar^2\Omega^2}}\right)^{-\frac{1}{2}}
  \frac{\sqrt{m^*_L \mu_C}}{2 \hbar k},
\end{equation}
where the polariton mass $m_L$ is once again replaced by the effective mass $m^*_L$. 
%where $\sqrt{m_L \mu_C}/2\hbar k$ corresponds to the asymptotic behavior for a conventional BEC. 
Hence, while the small Hopfield correction increases the speed of sound, it reduces the corresponding Bogoliubov amplitudes compared to those for a conventional Galilean-invariant BEC~\cite{pitaevskii2016bose}.

%Furthermore, we see that 
%We note, however, that 
At larger wavevectors, the impact of the reservoir %blueshift $\mu_T$ %modification 
on the spectrum \eqref{eq:spectrum} %due to $X_{\k}\neq X_0$ 
becomes more pronounced,
%at larger wavevectors when $X_{\k}^2$ increases towards 1, 
as illustrated in Fig.~\ref{fig:fig2}, since $X_{\k}^2$ shifts away from $X_0^2$ and increases towards 1 with increasing $k$.
We emphasize that this is different from previous theories of polariton condensates coexisting with a reservoir~\cite{Wouters2007,Byrnes2012}, which only find a modified Bogoliubov spectrum close to zero momentum. By contrast, our theory predicts a visible effect of the reservoir for the wavevectors accessible in the experiment, and there is no significant departure from the shape of the conventional Bogoliubov excitation branches near $k=0$.

\subsection{Photoluminescence of Bogoliubov excitations}
\label{sec:PLtheory}

\begin{figure}
    \centering
    \includegraphics[width=\columnwidth]{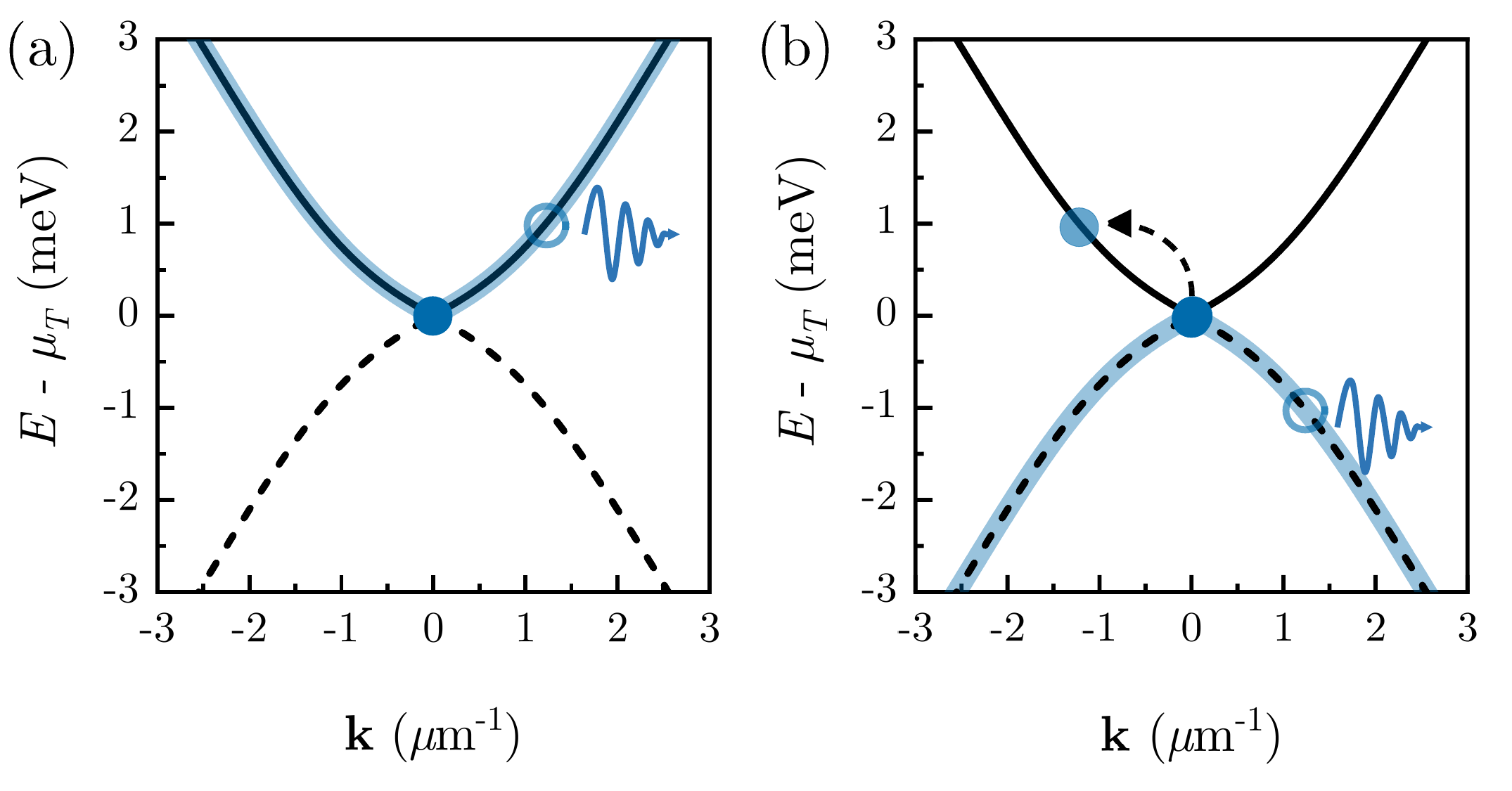}
    \caption{%Contributions to photoluminescence from the polariton excitation spectrum. 
    Schematics of the processes contributing to the photoluminescence spectrum at wavevector $\k$ and energy $E$ (indicated by an empty circle).
  %  There are two processes related to the the decay of one polariton. %at the branches to be 
 %   detected in the photoluminescence.
(a) Annihilation of a Bogoliubov excitation with photon emission from the normal branch. (b) Creation of a Bogoliubov excitation with photon emission from the ghost branch.  %Emission of a photon from a GB with simultaneous creation of a Bogoliubov excitation in the NB (of opposite $\k$).
The black solid and dashed lines correspond to the Bogoliubov excitation spectrum (normal branch) and its mirror image (ghost branch), respectively. The filled blue circle at $k=0$ represents the polariton condensate, and the blue shading highlights the possible luminescence signal associated with the two processes.}
    \label{fig:fig3}
\end{figure}

To connect the theoretical results with the experiment, we need to relate these to the observed photoluminescence. 
In practice, the photoluminescence spectrum is measured by detecting the photon emitted when a polariton is annihilated. We now calculate the photoluminescence within our model using Fermi's golden rule.

It is instructive to first consider the system
in the absence of polariton-polariton interactions.
%absence of condensate, when interactions can be neglected, 
In this case, the initial and final states for this process can be expressed in term of the single polariton Fock states. The photoluminescence spectrum can then be defined as $C_{\k}^2 I(\hbar \omega,\k)$ with $C_{\k}^2$ the polariton photonic fraction and 
\begin{align}\label{eq:luminonint}
I(\hbar\omega, \k)& \propto 
 \sum_{n_\k} p_{n_\k}\abs*{ \bra{n_\k-1,1}\hat{a}^\dagger_\k\hat{L}_\k\ket{n_\k,0} }^2\delta(\hbar\omega-E_\k^L),
\end{align}
where $\hat{a}^\dagger_\k$ is the emitted photon creation operator (note that for notational simplicity, we denote this only by its planar momentum). Here, $\ket{n_\k,0}$ corresponds to an initial state with $n_\k$ polaritons with wavevector $\k$ and zero emitted photons, and $p_{n_\k}$ is the probability to find this occupation. Equation~\eqref{eq:luminonint} can also be written as
\begin{align}\label{eq:luminonint2}
I(\hbar\omega, \k)& \propto 
 \langle\hat{L}^\dag_\k\hat{L}_\k\rangle\delta(\hbar\omega-E_\k^L).
\end{align}
where $\langle\hat{L}^\dag_\k\hat{L}_\k\rangle=\sum_{n_\k} p_{n_\k} n_\k$.
This shows that, in the absence of interactions, we only have a single branch in the photoluminescence spectrum.

A similar reasoning can be used to interpret the photoluminescence in the presence of interactions in the condensate. In the Bogoliubov approximation, the transformation in Eq.~\eqref{eq:Lannihilate} shows that the annihilation of a polariton at wavevector $\k$ is associated with two different processes: the annihilation of a collective excitation at $\k$ or the creation of an excitation at $-\k$, as illustrated in Fig.~\ref{fig:fig3}.
Both processes contribute to the photoluminescence spectrum as
\begin{align}\nonumber
 I_{B}(&\hbar\omega, \k) \propto u_\k^2\sum_{n_\k}  p_{n_\k}\abs*{\bra{n_\k-1,1}\hat{a}^\dagger_\k\hat{\mathcal{L}}_\k\ket{n_\k,0}}^2\delta(\hbar\omega-\epsilon_\k)
 \\ 
 &+v_\k^2\sum_{n_{-\k}} p_{n_{-\k}}\abs*{\bra{n_{-\k}+1,1}\hat{a}^\dagger_\k\hat{\mathcal{L}}_{-\k}^{\dagger}\ket{n_{-\k},0}}^2\delta(\hbar\omega+\epsilon_{-\k}),
\end{align}
where $\ket{n_\k,0}$ now denotes an initial Fock state with $n_\k$ Bogoliubov excitations and zero photons.
Thus, the photoluminescence spectrum can be expressed as:
\begin{align}\label{eq:PL}
I_{B}(\hbar\omega, \k) &\propto  u_\k^2 \langle\hat{\mathcal{L}}_\k^{\dagger}\hat{\mathcal{L}}_\k\rangle \delta(\hbar\omega-\epsilon_{\k})
\\ \nonumber
&+  v_\k^2  ( \langle\hat{\mathcal{L}}_{-\k}^{\dagger}\hat{\mathcal{L}}_{-\k}\rangle+1)  \delta(\hbar\omega+\epsilon_{-\k}).
\end{align}
We can see that it is composed of two branches symmetric with respect to the condensate chemical potential (which is defined as the origin here). We denote the positive one as the normal branch (NB) and the negative one as the ghost branch (GB) \cite{Pieczarka2015}. Note that the ghost branch is not a real excitation branch of the system: 
As depicted in Fig.~\ref{fig:fig3}(b), it corresponds to the emitted photons associated with the creation of real Bogoliubov excitations in the polariton system, and as such it is related to the photoluminescence process itself (i.e., to the photons escaping %outcoupling %coupling to the outside of 
the cavity).
\ob{Moreover, it is an intrinsic feature of an interacting condensate and it does not require the presence of a reservoir.}
In the absence of polariton-polariton interactions ($\mu_C=0$), we have $v_\k^2=0$ such that only the normal branch emits, in agreement with Eq.~\eqref{eq:luminonint2}.

The NB and GB occupations can be expressed as
\begin{subequations}\label{eq:NB_GB}
\begin{align}
&N_{NB,\k}= u_\k^2  \langle\hat{\mathcal{L}}_\k^{\dagger}\hat{\mathcal{L}}_\k\rangle,  \\
&N_{GB,\k}=  v_\k^2 (\langle\hat{\mathcal{L}}_{-\k}^{\dagger}\hat{\mathcal{L}}_{-\k}\rangle+1).
\end{align}
\end{subequations}
For large wavevectors with respect to the inverse healing length $1/\xi=\sqrt{m_L \mu_C}/\hbar$ we have $u_\k^2\rightarrow1$, and therefore $N_{NB,\k}\simeq  \langle\hat{\mathcal{L}}_\k^{\dagger}\hat{\mathcal{L}}_\k\rangle $ %and $N_{GB,\k}$ takes the form
while
\begin{eqnarray}\label{eq:NGB}
N_{GB,\k}\simeq v_\k^2 \left(N_{NB,-\k}+1\right).
\end{eqnarray}
Therefore, the ratio $N_{GB,\k}/ (N_{NB,-\k}+1)$ allows one to extract the coefficient $v_\k^2$ directly from the occupation number of the two branches which can be measured experimentally \cite{Pieczarka2020}.

Equations~\eqref{eq:NB_GB} and \eqref{eq:NGB} are general expressions that apply to any uniform system in dynamical equilibrium.
For the case of %a uniform system in 
thermal equilibrium with a well-defined temperature $T$, one would have $\langle\hat{\mathcal{L}}_\k^{\dagger}\hat{\mathcal{L}}_\k\rangle=\langle\hat{\mathcal{L}}_{-\k}^{\dagger}\hat{\mathcal{L}}_{-\k}\rangle=(\exp(\epsilon_\k/k_\text{B}T)-1)^{-1}$ in Eq.~\eqref{eq:NB_GB}. At $T=0$, %absolute zero temperature 
this would give $N_{NB,\k}^{\text{th.}}=0$ while $N_{GB,\k}^{\text{th.}}= v_\k^2$ because of quantum depletion \cite{pitaevskii2016bose}, and only the ghost branch would be observable.

  \section{Experimental results}

We now present our experimental results which are obtained from an analysis of the measured photoluminescence based on the theory introduced above.
We begin with %the 
an introduction of the experimental setup and method in Sec.~\ref{sec:ExpDet}. Then, we present the measurement of the polariton condensate and dark reservoir densities in Sec.~\ref{sec:dens}, and finally we present the extraction of the Bogoliubov amplitude from the measured momentum resolved occupations of the excitation branches in Sec.~\ref{sec:occ}.

  \begin{figure}
    \centering
    \includegraphics[width=\columnwidth]{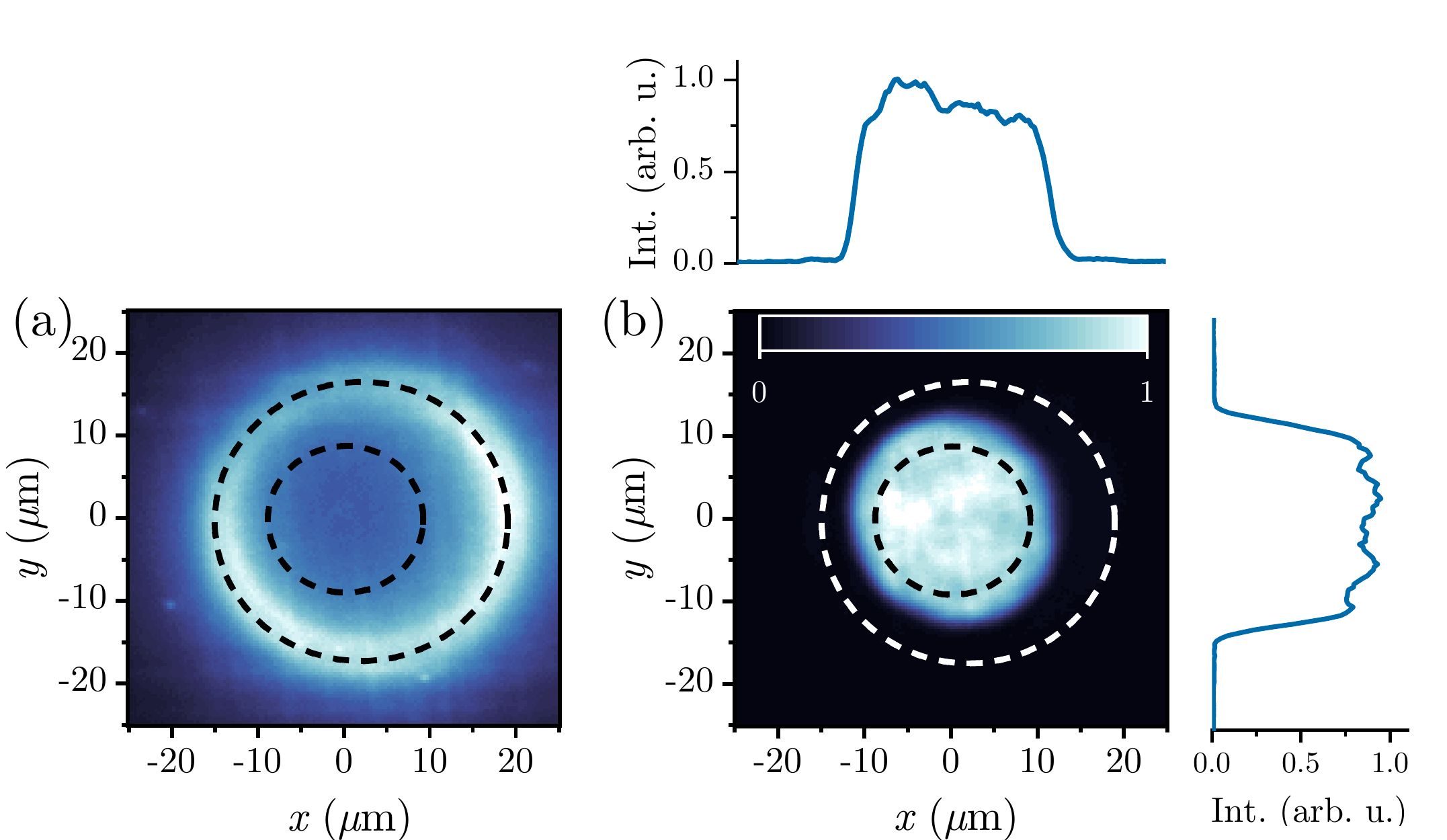}
    \caption{(a) Real-space image of the polariton emission at very low excitation power. In this regime, the emission is concentrated around the circular laser excitation profile. The mean polariton density inside the trap is $0.04~\mu\text{m}^{-2}$. (b) Real-space image of the spatial distribution of the high-density polariton condensate in the Thomas-Fermi regime ($n_0 =  1390~\mu\text{m}^{-2}$), showing a flat density distribution with sharp edges of the trap. The condensate density is spatially modulated due to the local disorder of the sample, as seen in cross sections at $y=0$ (top panel) and $x=0$ (right panel). The large dashed circle indicates the position of the laser excitation and the small dashed circle depicts the position of the spatial filter used to probe the flat part of the density.}
    \label{fig:fig4}
\end{figure}
  
  \subsection{Sample and methods}
  \label{sec:ExpDet}
 % We investigate experimentally the excitation spectrum of a high-density condensate in a trapped geometry (Fig.~\ref{fig:fig2}), which was successfully employed in our previous works \cite{Pieczarka2020,bieganska2020collective,Estrecho2021}. In the present work, we have developed an improved experiment, which is described below.
  
Our sample is an ultra-high-quality GaAs-based microcavity, ensuring a long cavity photon lifetime of at least 100~ps \cite{Estrecho2021,Steger2013}. The active region of the $3\lambda/2$ cavity consists of $\mathcal{N} = 12$ GaAs/AlAs quantum wells (7 nm thick), placed at the antinodes of the confined optical field. The high-reflective mirrors are composed of multilayer AlAs/AlGaAs distributed Bragg reflectors. QW excitons strongly couple to the cavity photons with a Rabi splitting $\hbar \Omega = 15.9~\text{meV}$. The data presented in the manuscript is taken for a photon-exciton detuning $\delta = E^C_0-E^X_0 \approx 1.39~\text{meV}$, corresponding to an exciton fraction $X_0^2\approx 0.543$. Further details on this sample can be found elsewhere \cite{Nelsen2013,Pieczarka2020}.

We investigate polaritons in the regime of bosonic condensation in an optically induced potential trap \cite{Pieczarka2019,Estrecho2021,Paschos2020,Orfanakis2021,Askitopoulos2013} --- see Fig.~\ref{fig:fig4}. This is done by focusing the pump laser (a single-mode continuous-wave Ti:Sapphire laser) into a ring shape on the sample via an axicon lens in between a pair of lenses in a confocal configuration~\cite{Pieczarka2019}. The nonresonant laser, tuned to the second reflectivity minimum of the microcavity around $1.715~\text{eV}$, creates high-energy electron-hole pairs, which relax into excitons and polaritons under the pump spot. %
 Simultaneously, the accumulated carriers and excitons in the pump region interact repulsively with polaritons, forming an effective circular trap within which polariton condensation occurs. 
The use of an optical trap geometry ensures a minimal overlap between the condensate and the pumping region, reducing potential decoherence effects \cite{Orfanakis2021,Askitopoulos2013}.
 In the configuration used here, the trap diameter is approximately $34~\mu\text{m}$.

  The photoluminescence of the polariton condensate is collected via a high numerical aperture objective ($\rm{NA} = 0.5$) and the real-space (near-field) or momentum-space (far-field) photon emission %distributions 
  are focused via a set of four lenses onto the slit of the imaging monochromator. The condensate is prepared in a high-density Thomas-Fermi (TF) regime \cite{Estrecho2019,Pieczarka2020}, where effects of polariton interactions and spatial hole-burning lead to a nearly homogeneous condensate with sharp edges, as seen in Fig.~\ref{fig:fig4}(b). Since the TF condensate wavefunction directly reflects the actual shape of the effective trapping potential, one observes a small modulation %of the homogeneous part 
  of the condensate density due to local sample disorder. To ensure that the local density approximation applies to our measurements, we only probe the central part of the condensate, away from the trap edges. 
  %Hence, ensuring the rightness of the TF approximation in the analysis, 
  %we probe only the central part of the condensate. 
  This is achieved with a circular spatial filter of diameter about $18~\mu\text{m}$ in the real-space plane, as depicted in Fig.~\ref{fig:fig4}. We follow the signal integration method in momentum space employed in our previous works~\cite{Pieczarka2020, Pieczarka2019} to measure the mean condensate density $n_0$ in this homogeneous region.
Throughout this paper, the quoted densities are those for %expressed in 
  a single spin component ($\uparrow$ or $\downarrow$) associated with the photon circular polarisation. As the condensates in our samples are nearly 100\% linearly polarized, we have $n_\uparrow \simeq n_\downarrow \equiv n_0$
  such that the total polariton density %in a linearly polarized condensate 
  is $n_{tot}=n_\uparrow +n_\downarrow \simeq 2 n_0$.
  %so we denote $n_\uparrow \simeq n_\downarrow \equiv n_0$ and $n_{tot} = 2 n_0$. 
  We can use this notation because of the negligible contribution of the singlet ($\uparrow\downarrow$) interactions in our multi-QW GaAs cavity~\cite{bieganska2020collective}. 

According to the arguments in Sec.~\ref{sec:PLtheory} and previous theoretical works~\cite{Szymanska2006,Byrnes2012,Hanai2018,Doan2020}, the collective excitation spectrum of a polariton condensate is visible in the photoluminescence from the cavity.  Indeed, signatures of collective excitations have been reported  
%This difficulty has been resolved 
in experiments with nonresonant pumping~\cite{Utsunomiya2008,Pieczarka2015,Pieczarka2020,Ballarini2020}, %and we also note that signatures of collective excitations have been reported 
and in the different regime of coherent excitation~\cite{Kohnle2011,Zajac2015,Stepanov2019}.
In practice, however,
%experimental realizations, 
the strong signal of the $k=0$ state dominates the luminescence and saturates the detection system, making it challenging to detect the weak signal from the excitation branches. 
In our experimental setup, we solve this problem by using an edge filter in momentum space to block the luminescence below a given wavevector $k_{\rm filter}$ \cite{Pieczarka2020,bieganska2020collective}, allowing us to resolve the weak signal originating from the excitation branches at $k>k_{\rm filter}$.
We also probe higher condensate densities than the recent experiments of Refs.~\cite{Ballarini2020,Stepanov2019,Pieczarka2020}.
%Additionally, the trapped geometry ensures a minimal overlap with the pumping regions, reducing the possible detrimental effects of decoherence \cite{Orfanakis2021,Askitopoulos2013}.

\begin{figure}[t]
    \centering
    \includegraphics[width=0.95\columnwidth]{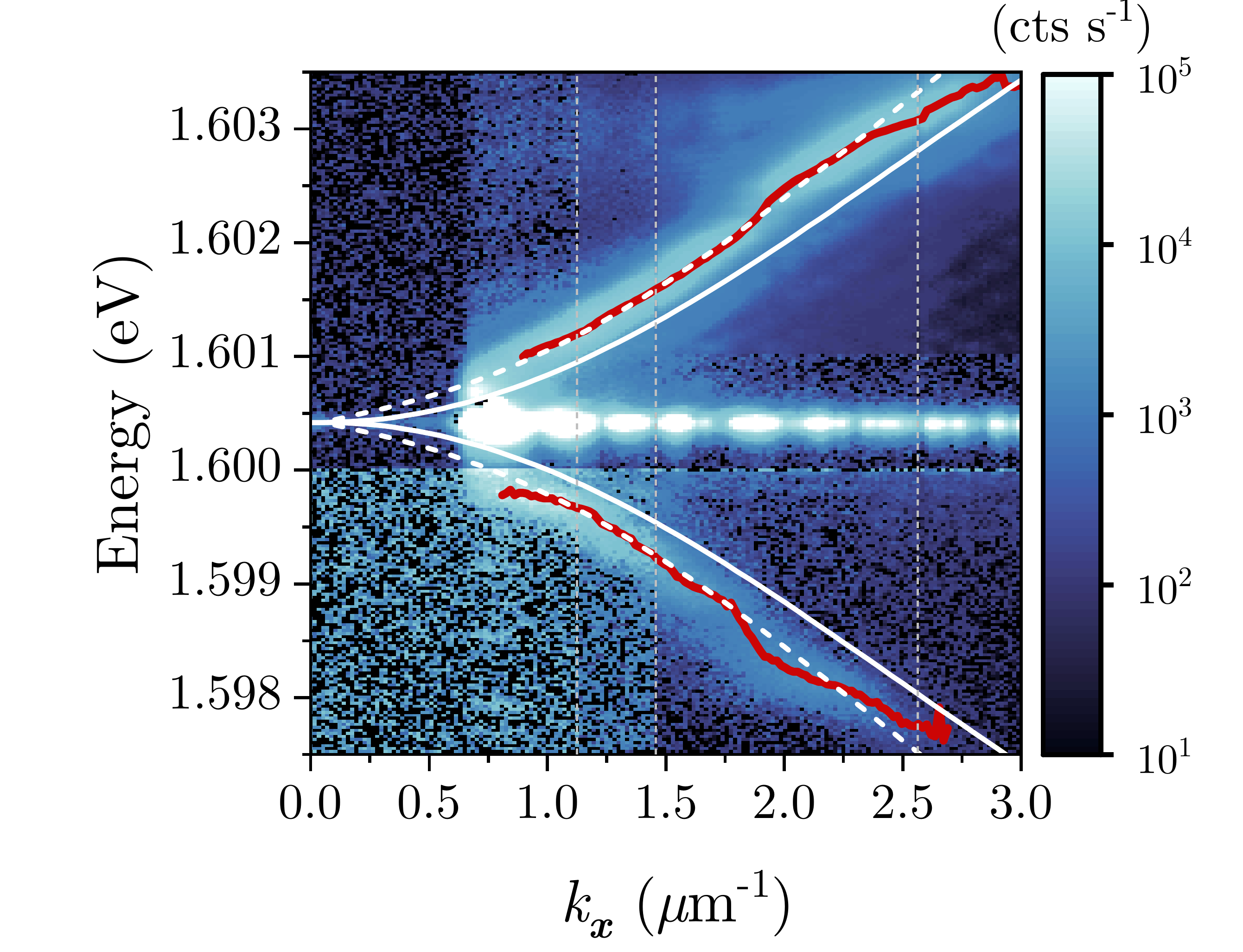}
    \caption{Momentum-space photoluminescence %of a Bogoliubov excitation 
    spectrum of a high-density polariton condensate recorded at $n_0 = 907~\mu\text{m}^{-2}$. This image is obtained by stitching together spectra along the $x$ direction recorded with different positions of a momentum edge filter (indicated with the thin dotted vertical lines). Red thick lines represent the extracted intensity maxima of the branches.  The white dashed lines represent the collective excitation dispersion obtained from the fit to Eq.~\eqref{eq:spectrum} and its mirror image. The white solid lines correspond to the single polariton dispersion and its mirror image, both shifted by $\mu_T$. A strong Airy pattern originating from the diffracted condensate emission at $k=0$ is visible at $\sim1.6005~\text{eV}$. The color scale is logarithmic.}
    \label{fig:fig5}
\end{figure}

%In the present work, we have upgraded the detection part of the experiment as well.
The polariton photoluminescence spectra are collected with the momentum-edge filter placed at different places in $\k$-space. Additionally, we adjust the acquisition time in each configuration, enabling us to increase the total dynamical range of the detection --- see the data in Fig.~\ref{fig:fig5}.
%The experimental setup allows for setting up the minimal signal rejection area in the momentum space to be $|\k|<|\k_{\rm filter}|\approx 0.65 \mu\text{m}^{-1}$. 
%This value is forced
Because the remaining diffracted signal from the condensate can over-saturate the CCD, the minimal wavevector for the detection is $k_{\rm min}\approx 0.65~\mu\text{m}^{-1}$. 
%
%Although the low-momentum part of the spectrum is still impossible to be probed, we extended the measured $\k$-region to larger $\k$ values, beyond the inflection point of the polariton dispersion, where the effective mass approximation does not hold. 
%
In comparison to our previous work \cite{Pieczarka2020}, here we are able to measure the spectrum at larger wavevectors as well as access both positive and negative wavevectors along a given direction.
%
%Moreover, this configuration also enables us to measure the excitation spectrum for both positive and negative wavevectors satisfying $|\k|>|\k_{\rm filter}|$ along a given direction.
The latter capability is essential for the experimental extraction of the Bogoliubov amplitude presented in Sec.~\ref{sec:occ}.
All measurements in this work are performed along the $x$-axis (oriented as in Fig.~\ref{fig:fig4}), colinear with the monochromator slit.

\subsection{Polariton condensate with large reservoir}\label{sec:dens}

  A typical photoluminescence image is presented in Fig.~\ref{fig:fig5}.
The strong bright signal at $\sim 1.6005$~eV corresponds to the condensate emission diffracted at the edges of the spatial filter \cite{Pieczarka2020}. The spectral position of the condensate with respect to the original single-polariton energy allows us to measure the total blueshift, which corresponds to the effective chemical potential $\mu_T$ in Eq.~\eqref{eq:muT}.
%The lower polariton dispersion $E_{\mathbf{k}}^L$ and the Hopfield coefficient $X_\mathbf{k}^2$ are extracted separately by fitting the single-particle photoluminescence spectrum at very low pump power and using Eq.~(\ref{eq:polDispersion}).
   %
   To obtain the condensate self-interaction energy $\mu_C$, we first extract the peak positions of the two excitation branches by fitting the spectra at each $k_x$ with Lorentzian functions. The resulting peak dispersions are shown with red thick lines in Fig.~\ref{fig:fig5}. We then fit these dispersions with Eq.~(\ref{eq:spectrum}), inserting the value of $\mu_T$ previously obtained from the blueshift and  using the condensate self-interaction energy $\mu_C$ as a fitting parameter. 
    %
    %The result of the fit is presented with the dashed line in Fig.~\ref{fig:fig4}. The zero-density limit of the dispersion is also shown as solid white lines for comparison, highlighting the energy shifts in dispersions. 
    %
       In the fitting procedure, we use the smooth part of the measured excitation spectrum below $k_x\approx 2~\mu\text{m}^{-1}$. 
    The result of the fit is presented with the dashed white lines in Fig.~\ref{fig:fig5}. The single-particle limit of the dispersion is also shown with the solid white lines for comparison. % highlighting the energy shifts in dispersions. 

\begin{figure}[t!]
    \centering
    \includegraphics[width=0.95\columnwidth]{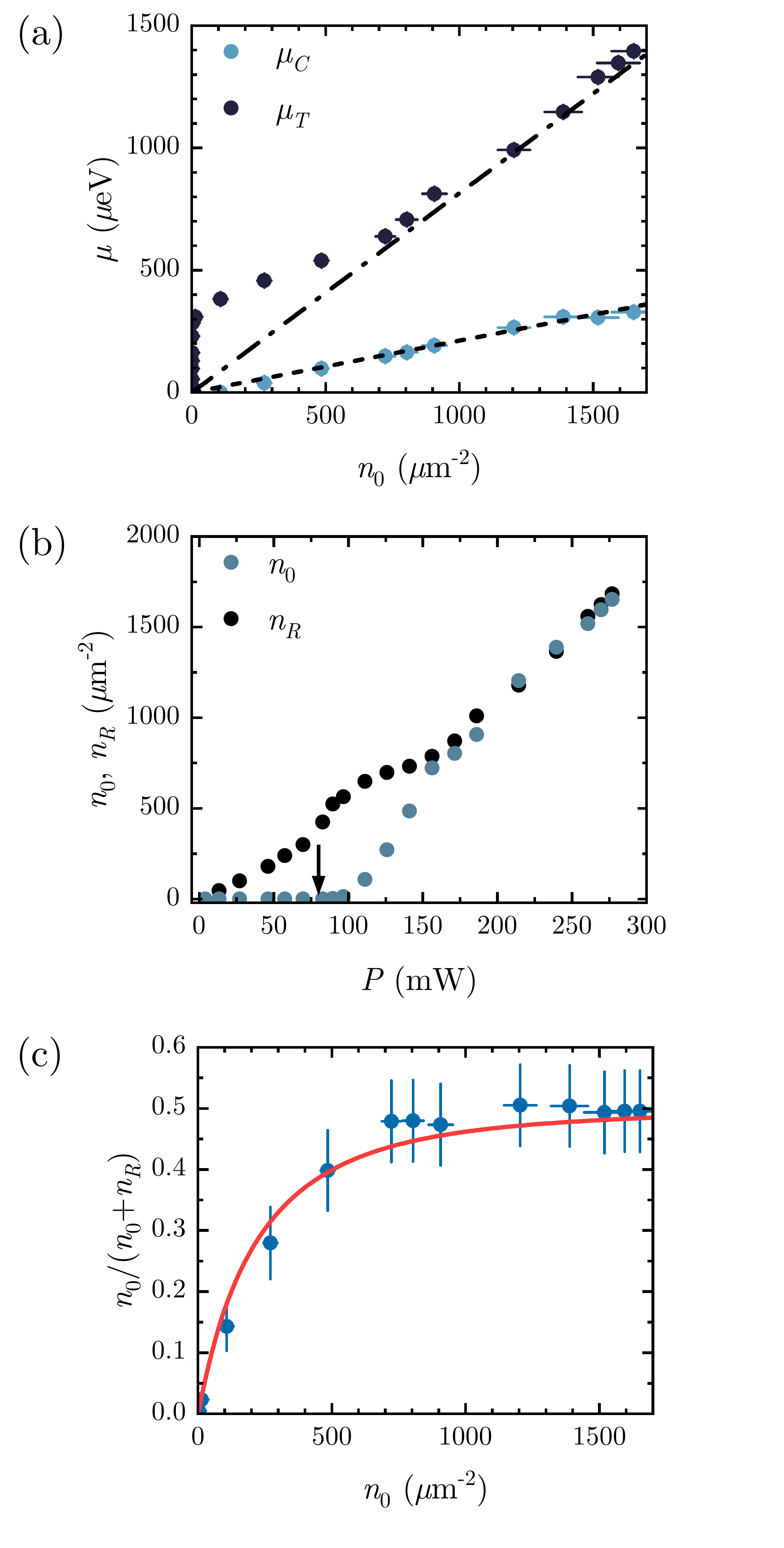}
    \caption{(a) Measured total blueshift $\mu_T$ and condensate self-interaction energy $\mu_C$ as a function of the polariton density. The dotted and dashed-dotted lines highlight the linear behaviors.
    %  The dotted line depicts the linear behaviour of the chemical potential. The dashed-dotted line indicates the linear asymptotic behavior of the total blueshift. 
(b) Condensate and reservoir densities ($n_0$, $n_R$), deduced from the data in (a) (see text),
as a function of the pumping power $P$. The threshold $P_{th}$ is indicated with an arrow. (c) Condensed fraction as a function of $n_0$. The red solid line is obtained from the model of Eqs.~\eqref{eq:rate} using experimentally relevant parameters $\gamma_R = 1/5000~$ps$^{-1}$, $\gamma_C = 1/135~$ps$^{-1}$, $\gamma =\gamma_R X_0^2 + \gamma_C (1-X_0^2)$, \ob{and using the fitted values} $R=W=2\times 10^{-8}~\mu$m$^4$ps$^{-1}$.}
    \label{fig:fig6}
\end{figure}

    We repeated the above procedure for several condensate densities, $n_0$, and the resulting measured total blueshift $\mu_T$ and condensate self-interaction energy $\mu_C$ are presented in Fig.~\ref{fig:fig6}(a).  
   The blueshift $\mu_T$ shows an interesting behaviour in which we can distinguish three stages: For very low polariton densities, below the condensation threshold, $\mu_T$ rises quickly and is dominated by the energy shift due to the reservoir~\cite{Pieczarka2019}. 
    Above the threshold, when $n_0$ is not too large, this rise slows down, which is correlated with the condensate growth. More surprisingly, for large densities $n_0\gtrsim 750~ \mu\text{m}^{-2}$,
    the slope for $\mu_T$ changes again and becomes approximately constant in our power range.
    By contrast, $\mu_C$ always grows linearly with a given slope $\mu_C = g_{\text{exp}}n_0$.
    This slope %$\mu_C = g_{\text{exp}}n_0$, which 
    corresponds to the effective polariton-polariton interaction strength $g_{\text{exp}}=g_{pp}X_0^4/\mathcal{N}= 0.239\pm0.008~\mu\text{eV}\mu m^2$, which agrees very well with previous measurements in our sample~\cite{Estrecho2019, Pieczarka2020}. 
    %The slope of $\mu_T$ is discussed below.  

Since $g_{pd}\simeq g_{pp}$ for our sample (see Sec.~\ref{sec:interactions}), 
we can estimate the reservoir densities $n_R$ using the expression for the total blueshift, Eq.~(\ref{eq:muT}), with the calibrated values of $n_0$ and the measured $g_{\text{exp}}$.
   %Essentially, we infer the reservoir density $n_R$.
   The result of this calculation is shown in Fig.~\ref{fig:fig6}(b), where both $n_0$ and $n_R$ are presented as a function of the pump power. 
   %
   %Figure \ref{fig:fig6}(c) shows 
   We also plot the condensate fraction $\rho = n_0/(n_0 + n_R)$ as a function of $n_0$ in Fig.~\ref{fig:fig6}(c).
   %
  % The extracted condensed fraction shows an interesting behaviour. 
  It grows rapidly from zero at low values of $n_0$, as the condensate forms above the threshold ($P_{th} \approx 76 \text{mW}$ in Fig.~\ref{fig:fig6}(b)), and increases with density until it saturates at around 50\% when $n_0 \gtrsim 750~\mu \text{m}^{-2}$.
   The saturation of $\rho$ is correlated with the change of slope of $\mu_T$ in Fig.~\ref{fig:fig6}(a) and is also visible in Fig.~\ref{fig:fig6}(b) at pump powers $P \gtrsim 170~\text{mW}$. %the densities dependence on the pumping power. %The slope 
Remarkably, this means that at large densities, the proportions of reservoir and condensed polaritons are approximately equal, $n_R\simeq n_0$. 
  
\begin{figure}
    \centering
    \includegraphics[width=0.95\columnwidth]{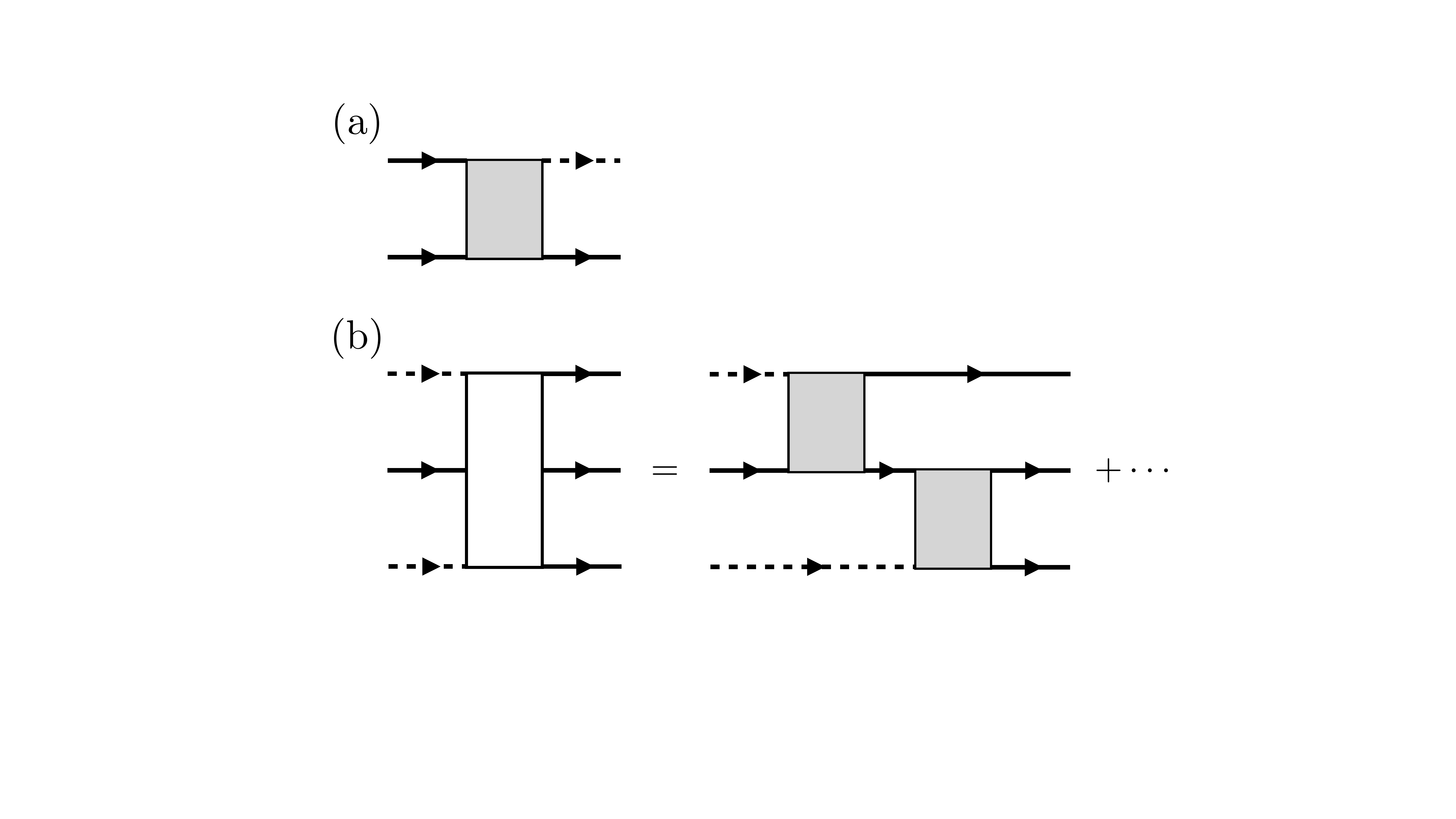}
    \caption{\ob{Diagrams contributing to (a) stimulated scattering into the condensate and (b) the upconversion of two polaritons to reservoir particles due to scattering with a third particle from the reservoir. Solid and dashed lines represent reservoir particles and condensed polaritons, respectively, while the gray squares correspond to repeated two-particle scattering processes. In (b), the diagram on the left is the sum of all possible contributions, while the diagram on the right is the simplest contribution in terms of two consecutive polariton-exciton scattering processes. 
    The lines on the left and right correspond to the initial and final states, respectively, while the central part of the diagram corresponds to the interaction matrix element that would enter Fermi's golden rule.}}
    \label{fig:fig7}
\end{figure}

   To our knowledge, the behavior we observe at large densities has not been reported in previous experiments. Furthermore, it differs from the predictions of commonly used phenomenological models for polariton condensation under continuous and incoherent pumping in which the reservoir density is locked to its value at the condensation threshold while the condensate density grows with the pump power \cite{Wouters2007,Smirnov2014,Bobrovska2015}. This suggests that there is a higher order process %, e.g., 
   involving the scattering between two condensed polaritons and a reservoir particle that transfers population from the condensate to the reservoir \ob{(Fig.~\ref{fig:fig7}). Similar three-body processes are commonly considered in cold atomic gases, where they play a dominant role in the loss of particles from an atomic condensate \cite{Bedaque2000,Burt1997,Makotyn2014}.}
    \ob{The high-density behavior can be captured by} the following set of phenomenological rate equations:
\begin{subequations}\label{eq:rate}
   \begin{align}
        \pdv{n_0}{t} & = - \gamma n_0 + R n_R^2 n_0 - W n_0^2 n_R, \\
        \pdv{n_R}{t} & = \mathcal{P} - \gamma_R n_R - R n_R^2 n_0 + W n_0^2 n_R .
    \end{align}  
\end{subequations}
Here, $\mathcal{P}$ is an effective reservoir pump rate at the center of the trap, originating in our experiment from a non-resonant pump that injects uncorrelated electron-hole pairs,
%\jfl{Here, $\mathcal{P}$ is an effective reservoir pump rate at the center of the trap, originating in our experiment from a non-resonant pump that injects uncorrelated electron-hole pairs, }
%    where $\mathcal{P}$ is an effective pump rate at the center of the trap, 
while $\gamma$ and $\gamma_R$ are the decay rates of the polariton condensate and excitonic reservoir, respectively.  %We note that $\mathcal{P}$ should be understood as an effective pump rate acting at the center of the trap.  
    The terms with coefficient $R$ correspond to the usual gain terms that populate the condensate via stimulated scattering of two particles from the reservoir. The additional $W$ terms encode the higher order scattering processes described above.
    In the steady state, the threshold pump power for condensation is $\mathcal{P}_{th} = \gamma_R \sqrt{\gamma/R}$, below which we have $n_0 = 0$ and $n_R = \mathcal{P}/\gamma_R$. For $\mathcal{P} > \mathcal{P}_{th}$, the condensate density grows and at large densities we eventually obtain $n_0/n_R \simeq R/W$, which agrees with the dependence in Fig.~\ref{fig:fig6}(b,c) if we take $R \sim W$. 
    The steady state solutions above threshold capture the saturation of the condensate fraction $\rho$ observed in the experiment, 
  %  Numerical solution of this model shows a very well correspondence to the experiment, 
  as shown in Fig.~\ref{fig:fig6}(c) with the solid line. Here we calculated $\rho$ using experimentally relevant decay rates $\gamma$, $\gamma_R$ and assuming $R=W$ in Eq.~\eqref{eq:rate}.

\ob{Figure~\ref{fig:fig7} illustrates the two- and three-body scattering processes contributing to $R$ and $W$. These in general depend on the momentum distribution of the reservoir, which we cannot measure, and furthermore a full calculation of the three-body diagram shown in panel (b) is a complicated multi-body scattering problem which takes us beyond the scope of this work.
However, since the Rabi splitting and the exciton binding energy scales are comparable in our sample}, we can estimate the expected size of the parameters $W$ and $R$ from simple dimensional analysis \ob{in terms of the exciton parameters. This is reasonable} if we assume that they are determined by polariton-polariton and polariton-exciton scattering events.
This \ob{estimate} leads to $\hbar R \sim \hbar W\sim \eb a_0^4$, 
in terms of the exciton binding energy, $\eb$, and the Bohr radius $a_0$. Using the experimental values $\eb\simeq 10 ~ \rm{meV}$ and  $a_0\simeq 10~ \rm{nm}$, we find $R\sim W\sim 1.5\cdot 10^{-7} ~\mu\rm{m}^4\rm{ ps}^{-1}$ which is within an order of magnitude of the value extracted from the data in Fig.~\ref{fig:fig6} \ob{($2 \cdot 10^{-8}~\mu\rm{m}^4\rm{ ps}^{-1}$)}. 
This suggests that these terms are consistent with those derived from excitonic scattering processes. %that expected for a 
%This shows that our extracted value of $W$ is within an order of magnitude of that expected for a few-body scattering process.

 The large reservoir density within the trap
    %The unexpectedly large reservoir contribution 
contrasts with the expectation that separating the pump region from the condensate in an optical trap geometry should result in a
%much lower
reservoir density negligible with respect to that of the condensate. 
    On the contrary, our observation indicates that excitonic particles forming the reservoir are able to move away from the pumped area, as measured in Ref.~\cite{Myers2018}, and that the threshold for condensation is reached once a substantial reservoir is present within the trap. While our multi-QW cavity is more complex, we note that the formation of an exciton gas at the center of a %the 
    trap induced by laser excitation has previously been reported in experiments with coupled GaAs quantum wells~\cite{Hammack2006}.

The large reservoir density observed here might be indicative of the dark exciton superpositions present in our multilayer cavity, as highlighted in Sec.~\ref{sec:Th}. Indeed, as the single-particle states are degenerate %and equivalent 
at large momenta, they must all be populated, and hence
%it would be incorrect to assume that 
the reservoir within the trap consists of both bright and dark superpositions.  
Since our measurements of $\mu_T$ and $\mu_C$ do not allow us to distinguish between the contribution to $n_R$ arising from the $\mathcal{N}-1$ dark superpositions %with $|\k|<q_0$ 
and the contribution from the commonly assumed bright ones occupying large momentum states with $k\gtrsim q_0$, future investigations in this direction would be beneficial for a better understanding of the role of the dark superpositions.
Interestingly, %we note that 
our extracted value for the condensate fraction $\rho\sim 0.5$ at large densities is comparable to the ones reported recently in different experiments performed in multi-QW cavities, such as under off-resonant pulsed excitation in our sample~\cite{Estrecho2021} and under resonant continuous-wave excitation in another experiment~\cite{Stepanov2019}.

\ob{We emphasize that the use of the generalized Bogoliubov spectrum, Eq.~\eqref{eq:spectrum}, with different $\mu_T$ and $\mu_C$ was crucial in the present analysis. Neither the observed Bogoliubov dispersions nor the amplitudes presented in the next section could have been reproduced if we had set $\mu_C$ equal to the observed blueshift $\mu_T$ as in the usual Bogoliubov theory.}

\begin{figure}[t!]%[htp]
    \centering
    \includegraphics[width=\columnwidth]{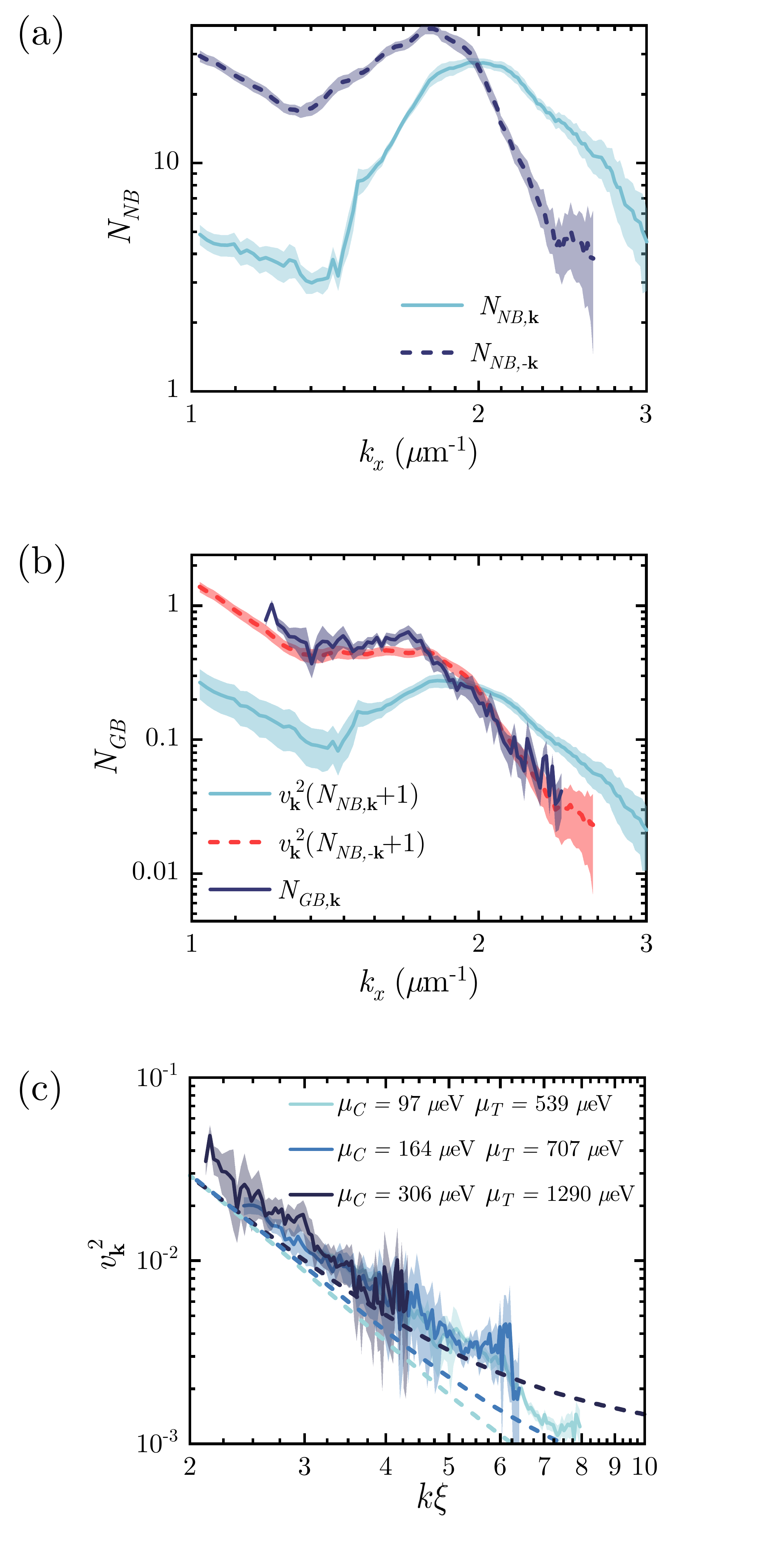}
    \caption{(a,b) Momentum-space occupations measured along the $x$-axis for $n_0 = 1519~\mu\text{m}^{-2}$ ($1/\xi\simeq 0.56~ \mu\rm{m}^{-1}$). In panel (a) the light-solid (dark-dashed) line represents $N_{NB,\k}$ ($N_{NB,-\k}$). In panel (b) the dark solid line shows $N_{GB,\k}$ while the light solid and red dashed lines correspond to $v_\k^2(N_{NB,\k}+1)$ and $v_\k^2(N_{NB,-\k}+1)$ respectively.  (c) Bogoliubov amplitude $v_\k^2$ for different condensate densities $n_0 = 485~\mu\text{m}^{-2}$,  $n_0 = 804~\mu\text{m}^{-2}$, $n_0 = 1519~\mu\text{m}^{-2}$. The solid lines correspond to the experimental measurement obtained from the ratio $N_{GB,\k}/(N_{NB,-\k}+1)$. The dashed lines show the theory predictions of Eq.~\eqref{eq:ukvks} with no free parameters. Here, the horizontal axis is rescaled using $\xi$. Shaded areas in all panels indicate the experimental error bars within the 95\% confidence interval.}
    \label{fig:fig8}
\end{figure}

\subsection{Momentum space occupation and extraction of Bogoliubov amplitudes} \label{sec:occ}

In addition to the reservoir and condensate densities, our experiment allows us to measure the occupation of the normal and ghost branches $N_{NB,\k}$ and $N_{GB,\k}$ following the method used in Ref.~\cite{Pieczarka2020}. In contrast to  Ref.~\cite{Pieczarka2020}, in the present experiment, we are able to analyse the signal from both positive and negative values of $\k$ %$k_x$ 
which gives us access to the Bogoliubov amplitudes via Eq.~\eqref{eq:NB_GB}.
%improve our interpretation.

The momentum occupation of the normal branch is presented in Fig.~\ref{fig:fig8}(a) for both positive and negative directions along the $x$-axis.  
%two opposite directions $\pm k_x$. 
We observe a general decrease of these occupations as $k$ increases and a hump at around $ 2~\mu {\rm m}^{-1}$. % which is a far from equilibrium distribution.
%The latter is due to the high-energy state around the bottleneck wavevectors as described earlier~\cite{Pieczarka2020}.
  The latter is related to the gain associated with the high-energy states in the ring pump region. This phenomenon can already be spotted in the raw data, such as that shown in Fig.~\ref{fig:fig5}, and was observed in a previous experiment performed at similar detuning \cite{Pieczarka2020}.

 In general, we observe a significant nonequilibrium occupation of the excitation branches, as shown in Fig.~\ref{fig:fig8}(a,b). This could be favored by the circular ring pump geometry, which allows additional parametric pair scattering processes to participate in the polariton relaxation within the trap compared %with respect 
 to a Gaussian pump spot \cite{Paschos2020}.
   Moreover, the momentum occupations are not symmetric with respect to $k=0$ and one has $N_{NB(GB),\k}\neq N_{NB(GB),-\k}$ [see Fig.~\ref{fig:fig8}(a)] which reflects the trap asymmetry arising from the cavity gradient 
   %of the sample 
   (a linear slope of the effective potential in the sample) and the imbalanced pump intensity across the ring excitation~\cite{Estrecho2021,Pieczarka2019,Myers2018}.

Nevertheless, we are able to extract the Bogoliubov amplitude $v_\k^2$ by correlating the signals of NB and GB from opposite sides of $\k$-space. 
To do so, we use the fact that the wavevectors probed are large with respect to the inverse healing length ($1/\xi\simeq 0.56~ \mu\rm{m}^{-1}$ for the data presented in Figs.~\ref{fig:fig8}(a,b)), which allows us to use Eq.~\eqref{eq:NGB} since $u_\k^2\simeq1$ when $k\xi > 1$.
As an example of the procedure, in Fig.~\ref{fig:fig8}(b), we have plotted $v_\k^2(N_{NB,-\k}+1)$ and $v_\k^2(N_{NB,\k}+1)$ where we have used the data from panel (a) for $N_{NB,\k}$, and where $v_\k^2$ is calculated using the measured values of $\mu_C$ and $ \mu_T$ in Eq.~\eqref{eq:ukvks}. We %can 
see that $v_\k^2(N_{NB,-\k}+1)$ matches remarkably %perfectly 
well with the measured values of $N_{GB,\k}$. However, this is not the case for $v_\k^2(N_{NB,\k}+1)$ because of the above-mentioned asymmetry related to the optical trap. 
  
  A direct comparison between the $v_\k^2$ extracted from the ratio of the measured occupations, Eq.~\eqref{eq:NGB}, and the theoretical prediction of Eq.~\eqref{eq:ukvks} is presented in Fig.~\ref{fig:fig8}(c) for different condensate densities $n_0$. We have used the measured values of $\mu_T$ and $\mu_C$ shown in Fig.~\ref{fig:fig6}(a) to plot the theoretical lines. 
  %First of all, one observes that the data measured at different $n_0$ shows an universal behaviour and falls into one curve plotted in the rescaled momentum axis.
    One observes that the three experimental curves superimpose in the region where they overlap when plotted with the rescaled momentum axis.
   None of these curves follow a $(k\xi)^{-4}$ power law as would be predicted in a conventional (Galilean-invariant) Bogoliubov theory.
   %with a parabolic single-particle kinetic energy.
   However, in the absence of any free parameters, one observes a good agreement between the predictions of our generalized theory and the experimental measurement.
   The small discrepancy between theory and experiment could be due to the spatial modulations in the condensate density arising from local disorder, as observed in Fig.~\ref{fig:fig4}.
\ob{In this case, the observed occupations of the different branches correspond to spatial averages over these modulations. Taking the average of Eq.~\eqref{eq:NGB} over position $\mathbf{r}$, we obtain:
\begin{align} \nonumber
    \left< N_{GB,\mathbf{k}} \right>_\mathbf{r} = \left<v^2_{\mathbf{k}} (N_{NB,-\mathbf{k}} + 1) \right>_\mathbf{r} > \left<v^2_{\mathbf{k}} \right>_\mathbf{r} \left<N_{NB,-\mathbf{k}} + 1 \right>_\mathbf{r} ,
\end{align}   
where the inequality is based on the reasonable assumption that the Bogoliubov amplitude and the NB occupation both vary in the same way with condensate density in the driven-dissipative polariton system. Thus, we find that the experimentally extracted Bogoliubov amplitude satisfies
\begin{align}
v^2_{\mathbf{k},exp} = \frac{\left<N_{GB,\mathbf{k}} \right>_\mathbf{r}} {\left<N_{NB,-\mathbf{k}} + 1 \right>_\mathbf{r}} > \left<v^2_{\mathbf{k}} \right>_\mathbf{r},      
\end{align}
which provides an explanation for why the experimental curves in Fig.~\ref{fig:fig8}(c) lie above the theoretical predictions.} 

Our measurement of the Bogoliubov amplitudes in a solid-state system complements previous fundamental experimental results obtained with ultracold atomic gases.
In contrast to the pioneering work of Ref.~\cite{Vogels2002}, we probe larger momenta where the Bogoliubov spectrum is no longer sound-like. 
Furthermore, unlike recent investigations of depletion in atomic condensates~\cite{Chang2016,Lopes2017}, our system is not in thermal equilibrium with a well-defined temperature, as evident from the highly  nonequilibrium occupations.

\section{Conclusions}

In conclusion, we have 
presented a generalized Bogoliubov theory for a polariton condensate in a microcavity embedding $\mathcal{N}$ quantum wells. 
%In a nutshell, 
The key differences from the conventional equilibrium Bogoliubov theory are: %that we account for 
(i) the presence of a large incoherent reservoir, a consequence of the polariton system's driven-dissipative nature; (ii) the non-parabolic polariton kinetic energy and the momentum dependence of the Hopfield coefficients (related to the lack of Galilean invariance); and (iii) the possibility of a %non-thermal and 
highly non-thermal distribution of %fact that 
Bogoliubov excitations. %are not necessarily thermalized. 
In particular, we assume that the polariton condensate has achieved dynamical equilibrium with the incoherent reservoir, which could be facilitated by interconversion between bright and dark excitonic superposition states. 
%the condensate and reservoir. 

%
We have used our theoretical framework to analyze the experimental photoluminescence spectrum of a polariton condensate formed in an optically induced trap. This allowed us to estimate the reservoir density and to highlight that it is not negligible within the trap, in agreement with previous observations \cite{Pieczarka2019,Pieczarka2020}. In addition, we have demonstrated that the condensate and reservoir densities become comparable at pump powers of $\sim 2 P_{th} $ and find that their ratio is locked at higher pump powers leading to a saturation of the condensate fraction $\rho$. This saturation implies that an effective upconversion from the condensate to the reservoir is taking place at large densities, as is well captured by our phenomenological rate equations.

Going further, we have measured the momentum resolved occupations, which allowed us to extract the Bogoliubov amplitudes directly from the experimental data. To our knowledge, this is the first measurement of Bogoliubov \textit{amplitudes} in a non-equilibrium condensate. %polariton system. 
We found a good agreement with the analytical expression obtained within our Bogoliubov theory.
This suggests that, in contrast to the nonequilibrium occupations, the Bogoliubov spectrum and amplitudes are not strongly affected by the complicated kinetics of the optically trapped polariton condensate.

Nontrivial questions remain open concerning the nature of the reservoir and whether or not its role should be reduced in single-layer microcavities where the dark superpositions are absent. Future experiments involving fewer quantum wells are required to address these remaining questions.
Furthermore, while the excitation geometry of the present experiment enabled us to unveil the upconversion from the condensate to the reservoir, there is no reason that forbids this process to occur under homogeneous excitation. Thus, it would be interesting to see if this mechanism, not captured in the commonly used phenomenological model introduced in Ref.~\cite{Wouters2007}, can play a role in the stability of polariton condensates recently studied theoretically~\cite{Smirnov2014,Bobrovska2014,Bobrovska2015,Dagvadorj2021} and experimentally~\cite{Bobrovska2018,Estrecho2018,Baboux2018}.
Our experimental results also clearly show that the interactions between polaritons and reservoir particles are not negligible and contribute to the blueshift. %It is worth noticing that 
In particular, these interactions are at the origin of the modulational instability in the above mentioned phenomenological model~\cite{Smirnov2014} and have been neglected in the recent theoretical studies investigating its mapping to a Kardar-Parisi-Zhang equation~\cite{Altman2015,Squizatto2018,Deligiannis2021}.
Finally, a better understanding of the role of the reservoir is also needed in the coherent excitation regime for which signatures of reservoir generation have been reported in several experiments \cite{Gavrilov2010,Klembt2015,Walker2017,Stepanov2019}.

\acknowledgements
This work was supported by the Australian Research Council Centre of Excellence in Future Low-Energy Electronics Technologies (CE170100039). J.L. and M.M.P. are also supported through Australian Research Council Future Fellowships FT160100244 and FT200100619, respectively. M.P. acknowledges support from the National Science Center in Poland, Grant No.~2018/30/E/ST7/00648 and the Foundation for Polish Science in the START program. The work of M.S., D.S., L.P. and K.W. was supported by the National Science Foundation (Grant No. DMR-2004570). L.P. and K.W. was additionally supported by and Betty Moore Foundation (Grant No.~GBMF-4420) and by the National Science Foundation MRSEC program through the Princeton Center for Complex Materials (Grant No.~DMR-0819860).
%We are grateful for fruitful discussions with ... 

  \appendix

\section{Derivation of the effective Hamiltonian }
\label{sec:Heff}

Here, we derive the effective Hamiltonian \eqref{eq:Heff}, that forms the basis of our calculation of the excitation spectrum for a polariton condensate in the presence of an incoherent excitonic reservoir in a $\mathcal{N}$-QW microcavity.
If we neglect the upper polariton and rewrite %the interaction in 
Eq.~\eqref{eq:interNB} in terms of the lower polariton operator and the $\mathcal{N}-1$ degenerate dark state operators, 
we arrive at
\begin{widetext}
\begin{align} \nonumber \label{eq:l-dark}
\hat{V} = \frac{ g_{}}{2\mathcal{N}\mathcal{A}}\sum_{\k_1\k_2\k_3\k_4}\delta_{\k_1+\k_2-\k_3-\k_4}\Bigg\{&X_{\k_1}X_{\k_2}X_{\k_3}X_{\k_4} \hat{L}_{\k_1}^{\dagger}\hat{L}_{\k_2}^{\dagger} \hat{L}_{\k_3}\hat{L}_{\k_4} 
+4 \sum_{\{l_a,l_b\}}^{\mathcal{N}-1}\delta_{\mathcal{M}_2}
%\delta_{\overline{(l_a-l_b)}_\mathcal{N}}
X_{\k_1}X_{\k_4} \hat{L}_{\k_1}^{\dagger}\hat{d}_{\k_2,l_a}^{\dagger} \hat{d}_{\k_3,l_b}\hat{L}_{\k_4} 
\nonumber \\ &
+\sum_{\{l_a,l_b\}}^{\mathcal{N}-1}\delta_{\mathcal{M}'_2}\left( X_{\k_1}X_{\k_2} \hat{L}_{\k_1}^{\dagger}\hat{L}_{\k_2}^{\dagger} \hat{d}_{\k_3,l_a}\hat{d}_{\k_4,l_b}+X_{\k_3}X_{\k_4} \hat{d}_{\k_1,l_a}^{\dagger}\hat{d}_{\k_2,l_b}^{\dagger} \hat{L}_{\k_3}\hat{L}_{\k_4}\right)
\nonumber \\ &
+2\sum_{\{l_a,l_b,l_c\}}^{\mathcal{N}-1}\delta_{\mathcal{M}_3'}X_{\k_1}\hat{L}_{\k_1}^{\dagger}\hat{d}_{\k_2,l_a}^{\dagger}\hat{d}_{\k_3,l_b}\hat{d}_{\k_4,l_c}
+2 \sum_{\{l_a,l_b,l_c\}}^{\mathcal{N}-1}\delta_{\mathcal{M}_3}X_{\mathbf{k}_4} \hat{d}_{\mathbf{k}_1,l_a}^\dagger \hat{d}_{\mathbf{k}_2,l_b}^{\dagger}\hat{d}_{\mathbf{k}_3,l_c}\hat{L}_{\mathbf{k}_4} \nonumber
\\
&+
\sum_{\{l_j\}}^{\mathcal{N}-1}\delta_{\mathcal{M}}\, \hat{d}_{\k_1,l_1}^{\dagger}\hat{d}_{\k_2,l_2}^{\dagger} \hat{d}_{\k_3,l_3}\hat{d}_{\k_4,l_4} \Bigg\},
\end{align}
\end{widetext}
where $\mathcal{M}=\text{Mod}{[l_1+ l_2- l_3 -l_4,\mathcal{N}}]$,  $\mathcal{M}_2=\text{Mod}{[l_a- l_b,\mathcal{N}}]$, $\mathcal{M}_2'=\text{Mod}{[l_a+ l_b,\mathcal{N}}]$,   $\mathcal{M}_3=\text{Mod}{[l_a- l_b-l_c,\mathcal{N}}]$,  $\mathcal{M}_3'=\text{Mod}{[l_a+ l_b-l_c,\mathcal{N}}]$.
We see that the Hamiltonian in this basis has terms involving zero, one, two, or four polariton states (or, equivalently, bright exciton states).

Our goal is to arrive at an effective low-energy model that describes the system in dynamical equilibrium. To this end, we note that the experiment probes momenta at which the single-particle energies of lower polaritons and the dark excitons are split. In the low-energy model, we will focus on the approximately elastic scattering processes that can take place in the system, namely polariton-polariton scattering (first term of Eq.~\eqref{eq:l-dark}), polariton-dark exciton scattering (second term), and dark exciton-dark exciton scattering (last term). However, if we simply ignore all other terms---corresponding to off-resonant processes---then we do not obtain the correct interaction strengths since, e.g., the polariton-polariton interaction can proceed via intermediate dark states according to the terms in the second line of Eq.~\eqref{eq:l-dark}. Taking this properly into account leads to modified polariton-polariton, polariton-exciton, and exciton-exciton interaction constants $g_{pp}$, $g_{pd}$, and $g_{dd}$, respectively~\cite{Bleu2020}. In our system where the Rabi splitting is large compared to the exciton binding energy, these effective interaction constants are expected to be dominated by the contribution from the Born approximation~\cite{li2021microscopic}.

To proceed, we furthermore treat the polaritons with $|\k|>q_0$ and all the dark exciton states as an incoherent semiclassical gas within the Hartree-Fock approximation. This consists of replacing the products of $\hat{d_l}$ operators by their average value
using
\begin{align}
\langle\hat{d}_{\k,l}^\dagger\hat{d}_{\k',l'}\rangle= N_{\k,l}   %= N_{l,\k_i}
\delta_{\k\k'}\delta_{ll'},
\end{align}
and
  \begin{align}
&\langle\hat{d}_{\k_1,l_1}^\dagger\hat{d}_{\k_2,l_2}^\dagger\hat{d}_{\k_3,l_3}\hat{d}_{\k_4,l_4}\rangle =\left(1-\delta_{\k_1\k_2}\delta_{l_1l_2}\right)
\\ \nonumber
& ~~  \times\left(\delta_{\k_1\k_4}\delta_{l_1l_4}\delta_{\k_2\k_3}\delta_{l_2l_3}+\delta_{\k_1\k_3}\delta_{l_1l_3}\delta_{\k_2\k_4}\delta_{l_2l_4}\right)N_{\k_1,l_1}N_{\k_2,l_2}
\\ \nonumber
&  ~~ +\delta_{\k_1\k_2}\delta_{l_1l_2}\delta_{\k_1\k_4}\delta_{l_1l_4}\delta_{\k_2\k_3}\delta_{l_2l_3} N_{\k_1,l_1}\left(N_{\k_1,l_1}-1\right),
\end{align}
with $N_{\k,l}\equiv \langle \hat{d}_{\k,l}^\dagger\hat{d}_{\k,l} \rangle$ the momentum occupations. 

The effective low-energy Hamiltonian then takes the form given in the main text
  \begin{align} \nonumber
\hat{H}_{\rm eff}&=E_{\text{res}}+\sum_{\k}\! E_{\mathbf{k}}^L \hat{L}_{\mathbf{k}}^{\dagger} \hat{L}_{\mathbf{k}} +\sum_{\k}\! 4 \frac{ g_{pd}}{2\mathcal{N}\mathcal{A}} X_\k^2 N_R \hat{L}_{\mathbf{k}}^{\dagger} \hat{L}_{\mathbf{k}}\\
\label{eq:Heff_supp}
&+\sum_{\k,\k',\q}\frac{ g_{pp}}{2\mathcal{N}\mathcal{A}} X_{\mathbf{k+q}}X_{\mathbf{k'-q}}X_{\mathbf{k'}}X_{\mathbf{k}} \hat{L}_{\mathbf{k+q}}^{\dagger}\hat{L}_{\mathbf{k'-q}}^{\dagger} \hat{L}_{\mathbf{k'}}\hat{L}_{\mathbf{k}} ,
  \end{align}
where the sum over a given momentum $\k$ implicitly assumes that all polariton operators act at momenta $|\k|<q_0$.
$E_{\text{res}}$ is the reservoir energy in the Hartree-Fock approximation which reads
\begin{align} \label{eq:EresFin}
E_{\text{res}}=\sum_{\substack{\k\\|\k|>q_0}}\! E_{\mathbf{k}}^X N_{\k,\mathcal{N}}\!\! +\sum_{\substack{\k}}\sum_{l=1}^{\mathcal{N}-1}  E_{\mathbf{k}}^X N_{\k,l}+\frac{ g_{dd}}{\mathcal{N}\mathcal{A}}N_R^2,
\end{align}
and $N_R$ %= \sum_{l=1}^{\mathcal{N}-1}\sum_{\k} N_{\k,l}+\sum_{\substack{\k\\|\k|>q_0}} N_{\k,\mathcal{N}}$ 
is the total number of reservoir particles defined in Eq.~\eqref{eq:Nr}. 

We conclude this section by noting that by introducing an effective low-energy Hamiltonian, we have removed the terms enabling parametric processes of the form $d_{\k,l_1}+d_{\k,l_2} \rightarrow L_0+d_{2\k,l_3}$ with $|\k|>q_0$. Such processes can contribute to the population of the lower polariton branch and play a role in the condensate formation as noticed in Ref.~\cite{Porras2002} but do not affect the spectrum once the system has achieved dynamical equilibrium.
%In a multilayer system at zero photon-exciton detuning, processes of the form $d_{0,l}+d_{0,\mathcal{N}-l} \rightarrow L_0+U_0$ can also exist if upper polaritons are involved.
This phenomenon is effectively introduced in the phenomenological rate equations \eqref{eq:rate} with the stimulated scattering term $\propto R n_R^2n_0$. 
Similarly, the higher order $W$ term in Eq.~\eqref{eq:rate} involves the  %interconversion 
terms in Eq.~\eqref{eq:l-dark} that convert a condensate polariton into a reservoir exciton, although in this case the scattering processes can proceed through virtual intermediate states.
%involve combinations of the opposite 

\bibliography{main}

\begin{thebibliography}{84}%
\makeatletter
\providecommand \@ifxundefined [1]{%
 \@ifx{#1\undefined}
}%
\providecommand \@ifnum [1]{%
 \ifnum #1\expandafter \@firstoftwo
 \else \expandafter \@secondoftwo
 \fi
}%
\providecommand \@ifx [1]{%
 \ifx #1\expandafter \@firstoftwo
 \else \expandafter \@secondoftwo
 \fi
}%
\providecommand \natexlab [1]{#1}%
\providecommand \enquote  [1]{``#1''}%
\providecommand \bibnamefont  [1]{#1}%
\providecommand \bibfnamefont [1]{#1}%
\providecommand \citenamefont [1]{#1}%
\providecommand \href@noop [0]{\@secondoftwo}%
\providecommand \href [0]{\begingroup \@sanitize@url \@href}%
\providecommand \@href[1]{\@@startlink{#1}\@@href}%
\providecommand \@@href[1]{\endgroup#1\@@endlink}%
\providecommand \@sanitize@url [0]{\catcode `\\12\catcode `\$12\catcode
  `\&12\catcode `\#12\catcode `\^12\catcode `\_12\catcode `\%12\relax}%
\providecommand \@@startlink[1]{}%
\providecommand \@@endlink[0]{}%
\providecommand \url  [0]{\begingroup\@sanitize@url \@url }%
\providecommand \@url [1]{\endgroup\@href {#1}{\urlprefix }}%
\providecommand \urlprefix  [0]{URL }%
\providecommand \Eprint [0]{\href }%
\providecommand \doibase [0]{http://dx.doi.org/}%
\providecommand \selectlanguage [0]{\@gobble}%
\providecommand \bibinfo  [0]{\@secondoftwo}%
\providecommand \bibfield  [0]{\@secondoftwo}%
\providecommand \translation [1]{[#1]}%
\providecommand \BibitemOpen [0]{}%
\providecommand \bibitemStop [0]{}%
\providecommand \bibitemNoStop [0]{.\EOS\space}%
\providecommand \EOS [0]{\spacefactor3000\relax}%
\providecommand \BibitemShut  [1]{\csname bibitem#1\endcsname}%
\let\auto@bib@innerbib\@empty
%</preamble>
\bibitem [{\citenamefont {Bogoliubov}(1947)}]{bogoliubov1947theory}%
  \BibitemOpen
  \bibfield  {author} {\bibinfo {author} {\bibfnamefont {N.}~\bibnamefont
  {Bogoliubov}},\ }\bibfield  {title} {\bibinfo {title} {\emph {On the theory
  of superfluidity}},\ }\href@noop {} {\bibfield  {journal} {\bibinfo
  {journal} {J. Phys}\ }\textbf {\bibinfo {volume} {11}},\ \bibinfo {pages}
  {23} (\bibinfo {year} {1947})}\BibitemShut {NoStop}%
\bibitem [{\citenamefont {Pitaevskii}\ and\ \citenamefont
  {Stringari}(2016)}]{pitaevskii2016bose}%
  \BibitemOpen
  \bibfield  {author} {\bibinfo {author} {\bibfnamefont {L.}~\bibnamefont
  {Pitaevskii}}\ and\ \bibinfo {author} {\bibfnamefont {S.}~\bibnamefont
  {Stringari}},\ }\href@noop {} {\emph {\bibinfo {title} {Bose-Einstein
  condensation and superfluidity}}},\ Vol.\ \bibinfo {volume} {164}\ (\bibinfo
  {publisher} {Oxford University Press},\ \bibinfo {year} {2016})\BibitemShut
  {NoStop}%
\bibitem [{\citenamefont {Anderson}\ \emph {et~al.}(1995)\citenamefont
  {Anderson}, \citenamefont {Ensher}, \citenamefont {Matthews}, \citenamefont
  {Wieman},\ and\ \citenamefont {Cornell}}]{anderson1995observation}%
  \BibitemOpen
  \bibfield  {author} {\bibinfo {author} {\bibfnamefont {M.~H.}\ \bibnamefont
  {Anderson}}, \bibinfo {author} {\bibfnamefont {J.~R.}\ \bibnamefont
  {Ensher}}, \bibinfo {author} {\bibfnamefont {M.~R.}\ \bibnamefont
  {Matthews}}, \bibinfo {author} {\bibfnamefont {C.~E.}\ \bibnamefont
  {Wieman}}, \ and\ \bibinfo {author} {\bibfnamefont {E.~A.}\ \bibnamefont
  {Cornell}},\ }\bibfield  {title} {\bibinfo {title} {\emph {Observation of
  Bose-Einstein Condensation in a Dilute Atomic Vapor}},\ }\href {\doibase
  10.1126/science.269.5221.198} {\bibfield  {journal} {\bibinfo  {journal}
  {Science}\ }\textbf {\bibinfo {volume} {269}},\ \bibinfo {pages} {198}
  (\bibinfo {year} {1995})}\BibitemShut {NoStop}%
\bibitem [{\citenamefont {Bradley}\ \emph {et~al.}(1995)\citenamefont
  {Bradley}, \citenamefont {Sackett}, \citenamefont {Tollett},\ and\
  \citenamefont {Hulet}}]{Bradley1995}%
  \BibitemOpen
  \bibfield  {author} {\bibinfo {author} {\bibfnamefont {C.~C.}\ \bibnamefont
  {Bradley}}, \bibinfo {author} {\bibfnamefont {C.~A.}\ \bibnamefont
  {Sackett}}, \bibinfo {author} {\bibfnamefont {J.~J.}\ \bibnamefont
  {Tollett}}, \ and\ \bibinfo {author} {\bibfnamefont {R.~G.}\ \bibnamefont
  {Hulet}},\ }\bibfield  {title} {\bibinfo {title} {\emph {Evidence of
  Bose-Einstein Condensation in an Atomic Gas with Attractive Interactions}},\
  }\href {\doibase 10.1103/PhysRevLett.75.1687} {\bibfield  {journal} {\bibinfo
   {journal} {Phys. Rev. Lett.}\ }\textbf {\bibinfo {volume} {75}},\ \bibinfo
  {pages} {1687} (\bibinfo {year} {1995})}\BibitemShut {NoStop}%
\bibitem [{\citenamefont {Davis}\ \emph {et~al.}(1995)\citenamefont {Davis},
  \citenamefont {Mewes}, \citenamefont {Andrews}, \citenamefont {van Druten},
  \citenamefont {Durfee}, \citenamefont {Kurn},\ and\ \citenamefont
  {Ketterle}}]{Davis1995}%
  \BibitemOpen
  \bibfield  {author} {\bibinfo {author} {\bibfnamefont {K.~B.}\ \bibnamefont
  {Davis}}, \bibinfo {author} {\bibfnamefont {M.~O.}\ \bibnamefont {Mewes}},
  \bibinfo {author} {\bibfnamefont {M.~R.}\ \bibnamefont {Andrews}}, \bibinfo
  {author} {\bibfnamefont {N.~J.}\ \bibnamefont {van Druten}}, \bibinfo
  {author} {\bibfnamefont {D.~S.}\ \bibnamefont {Durfee}}, \bibinfo {author}
  {\bibfnamefont {D.~M.}\ \bibnamefont {Kurn}}, \ and\ \bibinfo {author}
  {\bibfnamefont {W.}~\bibnamefont {Ketterle}},\ }\bibfield  {title} {\bibinfo
  {title} {\emph {Bose-Einstein Condensation in a Gas of Sodium Atoms}},\
  }\href {\doibase 10.1103/PhysRevLett.75.3969} {\bibfield  {journal} {\bibinfo
   {journal} {Phys. Rev. Lett.}\ }\textbf {\bibinfo {volume} {75}},\ \bibinfo
  {pages} {3969} (\bibinfo {year} {1995})}\BibitemShut {NoStop}%
\bibitem [{\citenamefont {Steinhauer}\ \emph {et~al.}(2002)\citenamefont
  {Steinhauer}, \citenamefont {Ozeri}, \citenamefont {Katz},\ and\
  \citenamefont {Davidson}}]{Steinhauer2002}%
  \BibitemOpen
  \bibfield  {author} {\bibinfo {author} {\bibfnamefont {J.}~\bibnamefont
  {Steinhauer}}, \bibinfo {author} {\bibfnamefont {R.}~\bibnamefont {Ozeri}},
  \bibinfo {author} {\bibfnamefont {N.}~\bibnamefont {Katz}}, \ and\ \bibinfo
  {author} {\bibfnamefont {N.}~\bibnamefont {Davidson}},\ }\bibfield  {title}
  {\bibinfo {title} {\emph {{Excitation Spectrum of a Bose-Einstein
  Condensate}}},\ }\href {\doibase 10.1103/PhysRevLett.88.120407} {\bibfield
  {journal} {\bibinfo  {journal} {Phys. Rev. Lett.}\ }\textbf {\bibinfo
  {volume} {88}},\ \bibinfo {pages} {120407} (\bibinfo {year}
  {2002})}\BibitemShut {NoStop}%
\bibitem [{\citenamefont {Vogels}\ \emph {et~al.}(2002)\citenamefont {Vogels},
  \citenamefont {Xu}, \citenamefont {Raman}, \citenamefont {Abo-Shaeer},\ and\
  \citenamefont {Ketterle}}]{Vogels2002}%
  \BibitemOpen
  \bibfield  {author} {\bibinfo {author} {\bibfnamefont {J.~M.}\ \bibnamefont
  {Vogels}}, \bibinfo {author} {\bibfnamefont {K.}~\bibnamefont {Xu}}, \bibinfo
  {author} {\bibfnamefont {C.}~\bibnamefont {Raman}}, \bibinfo {author}
  {\bibfnamefont {J.~R.}\ \bibnamefont {Abo-Shaeer}}, \ and\ \bibinfo {author}
  {\bibfnamefont {W.}~\bibnamefont {Ketterle}},\ }\bibfield  {title} {\bibinfo
  {title} {\emph {Experimental Observation of the Bogoliubov Transformation for
  a Bose-Einstein Condensed Gas}},\ }\href {\doibase
  10.1103/PhysRevLett.88.060402} {\bibfield  {journal} {\bibinfo  {journal}
  {Phys. Rev. Lett.}\ }\textbf {\bibinfo {volume} {88}},\ \bibinfo {pages}
  {060402} (\bibinfo {year} {2002})}\BibitemShut {NoStop}%
\bibitem [{\citenamefont {Deng}\ \emph {et~al.}(2010)\citenamefont {Deng},
  \citenamefont {Haug},\ and\ \citenamefont {Yamamoto}}]{DengRMP2010}%
  \BibitemOpen
  \bibfield  {author} {\bibinfo {author} {\bibfnamefont {H.}~\bibnamefont
  {Deng}}, \bibinfo {author} {\bibfnamefont {H.}~\bibnamefont {Haug}}, \ and\
  \bibinfo {author} {\bibfnamefont {Y.}~\bibnamefont {Yamamoto}},\ }\bibfield
  {title} {\bibinfo {title} {\emph {Exciton-polariton Bose-Einstein
  condensation}},\ }\href {\doibase 10.1103/RevModPhys.82.1489} {\bibfield
  {journal} {\bibinfo  {journal} {Rev. Mod. Phys.}\ }\textbf {\bibinfo {volume}
  {82}},\ \bibinfo {pages} {1489} (\bibinfo {year} {2010})}\BibitemShut
  {NoStop}%
\bibitem [{\citenamefont {Carusotto}\ and\ \citenamefont
  {Ciuti}(2013)}]{RMP2013QFL}%
  \BibitemOpen
  \bibfield  {author} {\bibinfo {author} {\bibfnamefont {I.}~\bibnamefont
  {Carusotto}}\ and\ \bibinfo {author} {\bibfnamefont {C.}~\bibnamefont
  {Ciuti}},\ }\bibfield  {title} {\bibinfo {title} {\emph {Quantum fluids of
  light}},\ }\href {\doibase 10.1103/RevModPhys.85.299} {\bibfield  {journal}
  {\bibinfo  {journal} {Rev. Mod. Phys.}\ }\textbf {\bibinfo {volume} {85}},\
  \bibinfo {pages} {299} (\bibinfo {year} {2013})}\BibitemShut {NoStop}%
\bibitem [{\citenamefont {Kavokin}\ \emph {et~al.}(2017)\citenamefont
  {Kavokin}, \citenamefont {Baumberg}, \citenamefont {Malpuech},\ and\
  \citenamefont {Laussy}}]{kavokin2017microcavities}%
  \BibitemOpen
  \bibfield  {author} {\bibinfo {author} {\bibfnamefont {A.~V.}\ \bibnamefont
  {Kavokin}}, \bibinfo {author} {\bibfnamefont {J.}~\bibnamefont {Baumberg}},
  \bibinfo {author} {\bibfnamefont {G.}~\bibnamefont {Malpuech}}, \ and\
  \bibinfo {author} {\bibfnamefont {F.}~\bibnamefont {Laussy}},\ }\href@noop {}
  {\emph {\bibinfo {title} {{Microcavities}}}},\ \bibinfo {edition} {2nd}\ ed.\
  (\bibinfo  {publisher} {Oxford University Press},\ \bibinfo {address}
  {Oxford},\ \bibinfo {year} {2017})\BibitemShut {NoStop}%
\bibitem [{\citenamefont {Kasprzak}\ \emph {et~al.}(2006)\citenamefont
  {Kasprzak}, \citenamefont {Richard}, \citenamefont {Kundermann},
  \citenamefont {Baas}, \citenamefont {Jeambrun}, \citenamefont {Keeling},
  \citenamefont {Marchetti}, \citenamefont {Szymańska}, \citenamefont
  {André}, \citenamefont {Staehli}, \citenamefont {Savona}, \citenamefont
  {Littlewood}, \citenamefont {Deveaud},\ and\ \citenamefont
  {Dang}}]{kasprzak2006bose}%
  \BibitemOpen
  \bibfield  {author} {\bibinfo {author} {\bibfnamefont {J.}~\bibnamefont
  {Kasprzak}}, \bibinfo {author} {\bibfnamefont {M.}~\bibnamefont {Richard}},
  \bibinfo {author} {\bibfnamefont {S.}~\bibnamefont {Kundermann}}, \bibinfo
  {author} {\bibfnamefont {A.}~\bibnamefont {Baas}}, \bibinfo {author}
  {\bibfnamefont {P.}~\bibnamefont {Jeambrun}}, \bibinfo {author}
  {\bibfnamefont {J.~M.~J.}\ \bibnamefont {Keeling}}, \bibinfo {author}
  {\bibfnamefont {F.~M.}\ \bibnamefont {Marchetti}}, \bibinfo {author}
  {\bibfnamefont {M.~H.}\ \bibnamefont {Szymańska}}, \bibinfo {author}
  {\bibfnamefont {R.}~\bibnamefont {André}}, \bibinfo {author} {\bibfnamefont
  {J.~L.}\ \bibnamefont {Staehli}}, \bibinfo {author} {\bibfnamefont
  {V.}~\bibnamefont {Savona}}, \bibinfo {author} {\bibfnamefont {P.~B.}\
  \bibnamefont {Littlewood}}, \bibinfo {author} {\bibfnamefont
  {B.}~\bibnamefont {Deveaud}}, \ and\ \bibinfo {author} {\bibfnamefont
  {L.~S.}\ \bibnamefont {Dang}},\ }\bibfield  {title} {\bibinfo {title} {\emph
  {Bose-Einstein condensation of exciton polaritons}},\ }\href
  {https://doi.org/10.1038/nature05131} {\bibfield  {journal} {\bibinfo
  {journal} {Nature}\ }\textbf {\bibinfo {volume} {443}},\ \bibinfo {pages}
  {409} (\bibinfo {year} {2006})}\BibitemShut {NoStop}%
\bibitem [{\citenamefont {Balili}\ \emph {et~al.}(2007)\citenamefont {Balili},
  \citenamefont {Hartwell}, \citenamefont {Snoke}, \citenamefont {Pfeiffer},\
  and\ \citenamefont {West}}]{balili2007bose}%
  \BibitemOpen
  \bibfield  {author} {\bibinfo {author} {\bibfnamefont {R.}~\bibnamefont
  {Balili}}, \bibinfo {author} {\bibfnamefont {V.}~\bibnamefont {Hartwell}},
  \bibinfo {author} {\bibfnamefont {D.}~\bibnamefont {Snoke}}, \bibinfo
  {author} {\bibfnamefont {L.}~\bibnamefont {Pfeiffer}}, \ and\ \bibinfo
  {author} {\bibfnamefont {K.}~\bibnamefont {West}},\ }\bibfield  {title}
  {\bibinfo {title} {\emph {Bose-Einstein Condensation of Microcavity
  Polaritons in a Trap}},\ }\href
  {http://science.sciencemag.org/content/316/5827/1007.abstract} {\bibfield
  {journal} {\bibinfo  {journal} {Science}\ }\textbf {\bibinfo {volume}
  {316}},\ \bibinfo {pages} {1007} (\bibinfo {year} {2007})}\BibitemShut
  {NoStop}%
\bibitem [{\citenamefont {Baumberg}\ \emph {et~al.}(2008)\citenamefont
  {Baumberg}, \citenamefont {Kavokin}, \citenamefont {Christopoulos},
  \citenamefont {Grundy}, \citenamefont {Butt\'e}, \citenamefont {Christmann},
  \citenamefont {Solnyshkov}, \citenamefont {Malpuech}, \citenamefont
  {Baldassarri H\"oger~von H\"ogersthal}, \citenamefont {Feltin}, \citenamefont
  {Carlin},\ and\ \citenamefont {Grandjean}}]{Baumberg2008}%
  \BibitemOpen
  \bibfield  {author} {\bibinfo {author} {\bibfnamefont {J.~J.}\ \bibnamefont
  {Baumberg}}, \bibinfo {author} {\bibfnamefont {A.~V.}\ \bibnamefont
  {Kavokin}}, \bibinfo {author} {\bibfnamefont {S.}~\bibnamefont
  {Christopoulos}}, \bibinfo {author} {\bibfnamefont {A.~J.~D.}\ \bibnamefont
  {Grundy}}, \bibinfo {author} {\bibfnamefont {R.}~\bibnamefont {Butt\'e}},
  \bibinfo {author} {\bibfnamefont {G.}~\bibnamefont {Christmann}}, \bibinfo
  {author} {\bibfnamefont {D.~D.}\ \bibnamefont {Solnyshkov}}, \bibinfo
  {author} {\bibfnamefont {G.}~\bibnamefont {Malpuech}}, \bibinfo {author}
  {\bibfnamefont {G.}~\bibnamefont {Baldassarri H\"oger~von H\"ogersthal}},
  \bibinfo {author} {\bibfnamefont {E.}~\bibnamefont {Feltin}}, \bibinfo
  {author} {\bibfnamefont {J.-F.}\ \bibnamefont {Carlin}}, \ and\ \bibinfo
  {author} {\bibfnamefont {N.}~\bibnamefont {Grandjean}},\ }\bibfield  {title}
  {\bibinfo {title} {\emph {Spontaneous Polarization Buildup in a
  Room-Temperature Polariton Laser}},\ }\href {\doibase
  10.1103/PhysRevLett.101.136409} {\bibfield  {journal} {\bibinfo  {journal}
  {Phys. Rev. Lett.}\ }\textbf {\bibinfo {volume} {101}},\ \bibinfo {pages}
  {136409} (\bibinfo {year} {2008})}\BibitemShut {NoStop}%
\bibitem [{\citenamefont {Li}\ \emph {et~al.}(2013)\citenamefont {Li},
  \citenamefont {Orosz}, \citenamefont {Kamoun}, \citenamefont {Bouchoule},
  \citenamefont {Brimont}, \citenamefont {Disseix}, \citenamefont {Guillet},
  \citenamefont {Lafosse}, \citenamefont {Leroux}, \citenamefont {Leymarie},
  \citenamefont {Mexis}, \citenamefont {Mihailovic}, \citenamefont
  {Patriarche}, \citenamefont {R\'everet}, \citenamefont {Solnyshkov},
  \citenamefont {Zuniga-Perez},\ and\ \citenamefont {Malpuech}}]{Li2013}%
  \BibitemOpen
  \bibfield  {author} {\bibinfo {author} {\bibfnamefont {F.}~\bibnamefont
  {Li}}, \bibinfo {author} {\bibfnamefont {L.}~\bibnamefont {Orosz}}, \bibinfo
  {author} {\bibfnamefont {O.}~\bibnamefont {Kamoun}}, \bibinfo {author}
  {\bibfnamefont {S.}~\bibnamefont {Bouchoule}}, \bibinfo {author}
  {\bibfnamefont {C.}~\bibnamefont {Brimont}}, \bibinfo {author} {\bibfnamefont
  {P.}~\bibnamefont {Disseix}}, \bibinfo {author} {\bibfnamefont
  {T.}~\bibnamefont {Guillet}}, \bibinfo {author} {\bibfnamefont
  {X.}~\bibnamefont {Lafosse}}, \bibinfo {author} {\bibfnamefont
  {M.}~\bibnamefont {Leroux}}, \bibinfo {author} {\bibfnamefont
  {J.}~\bibnamefont {Leymarie}}, \bibinfo {author} {\bibfnamefont
  {M.}~\bibnamefont {Mexis}}, \bibinfo {author} {\bibfnamefont
  {M.}~\bibnamefont {Mihailovic}}, \bibinfo {author} {\bibfnamefont
  {G.}~\bibnamefont {Patriarche}}, \bibinfo {author} {\bibfnamefont
  {F.}~\bibnamefont {R\'everet}}, \bibinfo {author} {\bibfnamefont
  {D.}~\bibnamefont {Solnyshkov}}, \bibinfo {author} {\bibfnamefont
  {J.}~\bibnamefont {Zuniga-Perez}}, \ and\ \bibinfo {author} {\bibfnamefont
  {G.}~\bibnamefont {Malpuech}},\ }\bibfield  {title} {\bibinfo {title} {\emph
  {From Excitonic to Photonic Polariton Condensate in a ZnO-Based
  Microcavity}},\ }\href {\doibase 10.1103/PhysRevLett.110.196406} {\bibfield
  {journal} {\bibinfo  {journal} {Phys. Rev. Lett.}\ }\textbf {\bibinfo
  {volume} {110}},\ \bibinfo {pages} {196406} (\bibinfo {year}
  {2013})}\BibitemShut {NoStop}%
\bibitem [{\citenamefont {Plumhof}\ \emph {et~al.}(2014)\citenamefont
  {Plumhof}, \citenamefont {St{\"{o}}ferle}, \citenamefont {Mai}, \citenamefont
  {Scherf},\ and\ \citenamefont {Mahrt}}]{Plumhof2014}%
  \BibitemOpen
  \bibfield  {author} {\bibinfo {author} {\bibfnamefont {J.~D.}\ \bibnamefont
  {Plumhof}}, \bibinfo {author} {\bibfnamefont {T.}~\bibnamefont
  {St{\"{o}}ferle}}, \bibinfo {author} {\bibfnamefont {L.}~\bibnamefont {Mai}},
  \bibinfo {author} {\bibfnamefont {U.}~\bibnamefont {Scherf}}, \ and\ \bibinfo
  {author} {\bibfnamefont {R.~F.}\ \bibnamefont {Mahrt}},\ }\bibfield  {title}
  {\bibinfo {title} {\emph {{Room-temperature Bose-Einstein condensation of
  cavity exciton-polaritons in a polymer}}},\ }\href {\doibase
  10.1038/nmat3825} {\bibfield  {journal} {\bibinfo  {journal} {Nat. Mater.}\
  }\textbf {\bibinfo {volume} {13}},\ \bibinfo {pages} {247} (\bibinfo {year}
  {2014})}\BibitemShut {NoStop}%
\bibitem [{\citenamefont {Su}\ \emph {et~al.}(2017)\citenamefont {Su},
  \citenamefont {Diederichs}, \citenamefont {Wang}, \citenamefont {Liew},
  \citenamefont {Zhao}, \citenamefont {Liu}, \citenamefont {Xu}, \citenamefont
  {Chen},\ and\ \citenamefont {Xiong}}]{Su2017}%
  \BibitemOpen
  \bibfield  {author} {\bibinfo {author} {\bibfnamefont {R.}~\bibnamefont
  {Su}}, \bibinfo {author} {\bibfnamefont {C.}~\bibnamefont {Diederichs}},
  \bibinfo {author} {\bibfnamefont {J.}~\bibnamefont {Wang}}, \bibinfo {author}
  {\bibfnamefont {T.~C.~H.}\ \bibnamefont {Liew}}, \bibinfo {author}
  {\bibfnamefont {J.}~\bibnamefont {Zhao}}, \bibinfo {author} {\bibfnamefont
  {S.}~\bibnamefont {Liu}}, \bibinfo {author} {\bibfnamefont {W.}~\bibnamefont
  {Xu}}, \bibinfo {author} {\bibfnamefont {Z.}~\bibnamefont {Chen}}, \ and\
  \bibinfo {author} {\bibfnamefont {Q.}~\bibnamefont {Xiong}},\ }\bibfield
  {title} {\bibinfo {title} {\emph {Room-Temperature Polariton Lasing in
  All-Inorganic Perovskite Nanoplatelets}},\ }\href {\doibase
  10.1021/acs.nanolett.7b01956} {\bibfield  {journal} {\bibinfo  {journal}
  {Nano Letters}\ }\textbf {\bibinfo {volume} {17}},\ \bibinfo {pages} {3982}
  (\bibinfo {year} {2017})}\BibitemShut {NoStop}%
\bibitem [{\citenamefont {Schneider}\ \emph {et~al.}(2013)\citenamefont
  {Schneider}, \citenamefont {Rahimi-Iman}, \citenamefont {Kim}, \citenamefont
  {Fischer}, \citenamefont {Savenko}, \citenamefont {Amthor}, \citenamefont
  {Lermer}, \citenamefont {Wolf}, \citenamefont {Worschech}, \citenamefont
  {Kulakovskii}, \citenamefont {Shelykh}, \citenamefont {Kamp}, \citenamefont
  {Reitzenstein}, \citenamefont {Forchel}, \citenamefont {Yamamoto},\ and\
  \citenamefont {H{\"{o}}fling}}]{Schneider2013}%
  \BibitemOpen
  \bibfield  {author} {\bibinfo {author} {\bibfnamefont {C.}~\bibnamefont
  {Schneider}}, \bibinfo {author} {\bibfnamefont {A.}~\bibnamefont
  {Rahimi-Iman}}, \bibinfo {author} {\bibfnamefont {N.~Y.}\ \bibnamefont
  {Kim}}, \bibinfo {author} {\bibfnamefont {J.}~\bibnamefont {Fischer}},
  \bibinfo {author} {\bibfnamefont {I.~G.}\ \bibnamefont {Savenko}}, \bibinfo
  {author} {\bibfnamefont {M.}~\bibnamefont {Amthor}}, \bibinfo {author}
  {\bibfnamefont {M.}~\bibnamefont {Lermer}}, \bibinfo {author} {\bibfnamefont
  {A.}~\bibnamefont {Wolf}}, \bibinfo {author} {\bibfnamefont {L.}~\bibnamefont
  {Worschech}}, \bibinfo {author} {\bibfnamefont {V.~D.}\ \bibnamefont
  {Kulakovskii}}, \bibinfo {author} {\bibfnamefont {I.~A.}\ \bibnamefont
  {Shelykh}}, \bibinfo {author} {\bibfnamefont {M.}~\bibnamefont {Kamp}},
  \bibinfo {author} {\bibfnamefont {S.}~\bibnamefont {Reitzenstein}}, \bibinfo
  {author} {\bibfnamefont {A.}~\bibnamefont {Forchel}}, \bibinfo {author}
  {\bibfnamefont {Y.}~\bibnamefont {Yamamoto}}, \ and\ \bibinfo {author}
  {\bibfnamefont {S.}~\bibnamefont {H{\"{o}}fling}},\ }\bibfield  {title}
  {\bibinfo {title} {\emph {{An electrically pumped polariton laser}}},\ }\href
  {\doibase 10.1038/nature12036} {\bibfield  {journal} {\bibinfo  {journal}
  {Nature}\ }\textbf {\bibinfo {volume} {497}},\ \bibinfo {pages} {348}
  (\bibinfo {year} {2013})}\BibitemShut {NoStop}%
\bibitem [{\citenamefont {Bhattacharya}\ \emph {et~al.}(2013)\citenamefont
  {Bhattacharya}, \citenamefont {Xiao}, \citenamefont {Das}, \citenamefont
  {Bhowmick},\ and\ \citenamefont {Heo}}]{Bhattacharya2013}%
  \BibitemOpen
  \bibfield  {author} {\bibinfo {author} {\bibfnamefont {P.}~\bibnamefont
  {Bhattacharya}}, \bibinfo {author} {\bibfnamefont {B.}~\bibnamefont {Xiao}},
  \bibinfo {author} {\bibfnamefont {A.}~\bibnamefont {Das}}, \bibinfo {author}
  {\bibfnamefont {S.}~\bibnamefont {Bhowmick}}, \ and\ \bibinfo {author}
  {\bibfnamefont {J.}~\bibnamefont {Heo}},\ }\bibfield  {title} {\bibinfo
  {title} {\emph {Solid State Electrically Injected Exciton-Polariton Laser}},\
  }\href {\doibase 10.1103/PhysRevLett.110.206403} {\bibfield  {journal}
  {\bibinfo  {journal} {Phys. Rev. Lett.}\ }\textbf {\bibinfo {volume} {110}},\
  \bibinfo {pages} {206403} (\bibinfo {year} {2013})}\BibitemShut {NoStop}%
\bibitem [{\citenamefont {Ferrier}\ \emph {et~al.}(2011)\citenamefont
  {Ferrier}, \citenamefont {Wertz}, \citenamefont {Johne}, \citenamefont
  {Solnyshkov}, \citenamefont {Senellart}, \citenamefont {Sagnes},
  \citenamefont {Lema\^{\i}tre}, \citenamefont {Malpuech},\ and\ \citenamefont
  {Bloch}}]{Ferrier2011}%
  \BibitemOpen
  \bibfield  {author} {\bibinfo {author} {\bibfnamefont {L.}~\bibnamefont
  {Ferrier}}, \bibinfo {author} {\bibfnamefont {E.}~\bibnamefont {Wertz}},
  \bibinfo {author} {\bibfnamefont {R.}~\bibnamefont {Johne}}, \bibinfo
  {author} {\bibfnamefont {D.~D.}\ \bibnamefont {Solnyshkov}}, \bibinfo
  {author} {\bibfnamefont {P.}~\bibnamefont {Senellart}}, \bibinfo {author}
  {\bibfnamefont {I.}~\bibnamefont {Sagnes}}, \bibinfo {author} {\bibfnamefont
  {A.}~\bibnamefont {Lema\^{\i}tre}}, \bibinfo {author} {\bibfnamefont
  {G.}~\bibnamefont {Malpuech}}, \ and\ \bibinfo {author} {\bibfnamefont
  {J.}~\bibnamefont {Bloch}},\ }\bibfield  {title} {\bibinfo {title} {\emph
  {Interactions in Confined Polariton Condensates}},\ }\href {\doibase
  10.1103/PhysRevLett.106.126401} {\bibfield  {journal} {\bibinfo  {journal}
  {Phys. Rev. Lett.}\ }\textbf {\bibinfo {volume} {106}},\ \bibinfo {pages}
  {126401} (\bibinfo {year} {2011})}\BibitemShut {NoStop}%
\bibitem [{\citenamefont {Schmidt}\ \emph {et~al.}(2019)\citenamefont
  {Schmidt}, \citenamefont {Berger}, \citenamefont {Kahlert}, \citenamefont
  {Bayer}, \citenamefont {Schneider}, \citenamefont {H\"ofling}, \citenamefont
  {Sedov}, \citenamefont {Kavokin},\ and\ \citenamefont
  {A\ss{}mann}}]{Schmidt2019}%
  \BibitemOpen
  \bibfield  {author} {\bibinfo {author} {\bibfnamefont {D.}~\bibnamefont
  {Schmidt}}, \bibinfo {author} {\bibfnamefont {B.}~\bibnamefont {Berger}},
  \bibinfo {author} {\bibfnamefont {M.}~\bibnamefont {Kahlert}}, \bibinfo
  {author} {\bibfnamefont {M.}~\bibnamefont {Bayer}}, \bibinfo {author}
  {\bibfnamefont {C.}~\bibnamefont {Schneider}}, \bibinfo {author}
  {\bibfnamefont {S.}~\bibnamefont {H\"ofling}}, \bibinfo {author}
  {\bibfnamefont {E.~S.}\ \bibnamefont {Sedov}}, \bibinfo {author}
  {\bibfnamefont {A.~V.}\ \bibnamefont {Kavokin}}, \ and\ \bibinfo {author}
  {\bibfnamefont {M.}~\bibnamefont {A\ss{}mann}},\ }\bibfield  {title}
  {\bibinfo {title} {\emph {Tracking Dark Excitons with Exciton Polaritons in
  Semiconductor Microcavities}},\ }\href {\doibase
  10.1103/PhysRevLett.122.047403} {\bibfield  {journal} {\bibinfo  {journal}
  {Phys. Rev. Lett.}\ }\textbf {\bibinfo {volume} {122}},\ \bibinfo {pages}
  {047403} (\bibinfo {year} {2019})}\BibitemShut {NoStop}%
\bibitem [{\citenamefont {Brichkin}\ \emph {et~al.}(2011)\citenamefont
  {Brichkin}, \citenamefont {Novikov}, \citenamefont {Larionov}, \citenamefont
  {Kulakovskii}, \citenamefont {Glazov}, \citenamefont {Schneider},
  \citenamefont {H{\"{o}}fling}, \citenamefont {Kamp},\ and\ \citenamefont
  {Forchel}}]{Brichkin2011}%
  \BibitemOpen
  \bibfield  {author} {\bibinfo {author} {\bibfnamefont {A.~S.}\ \bibnamefont
  {Brichkin}}, \bibinfo {author} {\bibfnamefont {S.~I.}\ \bibnamefont
  {Novikov}}, \bibinfo {author} {\bibfnamefont {A.~V.}\ \bibnamefont
  {Larionov}}, \bibinfo {author} {\bibfnamefont {V.~D.}\ \bibnamefont
  {Kulakovskii}}, \bibinfo {author} {\bibfnamefont {M.~M.}\ \bibnamefont
  {Glazov}}, \bibinfo {author} {\bibfnamefont {C.}~\bibnamefont {Schneider}},
  \bibinfo {author} {\bibfnamefont {S.}~\bibnamefont {H{\"{o}}fling}}, \bibinfo
  {author} {\bibfnamefont {M.}~\bibnamefont {Kamp}}, \ and\ \bibinfo {author}
  {\bibfnamefont {A.}~\bibnamefont {Forchel}},\ }\bibfield  {title} {\bibinfo
  {title} {\emph {{Effect of Coulomb interaction on exciton-polariton
  condensates in GaAs pillar microcavities}}},\ }\href {\doibase
  10.1103/PhysRevB.84.195301} {\bibfield  {journal} {\bibinfo  {journal} {Phys.
  Rev. B}\ }\textbf {\bibinfo {volume} {84}},\ \bibinfo {pages} {195301}
  (\bibinfo {year} {2011})}\BibitemShut {NoStop}%
\bibitem [{\citenamefont {Klaas}\ \emph {et~al.}(2017)\citenamefont {Klaas},
  \citenamefont {Mandal}, \citenamefont {Liew}, \citenamefont {Amthor},
  \citenamefont {Klembt}, \citenamefont {Worschech}, \citenamefont
  {Schneider},\ and\ \citenamefont {Höfling}}]{Klaas2017}%
  \BibitemOpen
  \bibfield  {author} {\bibinfo {author} {\bibfnamefont {M.}~\bibnamefont
  {Klaas}}, \bibinfo {author} {\bibfnamefont {S.}~\bibnamefont {Mandal}},
  \bibinfo {author} {\bibfnamefont {T.~C.~H.}\ \bibnamefont {Liew}}, \bibinfo
  {author} {\bibfnamefont {M.}~\bibnamefont {Amthor}}, \bibinfo {author}
  {\bibfnamefont {S.}~\bibnamefont {Klembt}}, \bibinfo {author} {\bibfnamefont
  {L.}~\bibnamefont {Worschech}}, \bibinfo {author} {\bibfnamefont
  {C.}~\bibnamefont {Schneider}}, \ and\ \bibinfo {author} {\bibfnamefont
  {S.}~\bibnamefont {Höfling}},\ }\bibfield  {title} {\bibinfo {title} {\emph
  {Optical probing of the Coulomb interactions of an electrically pumped
  polariton condensate}},\ }\href {\doibase 10.1063/1.4979836} {\bibfield
  {journal} {\bibinfo  {journal} {Appl. Phys. Lett.}\ }\textbf {\bibinfo
  {volume} {110}},\ \bibinfo {pages} {151103} (\bibinfo {year}
  {2017})}\BibitemShut {NoStop}%
\bibitem [{\citenamefont {Askitopoulos}\ \emph {et~al.}(2013)\citenamefont
  {Askitopoulos}, \citenamefont {Ohadi}, \citenamefont {Kavokin}, \citenamefont
  {Hatzopoulos}, \citenamefont {Savvidis},\ and\ \citenamefont
  {Lagoudakis}}]{Askitopoulos2013}%
  \BibitemOpen
  \bibfield  {author} {\bibinfo {author} {\bibfnamefont {A.}~\bibnamefont
  {Askitopoulos}}, \bibinfo {author} {\bibfnamefont {H.}~\bibnamefont {Ohadi}},
  \bibinfo {author} {\bibfnamefont {A.~V.}\ \bibnamefont {Kavokin}}, \bibinfo
  {author} {\bibfnamefont {Z.}~\bibnamefont {Hatzopoulos}}, \bibinfo {author}
  {\bibfnamefont {P.~G.}\ \bibnamefont {Savvidis}}, \ and\ \bibinfo {author}
  {\bibfnamefont {P.~G.}\ \bibnamefont {Lagoudakis}},\ }\bibfield  {title}
  {\bibinfo {title} {\emph {Polariton condensation in an optically induced
  two-dimensional potential}},\ }\href {\doibase 10.1103/PhysRevB.88.041308}
  {\bibfield  {journal} {\bibinfo  {journal} {Phys. Rev. B}\ }\textbf {\bibinfo
  {volume} {88}},\ \bibinfo {pages} {041308(R)} (\bibinfo {year}
  {2013})}\BibitemShut {NoStop}%
\bibitem [{\citenamefont {Cristofolini}\ \emph {et~al.}(2013)\citenamefont
  {Cristofolini}, \citenamefont {Dreismann}, \citenamefont {Christmann},
  \citenamefont {Franchetti}, \citenamefont {Berloff}, \citenamefont {Tsotsis},
  \citenamefont {Hatzopoulos}, \citenamefont {Savvidis},\ and\ \citenamefont
  {Baumberg}}]{Cristofolini2013}%
  \BibitemOpen
  \bibfield  {author} {\bibinfo {author} {\bibfnamefont {P.}~\bibnamefont
  {Cristofolini}}, \bibinfo {author} {\bibfnamefont {A.}~\bibnamefont
  {Dreismann}}, \bibinfo {author} {\bibfnamefont {G.}~\bibnamefont
  {Christmann}}, \bibinfo {author} {\bibfnamefont {G.}~\bibnamefont
  {Franchetti}}, \bibinfo {author} {\bibfnamefont {N.~G.}\ \bibnamefont
  {Berloff}}, \bibinfo {author} {\bibfnamefont {P.}~\bibnamefont {Tsotsis}},
  \bibinfo {author} {\bibfnamefont {Z.}~\bibnamefont {Hatzopoulos}}, \bibinfo
  {author} {\bibfnamefont {P.~G.}\ \bibnamefont {Savvidis}}, \ and\ \bibinfo
  {author} {\bibfnamefont {J.~J.}\ \bibnamefont {Baumberg}},\ }\bibfield
  {title} {\bibinfo {title} {\emph {Optical Superfluid Phase Transitions and
  Trapping of Polariton Condensates}},\ }\href {\doibase
  10.1103/PhysRevLett.110.186403} {\bibfield  {journal} {\bibinfo  {journal}
  {Phys. Rev. Lett.}\ }\textbf {\bibinfo {volume} {110}},\ \bibinfo {pages}
  {186403} (\bibinfo {year} {2013})}\BibitemShut {NoStop}%
\bibitem [{\citenamefont {Dall}\ \emph {et~al.}(2014)\citenamefont {Dall},
  \citenamefont {Fraser}, \citenamefont {Desyatnikov}, \citenamefont {Li},
  \citenamefont {Brodbeck}, \citenamefont {Kamp}, \citenamefont {Schneider},
  \citenamefont {H\"ofling},\ and\ \citenamefont {Ostrovskaya}}]{Dall2014}%
  \BibitemOpen
  \bibfield  {author} {\bibinfo {author} {\bibfnamefont {R.}~\bibnamefont
  {Dall}}, \bibinfo {author} {\bibfnamefont {M.~D.}\ \bibnamefont {Fraser}},
  \bibinfo {author} {\bibfnamefont {A.~S.}\ \bibnamefont {Desyatnikov}},
  \bibinfo {author} {\bibfnamefont {G.}~\bibnamefont {Li}}, \bibinfo {author}
  {\bibfnamefont {S.}~\bibnamefont {Brodbeck}}, \bibinfo {author}
  {\bibfnamefont {M.}~\bibnamefont {Kamp}}, \bibinfo {author} {\bibfnamefont
  {C.}~\bibnamefont {Schneider}}, \bibinfo {author} {\bibfnamefont
  {S.}~\bibnamefont {H\"ofling}}, \ and\ \bibinfo {author} {\bibfnamefont
  {E.~A.}\ \bibnamefont {Ostrovskaya}},\ }\bibfield  {title} {\bibinfo {title}
  {\emph {Creation of Orbital Angular Momentum States with Chiral Polaritonic
  Lenses}},\ }\href {\doibase 10.1103/PhysRevLett.113.200404} {\bibfield
  {journal} {\bibinfo  {journal} {Phys. Rev. Lett.}\ }\textbf {\bibinfo
  {volume} {113}},\ \bibinfo {pages} {200404} (\bibinfo {year}
  {2014})}\BibitemShut {NoStop}%
\bibitem [{\citenamefont {Paschos}\ \emph {et~al.}(2020)\citenamefont
  {Paschos}, \citenamefont {Tzimis}, \citenamefont {Tsintzos},\ and\
  \citenamefont {Savvidis}}]{Paschos2020}%
  \BibitemOpen
  \bibfield  {author} {\bibinfo {author} {\bibfnamefont {G.~G.}\ \bibnamefont
  {Paschos}}, \bibinfo {author} {\bibfnamefont {A.}~\bibnamefont {Tzimis}},
  \bibinfo {author} {\bibfnamefont {S.~I.}\ \bibnamefont {Tsintzos}}, \ and\
  \bibinfo {author} {\bibfnamefont {P.~G.}\ \bibnamefont {Savvidis}},\
  }\bibfield  {title} {\bibinfo {title} {\emph {{Polariton condensate trapping
  by parametric pair scattering}}},\ }\href {\doibase 10.1088/1361-648X/ab9267}
  {\bibfield  {journal} {\bibinfo  {journal} {J. Phys. Condens. Matter}\
  }\textbf {\bibinfo {volume} {32}},\ \bibinfo {pages} {6} (\bibinfo {year}
  {2020})}\BibitemShut {NoStop}%
\bibitem [{\citenamefont {Orfanakis}\ \emph {et~al.}(2021)\citenamefont
  {Orfanakis}, \citenamefont {Tzortzakakis}, \citenamefont {Petrosyan},
  \citenamefont {Savvidis},\ and\ \citenamefont {Ohadi}}]{Orfanakis2021}%
  \BibitemOpen
  \bibfield  {author} {\bibinfo {author} {\bibfnamefont {K.}~\bibnamefont
  {Orfanakis}}, \bibinfo {author} {\bibfnamefont {A.~F.}\ \bibnamefont
  {Tzortzakakis}}, \bibinfo {author} {\bibfnamefont {D.}~\bibnamefont
  {Petrosyan}}, \bibinfo {author} {\bibfnamefont {P.~G.}\ \bibnamefont
  {Savvidis}}, \ and\ \bibinfo {author} {\bibfnamefont {H.}~\bibnamefont
  {Ohadi}},\ }\bibfield  {title} {\bibinfo {title} {\emph {{Ultralong temporal
  coherence in optically trapped exciton-polariton condensates}}},\ }\href
  {\doibase 10.1103/PhysRevB.103.235313} {\bibfield  {journal} {\bibinfo
  {journal} {Phys. Rev. B}\ }\textbf {\bibinfo {volume} {103}},\ \bibinfo
  {pages} {235313} (\bibinfo {year} {2021})}\BibitemShut {NoStop}%
\bibitem [{\citenamefont {Keeling}\ \emph {et~al.}(2007)\citenamefont
  {Keeling}, \citenamefont {Marchetti}, \citenamefont {Szyma{\'{n}}ska},\ and\
  \citenamefont {Littlewood}}]{Keeling2007}%
  \BibitemOpen
  \bibfield  {author} {\bibinfo {author} {\bibfnamefont {J.}~\bibnamefont
  {Keeling}}, \bibinfo {author} {\bibfnamefont {F.~M.}\ \bibnamefont
  {Marchetti}}, \bibinfo {author} {\bibfnamefont {M.~H.}\ \bibnamefont
  {Szyma{\'{n}}ska}}, \ and\ \bibinfo {author} {\bibfnamefont {P.~B.}\
  \bibnamefont {Littlewood}},\ }\bibfield  {title} {\bibinfo {title} {\emph
  {Collective coherence in planar semiconductor microcavities}},\ }\href
  {\doibase 10.1088/0268-1242/22/5/r01} {\bibfield  {journal} {\bibinfo
  {journal} {Semiconductor Science and Technology}\ }\textbf {\bibinfo {volume}
  {22}},\ \bibinfo {pages} {R1} (\bibinfo {year} {2007})}\BibitemShut {NoStop}%
\bibitem [{\citenamefont {Szyma\ifmmode~\acute{n}\else \'{n}\fi{}ska}\ \emph
  {et~al.}(2006)\citenamefont {Szyma\ifmmode~\acute{n}\else \'{n}\fi{}ska},
  \citenamefont {Keeling},\ and\ \citenamefont {Littlewood}}]{Szymanska2006}%
  \BibitemOpen
  \bibfield  {author} {\bibinfo {author} {\bibfnamefont {M.~H.}\ \bibnamefont
  {Szyma\ifmmode~\acute{n}\else \'{n}\fi{}ska}}, \bibinfo {author}
  {\bibfnamefont {J.}~\bibnamefont {Keeling}}, \ and\ \bibinfo {author}
  {\bibfnamefont {P.~B.}\ \bibnamefont {Littlewood}},\ }\bibfield  {title}
  {\bibinfo {title} {\emph {Nonequilibrium Quantum Condensation in an
  Incoherently Pumped Dissipative System}},\ }\href {\doibase
  10.1103/PhysRevLett.96.230602} {\bibfield  {journal} {\bibinfo  {journal}
  {Phys. Rev. Lett.}\ }\textbf {\bibinfo {volume} {96}},\ \bibinfo {pages}
  {230602} (\bibinfo {year} {2006})}\BibitemShut {NoStop}%
\bibitem [{\citenamefont {Wouters}\ and\ \citenamefont
  {Carusotto}(2007)}]{Wouters2007}%
  \BibitemOpen
  \bibfield  {author} {\bibinfo {author} {\bibfnamefont {M.}~\bibnamefont
  {Wouters}}\ and\ \bibinfo {author} {\bibfnamefont {I.}~\bibnamefont
  {Carusotto}},\ }\bibfield  {title} {\bibinfo {title} {\emph {Excitations in a
  Nonequilibrium Bose-Einstein Condensate of Exciton Polaritons}},\ }\href
  {\doibase 10.1103/PhysRevLett.99.140402} {\bibfield  {journal} {\bibinfo
  {journal} {Phys. Rev. Lett.}\ }\textbf {\bibinfo {volume} {99}},\ \bibinfo
  {pages} {140402} (\bibinfo {year} {2007})}\BibitemShut {NoStop}%
\bibitem [{\citenamefont {Hanai}\ \emph {et~al.}(2018)\citenamefont {Hanai},
  \citenamefont {Littlewood},\ and\ \citenamefont {Ohashi}}]{Hanai2018}%
  \BibitemOpen
  \bibfield  {author} {\bibinfo {author} {\bibfnamefont {R.}~\bibnamefont
  {Hanai}}, \bibinfo {author} {\bibfnamefont {P.~B.}\ \bibnamefont
  {Littlewood}}, \ and\ \bibinfo {author} {\bibfnamefont {Y.}~\bibnamefont
  {Ohashi}},\ }\bibfield  {title} {\bibinfo {title} {\emph {Photoluminescence
  and gain/absorption spectra of a driven-dissipative electron-hole-photon
  condensate}},\ }\href {\doibase 10.1103/PhysRevB.97.245302} {\bibfield
  {journal} {\bibinfo  {journal} {Phys. Rev. B}\ }\textbf {\bibinfo {volume}
  {97}},\ \bibinfo {pages} {245302} (\bibinfo {year} {2018})}\BibitemShut
  {NoStop}%
\bibitem [{\citenamefont {Byrnes}\ \emph {et~al.}(2012)\citenamefont {Byrnes},
  \citenamefont {Horikiri}, \citenamefont {Ishida}, \citenamefont {Fraser},\
  and\ \citenamefont {Yamamoto}}]{Byrnes2012}%
  \BibitemOpen
  \bibfield  {author} {\bibinfo {author} {\bibfnamefont {T.}~\bibnamefont
  {Byrnes}}, \bibinfo {author} {\bibfnamefont {T.}~\bibnamefont {Horikiri}},
  \bibinfo {author} {\bibfnamefont {N.}~\bibnamefont {Ishida}}, \bibinfo
  {author} {\bibfnamefont {M.}~\bibnamefont {Fraser}}, \ and\ \bibinfo {author}
  {\bibfnamefont {Y.}~\bibnamefont {Yamamoto}},\ }\bibfield  {title} {\bibinfo
  {title} {\emph {Negative Bogoliubov dispersion in exciton-polariton
  condensates}},\ }\href {\doibase 10.1103/PhysRevB.85.075130} {\bibfield
  {journal} {\bibinfo  {journal} {Phys. Rev. B}\ }\textbf {\bibinfo {volume}
  {85}},\ \bibinfo {pages} {075130} (\bibinfo {year} {2012})}\BibitemShut
  {NoStop}%
\bibitem [{\citenamefont {Solnyshkov}\ \emph {et~al.}(2014)\citenamefont
  {Solnyshkov}, \citenamefont {Ter\ifmmode~\mbox{\c{c}}\else \c{c}\fi{}as},
  \citenamefont {Dini},\ and\ \citenamefont {Malpuech}}]{Solnyshkov2014}%
  \BibitemOpen
  \bibfield  {author} {\bibinfo {author} {\bibfnamefont {D.~D.}\ \bibnamefont
  {Solnyshkov}}, \bibinfo {author} {\bibfnamefont {H.}~\bibnamefont
  {Ter\ifmmode~\mbox{\c{c}}\else \c{c}\fi{}as}}, \bibinfo {author}
  {\bibfnamefont {K.}~\bibnamefont {Dini}}, \ and\ \bibinfo {author}
  {\bibfnamefont {G.}~\bibnamefont {Malpuech}},\ }\bibfield  {title} {\bibinfo
  {title} {\emph {Hybrid Boltzmann--Gross-Pitaevskii theory of Bose-Einstein
  condensation and superfluidity in open driven-dissipative systems}},\ }\href
  {\doibase 10.1103/PhysRevA.89.033626} {\bibfield  {journal} {\bibinfo
  {journal} {Phys. Rev. A}\ }\textbf {\bibinfo {volume} {89}},\ \bibinfo
  {pages} {033626} (\bibinfo {year} {2014})}\BibitemShut {NoStop}%
\bibitem [{\citenamefont {Ballarini}\ \emph {et~al.}(2020)\citenamefont
  {Ballarini}, \citenamefont {Caputo}, \citenamefont {Dagvadorj}, \citenamefont
  {Juggins}, \citenamefont {Giorgi}, \citenamefont {Dominici}, \citenamefont
  {West}, \citenamefont {Pfeiffer}, \citenamefont {Gigli}, \citenamefont
  {Szymańska},\ and\ \citenamefont {Sanvitto}}]{Ballarini2020}%
  \BibitemOpen
  \bibfield  {author} {\bibinfo {author} {\bibfnamefont {D.}~\bibnamefont
  {Ballarini}}, \bibinfo {author} {\bibfnamefont {D.}~\bibnamefont {Caputo}},
  \bibinfo {author} {\bibfnamefont {G.}~\bibnamefont {Dagvadorj}}, \bibinfo
  {author} {\bibfnamefont {R.}~\bibnamefont {Juggins}}, \bibinfo {author}
  {\bibfnamefont {M.~D.}\ \bibnamefont {Giorgi}}, \bibinfo {author}
  {\bibfnamefont {L.}~\bibnamefont {Dominici}}, \bibinfo {author}
  {\bibfnamefont {K.}~\bibnamefont {West}}, \bibinfo {author} {\bibfnamefont
  {L.~N.}\ \bibnamefont {Pfeiffer}}, \bibinfo {author} {\bibfnamefont
  {G.}~\bibnamefont {Gigli}}, \bibinfo {author} {\bibfnamefont {M.~H.}\
  \bibnamefont {Szymańska}}, \ and\ \bibinfo {author} {\bibfnamefont
  {D.}~\bibnamefont {Sanvitto}},\ }\bibfield  {title} {\bibinfo {title} {\emph
  {Directional Goldstone waves in polariton condensates close to
  equilibrium}},\ }\href {https://doi.org/10.1038/s41467-019-13733-x}
  {\bibfield  {journal} {\bibinfo  {journal} {Nat. Commun.}\ }\textbf {\bibinfo
  {volume} {11}},\ \bibinfo {pages} {217} (\bibinfo {year} {2020})}\BibitemShut
  {NoStop}%
\bibitem [{\citenamefont {Utsunomiya}\ \emph {et~al.}(2008)\citenamefont
  {Utsunomiya}, \citenamefont {Tian}, \citenamefont {Roumpos}, \citenamefont
  {Lai}, \citenamefont {Kumada}, \citenamefont {Fujisawa}, \citenamefont
  {Kuwata-Gonokami}, \citenamefont {Löffler}, \citenamefont {Höfling},
  \citenamefont {Forchel},\ and\ \citenamefont {Yamamoto}}]{Utsunomiya2008}%
  \BibitemOpen
  \bibfield  {author} {\bibinfo {author} {\bibfnamefont {S.}~\bibnamefont
  {Utsunomiya}}, \bibinfo {author} {\bibfnamefont {L.}~\bibnamefont {Tian}},
  \bibinfo {author} {\bibfnamefont {G.}~\bibnamefont {Roumpos}}, \bibinfo
  {author} {\bibfnamefont {C.~W.}\ \bibnamefont {Lai}}, \bibinfo {author}
  {\bibfnamefont {N.}~\bibnamefont {Kumada}}, \bibinfo {author} {\bibfnamefont
  {T.}~\bibnamefont {Fujisawa}}, \bibinfo {author} {\bibfnamefont
  {M.}~\bibnamefont {Kuwata-Gonokami}}, \bibinfo {author} {\bibfnamefont
  {A.}~\bibnamefont {Löffler}}, \bibinfo {author} {\bibfnamefont
  {S.}~\bibnamefont {Höfling}}, \bibinfo {author} {\bibfnamefont
  {A.}~\bibnamefont {Forchel}}, \ and\ \bibinfo {author} {\bibfnamefont
  {Y.}~\bibnamefont {Yamamoto}},\ }\bibfield  {title} {\bibinfo {title} {\emph
  {Observation of Bogoliubov excitations in exciton-polariton condensates}},\
  }\href {https://doi.org/10.1038/nphys1034} {\bibfield  {journal} {\bibinfo
  {journal} {Nat. Phys.}\ }\textbf {\bibinfo {volume} {4}},\ \bibinfo {pages}
  {700} (\bibinfo {year} {2008})}\BibitemShut {NoStop}%
\bibitem [{\citenamefont {Kohnle}\ \emph {et~al.}(2011)\citenamefont {Kohnle},
  \citenamefont {L\'eger}, \citenamefont {Wouters}, \citenamefont {Richard},
  \citenamefont {Portella-Oberli},\ and\ \citenamefont
  {Deveaud-Pl\'edran}}]{Kohnle2011}%
  \BibitemOpen
  \bibfield  {author} {\bibinfo {author} {\bibfnamefont {V.}~\bibnamefont
  {Kohnle}}, \bibinfo {author} {\bibfnamefont {Y.}~\bibnamefont {L\'eger}},
  \bibinfo {author} {\bibfnamefont {M.}~\bibnamefont {Wouters}}, \bibinfo
  {author} {\bibfnamefont {M.}~\bibnamefont {Richard}}, \bibinfo {author}
  {\bibfnamefont {M.~T.}\ \bibnamefont {Portella-Oberli}}, \ and\ \bibinfo
  {author} {\bibfnamefont {B.}~\bibnamefont {Deveaud-Pl\'edran}},\ }\bibfield
  {title} {\bibinfo {title} {\emph {From Single Particle to Superfluid
  Excitations in a Dissipative Polariton Gas}},\ }\href {\doibase
  10.1103/PhysRevLett.106.255302} {\bibfield  {journal} {\bibinfo  {journal}
  {Phys. Rev. Lett.}\ }\textbf {\bibinfo {volume} {106}},\ \bibinfo {pages}
  {255302} (\bibinfo {year} {2011})}\BibitemShut {NoStop}%
\bibitem [{\citenamefont {Pieczarka}\ \emph {et~al.}(2015)\citenamefont
  {Pieczarka}, \citenamefont {Syperek}, \citenamefont {Dusanowski},
  \citenamefont {Misiewicz}, \citenamefont {Langer}, \citenamefont {Forchel},
  \citenamefont {Kamp}, \citenamefont {Schneider}, \citenamefont {H\"ofling},
  \citenamefont {Kavokin},\ and\ \citenamefont {Sek}}]{Pieczarka2015}%
  \BibitemOpen
  \bibfield  {author} {\bibinfo {author} {\bibfnamefont {M.}~\bibnamefont
  {Pieczarka}}, \bibinfo {author} {\bibfnamefont {M.}~\bibnamefont {Syperek}},
  \bibinfo {author} {\bibfnamefont {L.}~\bibnamefont {Dusanowski}}, \bibinfo
  {author} {\bibfnamefont {J.}~\bibnamefont {Misiewicz}}, \bibinfo {author}
  {\bibfnamefont {F.}~\bibnamefont {Langer}}, \bibinfo {author} {\bibfnamefont
  {A.}~\bibnamefont {Forchel}}, \bibinfo {author} {\bibfnamefont
  {M.}~\bibnamefont {Kamp}}, \bibinfo {author} {\bibfnamefont {C.}~\bibnamefont
  {Schneider}}, \bibinfo {author} {\bibfnamefont {S.}~\bibnamefont
  {H\"ofling}}, \bibinfo {author} {\bibfnamefont {A.}~\bibnamefont {Kavokin}},
  \ and\ \bibinfo {author} {\bibfnamefont {G.}~\bibnamefont {Sek}},\ }\bibfield
   {title} {\bibinfo {title} {\emph {Ghost Branch Photoluminescence From a
  Polariton Fluid Under Nonresonant Excitation}},\ }\href {\doibase
  10.1103/PhysRevLett.115.186401} {\bibfield  {journal} {\bibinfo  {journal}
  {Phys. Rev. Lett.}\ }\textbf {\bibinfo {volume} {115}},\ \bibinfo {pages}
  {186401} (\bibinfo {year} {2015})}\BibitemShut {NoStop}%
\bibitem [{\citenamefont {Zajac}\ and\ \citenamefont
  {Langbein}(2015)}]{Zajac2015}%
  \BibitemOpen
  \bibfield  {author} {\bibinfo {author} {\bibfnamefont {J.~M.}\ \bibnamefont
  {Zajac}}\ and\ \bibinfo {author} {\bibfnamefont {W.}~\bibnamefont
  {Langbein}},\ }\bibfield  {title} {\bibinfo {title} {\emph {Parametric
  scattering of microcavity polaritons into ghost branches}},\ }\href {\doibase
  10.1103/PhysRevB.92.165305} {\bibfield  {journal} {\bibinfo  {journal} {Phys.
  Rev. B}\ }\textbf {\bibinfo {volume} {92}},\ \bibinfo {pages} {165305}
  (\bibinfo {year} {2015})}\BibitemShut {NoStop}%
\bibitem [{\citenamefont {Horikiri}\ \emph {et~al.}(2017)\citenamefont
  {Horikiri}, \citenamefont {Byrnes}, \citenamefont {Kusudo}, \citenamefont
  {Ishida}, \citenamefont {Matsuo}, \citenamefont {Shikano}, \citenamefont
  {L\"offler}, \citenamefont {H\"ofling}, \citenamefont {Forchel},\ and\
  \citenamefont {Yamamoto}}]{Horikiri2017}%
  \BibitemOpen
  \bibfield  {author} {\bibinfo {author} {\bibfnamefont {T.}~\bibnamefont
  {Horikiri}}, \bibinfo {author} {\bibfnamefont {T.}~\bibnamefont {Byrnes}},
  \bibinfo {author} {\bibfnamefont {K.}~\bibnamefont {Kusudo}}, \bibinfo
  {author} {\bibfnamefont {N.}~\bibnamefont {Ishida}}, \bibinfo {author}
  {\bibfnamefont {Y.}~\bibnamefont {Matsuo}}, \bibinfo {author} {\bibfnamefont
  {Y.}~\bibnamefont {Shikano}}, \bibinfo {author} {\bibfnamefont
  {A.}~\bibnamefont {L\"offler}}, \bibinfo {author} {\bibfnamefont
  {S.}~\bibnamefont {H\"ofling}}, \bibinfo {author} {\bibfnamefont
  {A.}~\bibnamefont {Forchel}}, \ and\ \bibinfo {author} {\bibfnamefont
  {Y.}~\bibnamefont {Yamamoto}},\ }\bibfield  {title} {\bibinfo {title} {\emph
  {Highly excited exciton-polariton condensates}},\ }\href {\doibase
  10.1103/PhysRevB.95.245122} {\bibfield  {journal} {\bibinfo  {journal} {Phys.
  Rev. B}\ }\textbf {\bibinfo {volume} {95}},\ \bibinfo {pages} {245122}
  (\bibinfo {year} {2017})}\BibitemShut {NoStop}%
\bibitem [{\citenamefont {Stepanov}\ \emph {et~al.}(2019)\citenamefont
  {Stepanov}, \citenamefont {Amelio}, \citenamefont {Rousset}, \citenamefont
  {Bloch}, \citenamefont {Lemaître}, \citenamefont {Amo}, \citenamefont
  {Minguzzi}, \citenamefont {Carusotto},\ and\ \citenamefont
  {Richard}}]{Stepanov2019}%
  \BibitemOpen
  \bibfield  {author} {\bibinfo {author} {\bibfnamefont {P.}~\bibnamefont
  {Stepanov}}, \bibinfo {author} {\bibfnamefont {I.}~\bibnamefont {Amelio}},
  \bibinfo {author} {\bibfnamefont {J.-G.}\ \bibnamefont {Rousset}}, \bibinfo
  {author} {\bibfnamefont {J.}~\bibnamefont {Bloch}}, \bibinfo {author}
  {\bibfnamefont {A.}~\bibnamefont {Lemaître}}, \bibinfo {author}
  {\bibfnamefont {A.}~\bibnamefont {Amo}}, \bibinfo {author} {\bibfnamefont
  {A.}~\bibnamefont {Minguzzi}}, \bibinfo {author} {\bibfnamefont
  {I.}~\bibnamefont {Carusotto}}, \ and\ \bibinfo {author} {\bibfnamefont
  {M.}~\bibnamefont {Richard}},\ }\bibfield  {title} {\bibinfo {title} {\emph
  {Dispersion relation of the collective excitations in a resonantly driven
  polariton fluid}},\ }\href {https://doi.org/10.1038/s41467-019-11886-3}
  {\bibfield  {journal} {\bibinfo  {journal} {Nat. Commun.}\ }\textbf {\bibinfo
  {volume} {10}},\ \bibinfo {pages} {3869} (\bibinfo {year}
  {2019})}\BibitemShut {NoStop}%
\bibitem [{\citenamefont {Pieczarka}\ \emph {et~al.}(2020)\citenamefont
  {Pieczarka}, \citenamefont {Estrecho}, \citenamefont {Boozarjmehr},
  \citenamefont {Bleu}, \citenamefont {Steger}, \citenamefont {West},
  \citenamefont {Pfeiffer}, \citenamefont {Snoke}, \citenamefont {Levinsen},
  \citenamefont {Parish}, \citenamefont {Truscott},\ and\ \citenamefont
  {Ostrovskaya}}]{Pieczarka2020}%
  \BibitemOpen
  \bibfield  {author} {\bibinfo {author} {\bibfnamefont {M.}~\bibnamefont
  {Pieczarka}}, \bibinfo {author} {\bibfnamefont {E.}~\bibnamefont {Estrecho}},
  \bibinfo {author} {\bibfnamefont {M.}~\bibnamefont {Boozarjmehr}}, \bibinfo
  {author} {\bibfnamefont {O.}~\bibnamefont {Bleu}}, \bibinfo {author}
  {\bibfnamefont {M.}~\bibnamefont {Steger}}, \bibinfo {author} {\bibfnamefont
  {K.}~\bibnamefont {West}}, \bibinfo {author} {\bibfnamefont {L.~N.}\
  \bibnamefont {Pfeiffer}}, \bibinfo {author} {\bibfnamefont {D.~W.}\
  \bibnamefont {Snoke}}, \bibinfo {author} {\bibfnamefont {J.}~\bibnamefont
  {Levinsen}}, \bibinfo {author} {\bibfnamefont {M.~M.}\ \bibnamefont
  {Parish}}, \bibinfo {author} {\bibfnamefont {A.~G.}\ \bibnamefont
  {Truscott}}, \ and\ \bibinfo {author} {\bibfnamefont {E.~A.}\ \bibnamefont
  {Ostrovskaya}},\ }\bibfield  {title} {\bibinfo {title} {\emph {Observation of
  quantum depletion in a non-equilibrium exciton-polariton condensate}},\
  }\href {https://doi.org/10.1038/s41467-019-14243-6} {\bibfield  {journal}
  {\bibinfo  {journal} {Nat. Commun.}\ }\textbf {\bibinfo {volume} {11}},\
  \bibinfo {pages} {429} (\bibinfo {year} {2020})}\BibitemShut {NoStop}%
\bibitem [{\citenamefont {Biega\ifmmode~\acute{n}\else \'{n}\fi{}ska}\ \emph
  {et~al.}(2021)\citenamefont {Biega\ifmmode~\acute{n}\else \'{n}\fi{}ska},
  \citenamefont {Pieczarka}, \citenamefont {Estrecho}, \citenamefont {Steger},
  \citenamefont {Snoke}, \citenamefont {West}, \citenamefont {Pfeiffer},
  \citenamefont {Syperek}, \citenamefont {Truscott},\ and\ \citenamefont
  {Ostrovskaya}}]{bieganska2020collective}%
  \BibitemOpen
  \bibfield  {author} {\bibinfo {author} {\bibfnamefont {D.}~\bibnamefont
  {Biega\ifmmode~\acute{n}\else \'{n}\fi{}ska}}, \bibinfo {author}
  {\bibfnamefont {M.}~\bibnamefont {Pieczarka}}, \bibinfo {author}
  {\bibfnamefont {E.}~\bibnamefont {Estrecho}}, \bibinfo {author}
  {\bibfnamefont {M.}~\bibnamefont {Steger}}, \bibinfo {author} {\bibfnamefont
  {D.~W.}\ \bibnamefont {Snoke}}, \bibinfo {author} {\bibfnamefont
  {K.}~\bibnamefont {West}}, \bibinfo {author} {\bibfnamefont {L.~N.}\
  \bibnamefont {Pfeiffer}}, \bibinfo {author} {\bibfnamefont {M.}~\bibnamefont
  {Syperek}}, \bibinfo {author} {\bibfnamefont {A.~G.}\ \bibnamefont
  {Truscott}}, \ and\ \bibinfo {author} {\bibfnamefont {E.~A.}\ \bibnamefont
  {Ostrovskaya}},\ }\bibfield  {title} {\bibinfo {title} {\emph {Collective
  Excitations of Exciton-Polariton Condensates in a Synthetic Gauge Field}},\
  }\href {\doibase 10.1103/PhysRevLett.127.185301} {\bibfield  {journal}
  {\bibinfo  {journal} {Phys. Rev. Lett.}\ }\textbf {\bibinfo {volume} {127}},\
  \bibinfo {pages} {185301} (\bibinfo {year} {2021})}\BibitemShut {NoStop}%
\bibitem [{\citenamefont {Steger}\ \emph {et~al.}(2021)\citenamefont {Steger},
  \citenamefont {Hanai}, \citenamefont {Edelman}, \citenamefont {Littlewood},
  \citenamefont {Snoke}, \citenamefont {Beaumariage}, \citenamefont {Fluegel},
  \citenamefont {West}, \citenamefont {Pfeiffer},\ and\ \citenamefont
  {Mascarenhas}}]{Steger2021}%
  \BibitemOpen
  \bibfield  {author} {\bibinfo {author} {\bibfnamefont {M.}~\bibnamefont
  {Steger}}, \bibinfo {author} {\bibfnamefont {R.}~\bibnamefont {Hanai}},
  \bibinfo {author} {\bibfnamefont {A.~O.}\ \bibnamefont {Edelman}}, \bibinfo
  {author} {\bibfnamefont {P.~B.}\ \bibnamefont {Littlewood}}, \bibinfo
  {author} {\bibfnamefont {D.~W.}\ \bibnamefont {Snoke}}, \bibinfo {author}
  {\bibfnamefont {J.}~\bibnamefont {Beaumariage}}, \bibinfo {author}
  {\bibfnamefont {B.}~\bibnamefont {Fluegel}}, \bibinfo {author} {\bibfnamefont
  {K.}~\bibnamefont {West}}, \bibinfo {author} {\bibfnamefont {L.~N.}\
  \bibnamefont {Pfeiffer}}, \ and\ \bibinfo {author} {\bibfnamefont
  {A.}~\bibnamefont {Mascarenhas}},\ }\bibfield  {title} {\bibinfo {title}
  {\emph {{Direct observation of the quantum fluctuation driven amplitude mode
  in a microcavity polariton condensate}}},\ }\href {\doibase
  10.1103/physrevb.103.205125} {\bibfield  {journal} {\bibinfo  {journal}
  {Phys. Rev. B}\ }\textbf {\bibinfo {volume} {103}},\ \bibinfo {pages}
  {205125} (\bibinfo {year} {2021})}\BibitemShut {NoStop}%
\bibitem [{\citenamefont {Bleu}\ \emph {et~al.}(2020)\citenamefont {Bleu},
  \citenamefont {Li}, \citenamefont {Levinsen},\ and\ \citenamefont
  {Parish}}]{Bleu2020}%
  \BibitemOpen
  \bibfield  {author} {\bibinfo {author} {\bibfnamefont {O.}~\bibnamefont
  {Bleu}}, \bibinfo {author} {\bibfnamefont {G.}~\bibnamefont {Li}}, \bibinfo
  {author} {\bibfnamefont {J.}~\bibnamefont {Levinsen}}, \ and\ \bibinfo
  {author} {\bibfnamefont {M.~M.}\ \bibnamefont {Parish}},\ }\bibfield  {title}
  {\bibinfo {title} {\emph {Polariton interactions in microcavities with
  atomically thin semiconductor layers}},\ }\href {\doibase
  10.1103/PhysRevResearch.2.043185} {\bibfield  {journal} {\bibinfo  {journal}
  {Phys. Rev. Research}\ }\textbf {\bibinfo {volume} {2}},\ \bibinfo {pages}
  {043185} (\bibinfo {year} {2020})}\BibitemShut {NoStop}%
\bibitem [{Note1()}]{Note1}%
  \BibitemOpen
  \bibinfo {note} {A significant tunneling between QWs would result in multiple
  polariton branches~\cite {Ouellet2015}, a feature that is absent in our
  sample.}\BibitemShut {Stop}%
\bibitem [{Note2()}]{Note2}%
  \BibitemOpen
  \bibinfo {note} {For convenience, we denote these superpositions as bright or
  dark regardless of whether the momentum is within or outside the radiative
  region}\BibitemShut {NoStop}%
\bibitem [{\citenamefont {Keeling}\ and\ \citenamefont
  {Kéna-Cohen}(2020)}]{Keeling2020}%
  \BibitemOpen
  \bibfield  {author} {\bibinfo {author} {\bibfnamefont {J.}~\bibnamefont
  {Keeling}}\ and\ \bibinfo {author} {\bibfnamefont {S.}~\bibnamefont
  {Kéna-Cohen}},\ }\bibfield  {title} {\bibinfo {title} {\emph
  {Bose–Einstein Condensation of Exciton-Polaritons in Organic
  Microcavities}},\ }\href {\doibase 10.1146/annurev-physchem-010920-102509}
  {\bibfield  {journal} {\bibinfo  {journal} {Annu. Rev. Phys. Chem.}\ }\textbf
  {\bibinfo {volume} {71}},\ \bibinfo {pages} {435} (\bibinfo {year}
  {2020})}\BibitemShut {NoStop}%
\bibitem [{\citenamefont {Richard}\ \emph {et~al.}(2005)\citenamefont
  {Richard}, \citenamefont {Romestain}, \citenamefont {André},\ and\
  \citenamefont {Dang}}]{Richard2005}%
  \BibitemOpen
  \bibfield  {author} {\bibinfo {author} {\bibfnamefont {M.}~\bibnamefont
  {Richard}}, \bibinfo {author} {\bibfnamefont {R.}~\bibnamefont {Romestain}},
  \bibinfo {author} {\bibfnamefont {R.}~\bibnamefont {André}}, \ and\ \bibinfo
  {author} {\bibfnamefont {L.~S.}\ \bibnamefont {Dang}},\ }\bibfield  {title}
  {\bibinfo {title} {\emph {Consequences of strong coupling between excitons
  and microcavity leaky modes}},\ }\href {\doibase 10.1063/1.1861979}
  {\bibfield  {journal} {\bibinfo  {journal} {Applied Physics Letters}\
  }\textbf {\bibinfo {volume} {86}},\ \bibinfo {pages} {071916} (\bibinfo
  {year} {2005})}\BibitemShut {NoStop}%
\bibitem [{\citenamefont {Levinsen}\ and\ \citenamefont
  {Parish}(2015)}]{levinsen2015strongly}%
  \BibitemOpen
  \bibfield  {author} {\bibinfo {author} {\bibfnamefont {J.}~\bibnamefont
  {Levinsen}}\ and\ \bibinfo {author} {\bibfnamefont {M.~M.}\ \bibnamefont
  {Parish}},\ }\bibfield  {title} {\bibinfo {title} {\emph {{Strongly
  interacting two-dimensional Fermi gases}}},\ }\href
  {https://doi.org/10.1142/9789814667746_0001} {\bibfield  {journal} {\bibinfo
  {journal} {Annu. Rev. Cold Atoms Mol.}\ }\textbf {\bibinfo {volume} {3}},\
  \bibinfo {pages} {1} (\bibinfo {year} {2015})}\BibitemShut {NoStop}%
\bibitem [{\citenamefont {Tassone}\ and\ \citenamefont
  {Yamamoto}(1999)}]{tassone1999exciton}%
  \BibitemOpen
  \bibfield  {author} {\bibinfo {author} {\bibfnamefont {F.}~\bibnamefont
  {Tassone}}\ and\ \bibinfo {author} {\bibfnamefont {Y.}~\bibnamefont
  {Yamamoto}},\ }\bibfield  {title} {\bibinfo {title} {\emph {Exciton-exciton
  scattering dynamics in a semiconductor microcavity and stimulated scattering
  into polaritons}},\ }\href {\doibase 10.1103/PhysRevB.59.10830} {\bibfield
  {journal} {\bibinfo  {journal} {Phys. Rev. B}\ }\textbf {\bibinfo {volume}
  {59}},\ \bibinfo {pages} {10830} (\bibinfo {year} {1999})}\BibitemShut
  {NoStop}%
\bibitem [{\citenamefont {Levinsen}\ \emph {et~al.}(2019)\citenamefont
  {Levinsen}, \citenamefont {Li},\ and\ \citenamefont {Parish}}]{Levinsen2019}%
  \BibitemOpen
  \bibfield  {author} {\bibinfo {author} {\bibfnamefont {J.}~\bibnamefont
  {Levinsen}}, \bibinfo {author} {\bibfnamefont {G.}~\bibnamefont {Li}}, \ and\
  \bibinfo {author} {\bibfnamefont {M.~M.}\ \bibnamefont {Parish}},\ }\bibfield
   {title} {\bibinfo {title} {\emph {Microscopic description of
  exciton-polaritons in microcavities}},\ }\href {\doibase
  10.1103/PhysRevResearch.1.033120} {\bibfield  {journal} {\bibinfo  {journal}
  {Phys. Rev. Research}\ }\textbf {\bibinfo {volume} {1}},\ \bibinfo {pages}
  {033120} (\bibinfo {year} {2019})}\BibitemShut {NoStop}%
\bibitem [{\citenamefont {Li}\ \emph {et~al.}(2021)\citenamefont {Li},
  \citenamefont {Parish},\ and\ \citenamefont {Levinsen}}]{li2021microscopic}%
  \BibitemOpen
  \bibfield  {author} {\bibinfo {author} {\bibfnamefont {G.}~\bibnamefont
  {Li}}, \bibinfo {author} {\bibfnamefont {M.~M.}\ \bibnamefont {Parish}}, \
  and\ \bibinfo {author} {\bibfnamefont {J.}~\bibnamefont {Levinsen}},\
  }\bibfield  {title} {\bibinfo {title} {\emph {Microscopic calculation of
  polariton scattering in semiconductor microcavities}},\ }\href {\doibase
  10.1103/PhysRevB.104.245404} {\bibfield  {journal} {\bibinfo  {journal}
  {Phys. Rev. B}\ }\textbf {\bibinfo {volume} {104}},\ \bibinfo {pages}
  {245404} (\bibinfo {year} {2021})}\BibitemShut {NoStop}%
\bibitem [{\citenamefont {Estrecho}\ \emph {et~al.}(2019)\citenamefont
  {Estrecho}, \citenamefont {Gao}, \citenamefont {Bobrovska}, \citenamefont
  {Comber-Todd}, \citenamefont {Fraser}, \citenamefont {Steger}, \citenamefont
  {West}, \citenamefont {Pfeiffer}, \citenamefont {Levinsen}, \citenamefont
  {Parish}, \citenamefont {Liew}, \citenamefont {Matuszewski}, \citenamefont
  {Snoke}, \citenamefont {Truscott},\ and\ \citenamefont
  {Ostrovskaya}}]{Estrecho2019}%
  \BibitemOpen
  \bibfield  {author} {\bibinfo {author} {\bibfnamefont {E.}~\bibnamefont
  {Estrecho}}, \bibinfo {author} {\bibfnamefont {T.}~\bibnamefont {Gao}},
  \bibinfo {author} {\bibfnamefont {N.}~\bibnamefont {Bobrovska}}, \bibinfo
  {author} {\bibfnamefont {D.}~\bibnamefont {Comber-Todd}}, \bibinfo {author}
  {\bibfnamefont {M.~D.}\ \bibnamefont {Fraser}}, \bibinfo {author}
  {\bibfnamefont {M.}~\bibnamefont {Steger}}, \bibinfo {author} {\bibfnamefont
  {K.}~\bibnamefont {West}}, \bibinfo {author} {\bibfnamefont {L.~N.}\
  \bibnamefont {Pfeiffer}}, \bibinfo {author} {\bibfnamefont {J.}~\bibnamefont
  {Levinsen}}, \bibinfo {author} {\bibfnamefont {M.~M.}\ \bibnamefont
  {Parish}}, \bibinfo {author} {\bibfnamefont {T.~C.~H.}\ \bibnamefont {Liew}},
  \bibinfo {author} {\bibfnamefont {M.}~\bibnamefont {Matuszewski}}, \bibinfo
  {author} {\bibfnamefont {D.~W.}\ \bibnamefont {Snoke}}, \bibinfo {author}
  {\bibfnamefont {A.~G.}\ \bibnamefont {Truscott}}, \ and\ \bibinfo {author}
  {\bibfnamefont {E.~A.}\ \bibnamefont {Ostrovskaya}},\ }\bibfield  {title}
  {\bibinfo {title} {\emph {Direct measurement of polariton-polariton
  interaction strength in the Thomas-Fermi regime of exciton-polariton
  condensation}},\ }\href {\doibase 10.1103/physrevb.100.035306} {\bibfield
  {journal} {\bibinfo  {journal} {Phys. Rev. B}\ }\textbf {\bibinfo {volume}
  {100}},\ \bibinfo {pages} {035306} (\bibinfo {year} {2019})}\BibitemShut
  {NoStop}%
\bibitem [{\citenamefont {Delteil}\ \emph {et~al.}(2019)\citenamefont
  {Delteil}, \citenamefont {Fink}, \citenamefont {Schade}, \citenamefont
  {Höfling}, \citenamefont {Schneider},\ and\ \citenamefont
  {İmamoğlu}}]{delteil2019}%
  \BibitemOpen
  \bibfield  {author} {\bibinfo {author} {\bibfnamefont {A.}~\bibnamefont
  {Delteil}}, \bibinfo {author} {\bibfnamefont {T.}~\bibnamefont {Fink}},
  \bibinfo {author} {\bibfnamefont {A.}~\bibnamefont {Schade}}, \bibinfo
  {author} {\bibfnamefont {S.}~\bibnamefont {Höfling}}, \bibinfo {author}
  {\bibfnamefont {C.}~\bibnamefont {Schneider}}, \ and\ \bibinfo {author}
  {\bibfnamefont {A.}~\bibnamefont {İmamoğlu}},\ }\bibfield  {title}
  {\bibinfo {title} {\emph {Towards polariton blockade of confined
  exciton-polaritons}},\ }\href {https://doi.org/10.1038/s41563-019-0282-y}
  {\bibfield  {journal} {\bibinfo  {journal} {Nature Materials}\ }\textbf
  {\bibinfo {volume} {18}},\ \bibinfo {pages} {219} (\bibinfo {year}
  {2019})}\BibitemShut {NoStop}%
\bibitem [{\citenamefont {Muñoz-Matutano}\ \emph {et~al.}(2019)\citenamefont
  {Muñoz-Matutano}, \citenamefont {Wood}, \citenamefont {Johnsson},
  \citenamefont {Vidal}, \citenamefont {Baragiola}, \citenamefont {Reinhard},
  \citenamefont {Lemaître}, \citenamefont {Bloch}, \citenamefont {Amo},
  \citenamefont {Nogues}, \citenamefont {Besga}, \citenamefont {Richard},\ and\
  \citenamefont {Volz}}]{MunozMatutano2019}%
  \BibitemOpen
  \bibfield  {author} {\bibinfo {author} {\bibfnamefont {G.}~\bibnamefont
  {Muñoz-Matutano}}, \bibinfo {author} {\bibfnamefont {A.}~\bibnamefont
  {Wood}}, \bibinfo {author} {\bibfnamefont {M.}~\bibnamefont {Johnsson}},
  \bibinfo {author} {\bibfnamefont {X.}~\bibnamefont {Vidal}}, \bibinfo
  {author} {\bibfnamefont {B.~Q.}\ \bibnamefont {Baragiola}}, \bibinfo {author}
  {\bibfnamefont {A.}~\bibnamefont {Reinhard}}, \bibinfo {author}
  {\bibfnamefont {A.}~\bibnamefont {Lemaître}}, \bibinfo {author}
  {\bibfnamefont {J.}~\bibnamefont {Bloch}}, \bibinfo {author} {\bibfnamefont
  {A.}~\bibnamefont {Amo}}, \bibinfo {author} {\bibfnamefont {G.}~\bibnamefont
  {Nogues}}, \bibinfo {author} {\bibfnamefont {B.}~\bibnamefont {Besga}},
  \bibinfo {author} {\bibfnamefont {M.}~\bibnamefont {Richard}}, \ and\
  \bibinfo {author} {\bibfnamefont {T.}~\bibnamefont {Volz}},\ }\bibfield
  {title} {\bibinfo {title} {\emph {Emergence of quantum correlations from
  interacting fibre-cavity polaritons}},\ }\href
  {https://doi.org/10.1038/s41563-019-0281-z} {\bibfield  {journal} {\bibinfo
  {journal} {Nature Materials}\ }\textbf {\bibinfo {volume} {18}},\ \bibinfo
  {pages} {213} (\bibinfo {year} {2019})}\BibitemShut {NoStop}%
\bibitem [{\citenamefont {Yu}\ and\ \citenamefont {Cardona}(2010)}]{Yu2010}%
  \BibitemOpen
  \bibfield  {author} {\bibinfo {author} {\bibfnamefont {P.~Y.}\ \bibnamefont
  {Yu}}\ and\ \bibinfo {author} {\bibfnamefont {M.}~\bibnamefont {Cardona}},\
  }\href@noop {} {\emph {\bibinfo {title} {Fundamentals of Semiconductors}}},\
  \bibinfo {edition} {4th}\ ed.\ (\bibinfo  {publisher} {Springer},\ \bibinfo
  {address} {New York},\ \bibinfo {year} {2010})\BibitemShut {NoStop}%
\bibitem [{\citenamefont {Grudinina}\ \emph {et~al.}(2021)\citenamefont
  {Grudinina}, \citenamefont {Kurbakov}, \citenamefont {Lozovik},\ and\
  \citenamefont {Voronova}}]{Grudinina2021}%
  \BibitemOpen
  \bibfield  {author} {\bibinfo {author} {\bibfnamefont {A.~M.}\ \bibnamefont
  {Grudinina}}, \bibinfo {author} {\bibfnamefont {I.~L.}\ \bibnamefont
  {Kurbakov}}, \bibinfo {author} {\bibfnamefont {Y.~E.}\ \bibnamefont
  {Lozovik}}, \ and\ \bibinfo {author} {\bibfnamefont {N.~S.}\ \bibnamefont
  {Voronova}},\ }\bibfield  {title} {\bibinfo {title} {\emph
  {Finite-temperature Hartree-Fock-Bogoliubov theory for exciton-polaritons}},\
  }\href {\doibase 10.1103/PhysRevB.104.125301} {\bibfield  {journal} {\bibinfo
   {journal} {Phys. Rev. B}\ }\textbf {\bibinfo {volume} {104}},\ \bibinfo
  {pages} {125301} (\bibinfo {year} {2021})}\BibitemShut {NoStop}%
\bibitem [{\citenamefont {Estrecho}\ \emph {et~al.}(2021)\citenamefont
  {Estrecho}, \citenamefont {Pieczarka}, \citenamefont {Wurdack}, \citenamefont
  {Steger}, \citenamefont {West}, \citenamefont {Pfeiffer}, \citenamefont
  {Snoke}, \citenamefont {Truscott},\ and\ \citenamefont
  {Ostrovskaya}}]{Estrecho2021}%
  \BibitemOpen
  \bibfield  {author} {\bibinfo {author} {\bibfnamefont {E.}~\bibnamefont
  {Estrecho}}, \bibinfo {author} {\bibfnamefont {M.}~\bibnamefont {Pieczarka}},
  \bibinfo {author} {\bibfnamefont {M.}~\bibnamefont {Wurdack}}, \bibinfo
  {author} {\bibfnamefont {M.}~\bibnamefont {Steger}}, \bibinfo {author}
  {\bibfnamefont {K.}~\bibnamefont {West}}, \bibinfo {author} {\bibfnamefont
  {L.~N.}\ \bibnamefont {Pfeiffer}}, \bibinfo {author} {\bibfnamefont {D.~W.}\
  \bibnamefont {Snoke}}, \bibinfo {author} {\bibfnamefont {A.~G.}\ \bibnamefont
  {Truscott}}, \ and\ \bibinfo {author} {\bibfnamefont {E.~A.}\ \bibnamefont
  {Ostrovskaya}},\ }\bibfield  {title} {\bibinfo {title} {\emph {Low-Energy
  Collective Oscillations and Bogoliubov Sound in an Exciton-Polariton
  Condensate}},\ }\href {\doibase 10.1103/PhysRevLett.126.075301} {\bibfield
  {journal} {\bibinfo  {journal} {Phys. Rev. Lett.}\ }\textbf {\bibinfo
  {volume} {126}},\ \bibinfo {pages} {075301} (\bibinfo {year}
  {2021})}\BibitemShut {NoStop}%
\bibitem [{\citenamefont {Steger}\ \emph {et~al.}(2013)\citenamefont {Steger},
  \citenamefont {Liu}, \citenamefont {Nelsen}, \citenamefont {Gautham},
  \citenamefont {Snoke}, \citenamefont {Balili}, \citenamefont {Pfeiffer},\
  and\ \citenamefont {West}}]{Steger2013}%
  \BibitemOpen
  \bibfield  {author} {\bibinfo {author} {\bibfnamefont {M.}~\bibnamefont
  {Steger}}, \bibinfo {author} {\bibfnamefont {G.}~\bibnamefont {Liu}},
  \bibinfo {author} {\bibfnamefont {B.}~\bibnamefont {Nelsen}}, \bibinfo
  {author} {\bibfnamefont {C.}~\bibnamefont {Gautham}}, \bibinfo {author}
  {\bibfnamefont {D.~W.}\ \bibnamefont {Snoke}}, \bibinfo {author}
  {\bibfnamefont {R.}~\bibnamefont {Balili}}, \bibinfo {author} {\bibfnamefont
  {L.}~\bibnamefont {Pfeiffer}}, \ and\ \bibinfo {author} {\bibfnamefont
  {K.}~\bibnamefont {West}},\ }\bibfield  {title} {\bibinfo {title} {\emph
  {Long-range ballistic motion and coherent flow of long-lifetime
  polaritons}},\ }\href {\doibase 10.1103/PhysRevB.88.235314} {\bibfield
  {journal} {\bibinfo  {journal} {Phys. Rev. B}\ }\textbf {\bibinfo {volume}
  {88}},\ \bibinfo {pages} {235314} (\bibinfo {year} {2013})}\BibitemShut
  {NoStop}%
\bibitem [{\citenamefont {Nelsen}\ \emph {et~al.}(2013)\citenamefont {Nelsen},
  \citenamefont {Liu}, \citenamefont {Steger}, \citenamefont {Snoke},
  \citenamefont {Balili}, \citenamefont {West},\ and\ \citenamefont
  {Pfeiffer}}]{Nelsen2013}%
  \BibitemOpen
  \bibfield  {author} {\bibinfo {author} {\bibfnamefont {B.}~\bibnamefont
  {Nelsen}}, \bibinfo {author} {\bibfnamefont {G.}~\bibnamefont {Liu}},
  \bibinfo {author} {\bibfnamefont {M.}~\bibnamefont {Steger}}, \bibinfo
  {author} {\bibfnamefont {D.~W.}\ \bibnamefont {Snoke}}, \bibinfo {author}
  {\bibfnamefont {R.}~\bibnamefont {Balili}}, \bibinfo {author} {\bibfnamefont
  {K.}~\bibnamefont {West}}, \ and\ \bibinfo {author} {\bibfnamefont
  {L.}~\bibnamefont {Pfeiffer}},\ }\bibfield  {title} {\bibinfo {title} {\emph
  {Dissipationless Flow and Sharp Threshold of a Polariton Condensate with Long
  Lifetime}},\ }\href {\doibase 10.1103/PhysRevX.3.041015} {\bibfield
  {journal} {\bibinfo  {journal} {Phys. Rev. X}\ }\textbf {\bibinfo {volume}
  {3}},\ \bibinfo {pages} {041015} (\bibinfo {year} {2013})}\BibitemShut
  {NoStop}%
\bibitem [{\citenamefont {Pieczarka}\ \emph {et~al.}(2019)\citenamefont
  {Pieczarka}, \citenamefont {Boozarjmehr}, \citenamefont {Estrecho},
  \citenamefont {Yoon}, \citenamefont {Steger}, \citenamefont {West},
  \citenamefont {Pfeiffer}, \citenamefont {Nelson}, \citenamefont {Snoke},
  \citenamefont {Truscott},\ and\ \citenamefont {Ostrovskaya}}]{Pieczarka2019}%
  \BibitemOpen
  \bibfield  {author} {\bibinfo {author} {\bibfnamefont {M.}~\bibnamefont
  {Pieczarka}}, \bibinfo {author} {\bibfnamefont {M.}~\bibnamefont
  {Boozarjmehr}}, \bibinfo {author} {\bibfnamefont {E.}~\bibnamefont
  {Estrecho}}, \bibinfo {author} {\bibfnamefont {Y.}~\bibnamefont {Yoon}},
  \bibinfo {author} {\bibfnamefont {M.}~\bibnamefont {Steger}}, \bibinfo
  {author} {\bibfnamefont {K.}~\bibnamefont {West}}, \bibinfo {author}
  {\bibfnamefont {L.~N.}\ \bibnamefont {Pfeiffer}}, \bibinfo {author}
  {\bibfnamefont {K.~A.}\ \bibnamefont {Nelson}}, \bibinfo {author}
  {\bibfnamefont {D.~W.}\ \bibnamefont {Snoke}}, \bibinfo {author}
  {\bibfnamefont {A.~G.}\ \bibnamefont {Truscott}}, \ and\ \bibinfo {author}
  {\bibfnamefont {E.~A.}\ \bibnamefont {Ostrovskaya}},\ }\bibfield  {title}
  {\bibinfo {title} {\emph {Effect of optically induced potential on the energy
  of trapped exciton polaritons below the condensation threshold}},\ }\href
  {\doibase 10.1103/PhysRevB.100.085301} {\bibfield  {journal} {\bibinfo
  {journal} {Phys. Rev. B}\ }\textbf {\bibinfo {volume} {100}},\ \bibinfo
  {pages} {085301} (\bibinfo {year} {2019})}\BibitemShut {NoStop}%
\bibitem [{\citenamefont {Doan}\ \emph {et~al.}(2020)\citenamefont {Doan},
  \citenamefont {Thoai},\ and\ \citenamefont {Haug}}]{Doan2020}%
  \BibitemOpen
  \bibfield  {author} {\bibinfo {author} {\bibfnamefont {T.~D.}\ \bibnamefont
  {Doan}}, \bibinfo {author} {\bibfnamefont {D.~B.~T.}\ \bibnamefont {Thoai}},
  \ and\ \bibinfo {author} {\bibfnamefont {H.}~\bibnamefont {Haug}},\
  }\bibfield  {title} {\bibinfo {title} {\emph {Kinetics and luminescence of
  the excitations of a nonequilibrium polariton condensate}},\ }\href {\doibase
  10.1103/PhysRevB.102.165126} {\bibfield  {journal} {\bibinfo  {journal}
  {Phys. Rev. B}\ }\textbf {\bibinfo {volume} {102}},\ \bibinfo {pages}
  {165126} (\bibinfo {year} {2020})}\BibitemShut {NoStop}%
\bibitem [{\citenamefont {Smirnov}\ \emph {et~al.}(2014)\citenamefont
  {Smirnov}, \citenamefont {Smirnova}, \citenamefont {Ostrovskaya},\ and\
  \citenamefont {Kivshar}}]{Smirnov2014}%
  \BibitemOpen
  \bibfield  {author} {\bibinfo {author} {\bibfnamefont {L.~A.}\ \bibnamefont
  {Smirnov}}, \bibinfo {author} {\bibfnamefont {D.~A.}\ \bibnamefont
  {Smirnova}}, \bibinfo {author} {\bibfnamefont {E.~A.}\ \bibnamefont
  {Ostrovskaya}}, \ and\ \bibinfo {author} {\bibfnamefont {Y.~S.}\ \bibnamefont
  {Kivshar}},\ }\bibfield  {title} {\bibinfo {title} {\emph {{Dynamics and
  stability of dark solitons in exciton-polariton condensates}}},\ }\href
  {\doibase 10.1103/PhysRevB.89.235310} {\bibfield  {journal} {\bibinfo
  {journal} {Phys. Rev. B}\ }\textbf {\bibinfo {volume} {89}},\ \bibinfo
  {pages} {235310} (\bibinfo {year} {2014})}\BibitemShut {NoStop}%
\bibitem [{\citenamefont {Bobrovska}\ and\ \citenamefont
  {Matuszewski}(2015)}]{Bobrovska2015}%
  \BibitemOpen
  \bibfield  {author} {\bibinfo {author} {\bibfnamefont {N.}~\bibnamefont
  {Bobrovska}}\ and\ \bibinfo {author} {\bibfnamefont {M.}~\bibnamefont
  {Matuszewski}},\ }\bibfield  {title} {\bibinfo {title} {\emph {Adiabatic
  approximation and fluctuations in exciton-polariton condensates}},\ }\href
  {\doibase 10.1103/PhysRevB.92.035311} {\bibfield  {journal} {\bibinfo
  {journal} {Phys. Rev. B}\ }\textbf {\bibinfo {volume} {92}},\ \bibinfo
  {pages} {035311} (\bibinfo {year} {2015})}\BibitemShut {NoStop}%
\bibitem [{\citenamefont {Bedaque}\ \emph {et~al.}(2000)\citenamefont
  {Bedaque}, \citenamefont {Braaten},\ and\ \citenamefont
  {Hammer}}]{Bedaque2000}%
  \BibitemOpen
  \bibfield  {author} {\bibinfo {author} {\bibfnamefont {P.~F.}\ \bibnamefont
  {Bedaque}}, \bibinfo {author} {\bibfnamefont {E.}~\bibnamefont {Braaten}}, \
  and\ \bibinfo {author} {\bibfnamefont {H.-W.}\ \bibnamefont {Hammer}},\
  }\bibfield  {title} {\bibinfo {title} {\emph {Three-body Recombination in
  Bose Gases with Large Scattering Length}},\ }\href {\doibase
  10.1103/PhysRevLett.85.908} {\bibfield  {journal} {\bibinfo  {journal} {Phys.
  Rev. Lett.}\ }\textbf {\bibinfo {volume} {85}},\ \bibinfo {pages} {908}
  (\bibinfo {year} {2000})}\BibitemShut {NoStop}%
\bibitem [{\citenamefont {Burt}\ \emph {et~al.}(1997)\citenamefont {Burt},
  \citenamefont {Ghrist}, \citenamefont {Myatt}, \citenamefont {Holland},
  \citenamefont {Cornell},\ and\ \citenamefont {Wieman}}]{Burt1997}%
  \BibitemOpen
  \bibfield  {author} {\bibinfo {author} {\bibfnamefont {E.~A.}\ \bibnamefont
  {Burt}}, \bibinfo {author} {\bibfnamefont {R.~W.}\ \bibnamefont {Ghrist}},
  \bibinfo {author} {\bibfnamefont {C.~J.}\ \bibnamefont {Myatt}}, \bibinfo
  {author} {\bibfnamefont {M.~J.}\ \bibnamefont {Holland}}, \bibinfo {author}
  {\bibfnamefont {E.~A.}\ \bibnamefont {Cornell}}, \ and\ \bibinfo {author}
  {\bibfnamefont {C.~E.}\ \bibnamefont {Wieman}},\ }\bibfield  {title}
  {\bibinfo {title} {\emph {Coherence, Correlations, and Collisions: What One
  Learns about Bose-Einstein Condensates from Their Decay}},\ }\href {\doibase
  10.1103/PhysRevLett.79.337} {\bibfield  {journal} {\bibinfo  {journal} {Phys.
  Rev. Lett.}\ }\textbf {\bibinfo {volume} {79}},\ \bibinfo {pages} {337}
  (\bibinfo {year} {1997})}\BibitemShut {NoStop}%
\bibitem [{\citenamefont {Makotyn}\ \emph {et~al.}()\citenamefont {Makotyn},
  \citenamefont {Klauss}, \citenamefont {Goldberger}, \citenamefont {Cornell},\
  and\ \citenamefont {Jin}}]{Makotyn2014}%
  \BibitemOpen
  \bibfield  {author} {\bibinfo {author} {\bibfnamefont {P.}~\bibnamefont
  {Makotyn}}, \bibinfo {author} {\bibfnamefont {C.~E.}\ \bibnamefont {Klauss}},
  \bibinfo {author} {\bibfnamefont {D.~L.}\ \bibnamefont {Goldberger}},
  \bibinfo {author} {\bibfnamefont {E.~A.}\ \bibnamefont {Cornell}}, \ and\
  \bibinfo {author} {\bibfnamefont {D.~S.}\ \bibnamefont {Jin}},\ }\bibfield
  {title} {\bibinfo {title} {\emph {Universal dynamics of a degenerate unitary
  Bose gas}},\ }\href {\doibase 10.1038/nphys2850} {\bibfield  {journal}
  {\bibinfo  {journal} {Nature Physics}\ }\textbf {\bibinfo {volume} {10}},\
  \bibinfo {pages} {116}}\BibitemShut {NoStop}%
\bibitem [{\citenamefont {Myers}\ \emph {et~al.}(2018)\citenamefont {Myers},
  \citenamefont {Mukherjee}, \citenamefont {Beaumariage}, \citenamefont
  {Snoke}, \citenamefont {Steger}, \citenamefont {Pfeiffer},\ and\
  \citenamefont {West}}]{Myers2018}%
  \BibitemOpen
  \bibfield  {author} {\bibinfo {author} {\bibfnamefont {D.~M.}\ \bibnamefont
  {Myers}}, \bibinfo {author} {\bibfnamefont {S.}~\bibnamefont {Mukherjee}},
  \bibinfo {author} {\bibfnamefont {J.}~\bibnamefont {Beaumariage}}, \bibinfo
  {author} {\bibfnamefont {D.~W.}\ \bibnamefont {Snoke}}, \bibinfo {author}
  {\bibfnamefont {M.}~\bibnamefont {Steger}}, \bibinfo {author} {\bibfnamefont
  {L.~N.}\ \bibnamefont {Pfeiffer}}, \ and\ \bibinfo {author} {\bibfnamefont
  {K.}~\bibnamefont {West}},\ }\bibfield  {title} {\bibinfo {title} {\emph
  {Polariton-enhanced exciton transport}},\ }\href {\doibase
  10.1103/PhysRevB.98.235302} {\bibfield  {journal} {\bibinfo  {journal} {Phys.
  Rev. B}\ }\textbf {\bibinfo {volume} {98}},\ \bibinfo {pages} {235302}
  (\bibinfo {year} {2018})}\BibitemShut {NoStop}%
\bibitem [{\citenamefont {Hammack}\ \emph {et~al.}(2006)\citenamefont
  {Hammack}, \citenamefont {Griswold}, \citenamefont {Butov}, \citenamefont
  {Smallwood}, \citenamefont {Ivanov},\ and\ \citenamefont
  {Gossard}}]{Hammack2006}%
  \BibitemOpen
  \bibfield  {author} {\bibinfo {author} {\bibfnamefont {A.~T.}\ \bibnamefont
  {Hammack}}, \bibinfo {author} {\bibfnamefont {M.}~\bibnamefont {Griswold}},
  \bibinfo {author} {\bibfnamefont {L.~V.}\ \bibnamefont {Butov}}, \bibinfo
  {author} {\bibfnamefont {L.~E.}\ \bibnamefont {Smallwood}}, \bibinfo {author}
  {\bibfnamefont {A.~L.}\ \bibnamefont {Ivanov}}, \ and\ \bibinfo {author}
  {\bibfnamefont {A.~C.}\ \bibnamefont {Gossard}},\ }\bibfield  {title}
  {\bibinfo {title} {\emph {Trapping of Cold Excitons in Quantum Well
  Structures with Laser Light}},\ }\href {\doibase
  10.1103/PhysRevLett.96.227402} {\bibfield  {journal} {\bibinfo  {journal}
  {Phys. Rev. Lett.}\ }\textbf {\bibinfo {volume} {96}},\ \bibinfo {pages}
  {227402} (\bibinfo {year} {2006})}\BibitemShut {NoStop}%
\bibitem [{\citenamefont {Chang}\ \emph {et~al.}(2016)\citenamefont {Chang},
  \citenamefont {Bouton}, \citenamefont {Cayla}, \citenamefont {Qu},
  \citenamefont {Aspect}, \citenamefont {Westbrook},\ and\ \citenamefont
  {Cl\'ement}}]{Chang2016}%
  \BibitemOpen
  \bibfield  {author} {\bibinfo {author} {\bibfnamefont {R.}~\bibnamefont
  {Chang}}, \bibinfo {author} {\bibfnamefont {Q.}~\bibnamefont {Bouton}},
  \bibinfo {author} {\bibfnamefont {H.}~\bibnamefont {Cayla}}, \bibinfo
  {author} {\bibfnamefont {C.}~\bibnamefont {Qu}}, \bibinfo {author}
  {\bibfnamefont {A.}~\bibnamefont {Aspect}}, \bibinfo {author} {\bibfnamefont
  {C.~I.}\ \bibnamefont {Westbrook}}, \ and\ \bibinfo {author} {\bibfnamefont
  {D.}~\bibnamefont {Cl\'ement}},\ }\bibfield  {title} {\bibinfo {title} {\emph
  {Momentum-Resolved Observation of Thermal and Quantum Depletion in a Bose
  Gas}},\ }\href {\doibase 10.1103/PhysRevLett.117.235303} {\bibfield
  {journal} {\bibinfo  {journal} {Phys. Rev. Lett.}\ }\textbf {\bibinfo
  {volume} {117}},\ \bibinfo {pages} {235303} (\bibinfo {year}
  {2016})}\BibitemShut {NoStop}%
\bibitem [{\citenamefont {Lopes}\ \emph {et~al.}(2017)\citenamefont {Lopes},
  \citenamefont {Eigen}, \citenamefont {Navon}, \citenamefont {Cl\'ement},
  \citenamefont {Smith},\ and\ \citenamefont {Hadzibabic}}]{Lopes2017}%
  \BibitemOpen
  \bibfield  {author} {\bibinfo {author} {\bibfnamefont {R.}~\bibnamefont
  {Lopes}}, \bibinfo {author} {\bibfnamefont {C.}~\bibnamefont {Eigen}},
  \bibinfo {author} {\bibfnamefont {N.}~\bibnamefont {Navon}}, \bibinfo
  {author} {\bibfnamefont {D.}~\bibnamefont {Cl\'ement}}, \bibinfo {author}
  {\bibfnamefont {R.~P.}\ \bibnamefont {Smith}}, \ and\ \bibinfo {author}
  {\bibfnamefont {Z.}~\bibnamefont {Hadzibabic}},\ }\bibfield  {title}
  {\bibinfo {title} {\emph {Quantum Depletion of a Homogeneous Bose-Einstein
  Condensate}},\ }\href {\doibase 10.1103/PhysRevLett.119.190404} {\bibfield
  {journal} {\bibinfo  {journal} {Phys. Rev. Lett.}\ }\textbf {\bibinfo
  {volume} {119}},\ \bibinfo {pages} {190404} (\bibinfo {year}
  {2017})}\BibitemShut {NoStop}%
\bibitem [{\citenamefont {Bobrovska}\ \emph {et~al.}(2014)\citenamefont
  {Bobrovska}, \citenamefont {Ostrovskaya},\ and\ \citenamefont
  {Matuszewski}}]{Bobrovska2014}%
  \BibitemOpen
  \bibfield  {author} {\bibinfo {author} {\bibfnamefont {N.}~\bibnamefont
  {Bobrovska}}, \bibinfo {author} {\bibfnamefont {E.~A.}\ \bibnamefont
  {Ostrovskaya}}, \ and\ \bibinfo {author} {\bibfnamefont {M.}~\bibnamefont
  {Matuszewski}},\ }\bibfield  {title} {\bibinfo {title} {\emph {Stability and
  spatial coherence of nonresonantly pumped exciton-polariton condensates}},\
  }\href {\doibase 10.1103/PhysRevB.90.205304} {\bibfield  {journal} {\bibinfo
  {journal} {Phys. Rev. B}\ }\textbf {\bibinfo {volume} {90}},\ \bibinfo
  {pages} {205304} (\bibinfo {year} {2014})}\BibitemShut {NoStop}%
\bibitem [{\citenamefont {Dagvadorj}\ \emph {et~al.}(2021)\citenamefont
  {Dagvadorj}, \citenamefont {Kulczykowski}, \citenamefont
  {Szyma\ifmmode~\acute{n}\else \'{n}\fi{}ska},\ and\ \citenamefont
  {Matuszewski}}]{Dagvadorj2021}%
  \BibitemOpen
  \bibfield  {author} {\bibinfo {author} {\bibfnamefont {G.}~\bibnamefont
  {Dagvadorj}}, \bibinfo {author} {\bibfnamefont {M.}~\bibnamefont
  {Kulczykowski}}, \bibinfo {author} {\bibfnamefont {M.~H.}\ \bibnamefont
  {Szyma\ifmmode~\acute{n}\else \'{n}\fi{}ska}}, \ and\ \bibinfo {author}
  {\bibfnamefont {M.}~\bibnamefont {Matuszewski}},\ }\bibfield  {title}
  {\bibinfo {title} {\emph {First-order dissipative phase transition in an
  exciton-polariton condensate}},\ }\href {\doibase
  10.1103/PhysRevB.104.165301} {\bibfield  {journal} {\bibinfo  {journal}
  {Phys. Rev. B}\ }\textbf {\bibinfo {volume} {104}},\ \bibinfo {pages}
  {165301} (\bibinfo {year} {2021})}\BibitemShut {NoStop}%
\bibitem [{\citenamefont {Bobrovska}\ \emph {et~al.}(2018)\citenamefont
  {Bobrovska}, \citenamefont {Matuszewski}, \citenamefont {Daskalakis},
  \citenamefont {Maier},\ and\ \citenamefont
  {K{\'{e}}na-Cohen}}]{Bobrovska2018}%
  \BibitemOpen
  \bibfield  {author} {\bibinfo {author} {\bibfnamefont {N.}~\bibnamefont
  {Bobrovska}}, \bibinfo {author} {\bibfnamefont {M.}~\bibnamefont
  {Matuszewski}}, \bibinfo {author} {\bibfnamefont {K.~S.}\ \bibnamefont
  {Daskalakis}}, \bibinfo {author} {\bibfnamefont {S.~A.}\ \bibnamefont
  {Maier}}, \ and\ \bibinfo {author} {\bibfnamefont {S.}~\bibnamefont
  {K{\'{e}}na-Cohen}},\ }\bibfield  {title} {\bibinfo {title} {\emph
  {{Dynamical Instability of a Nonequilibrium Exciton-Polariton Condensate}}},\
  }\href {\doibase 10.1021/acsphotonics.7b00283} {\bibfield  {journal}
  {\bibinfo  {journal} {ACS Photonics}\ }\textbf {\bibinfo {volume} {5}},\
  \bibinfo {pages} {111} (\bibinfo {year} {2018})}\BibitemShut {NoStop}%
\bibitem [{\citenamefont {Estrecho}\ \emph {et~al.}(2018)\citenamefont
  {Estrecho}, \citenamefont {Gao}, \citenamefont {Bobrovska}, \citenamefont
  {Fraser}, \citenamefont {Steger}, \citenamefont {Pfeiffer}, \citenamefont
  {West}, \citenamefont {Liew}, \citenamefont {Matuszewski}, \citenamefont
  {Snoke}, \citenamefont {Truscott},\ and\ \citenamefont
  {Ostrovskaya}}]{Estrecho2018}%
  \BibitemOpen
  \bibfield  {author} {\bibinfo {author} {\bibfnamefont {E.}~\bibnamefont
  {Estrecho}}, \bibinfo {author} {\bibfnamefont {T.}~\bibnamefont {Gao}},
  \bibinfo {author} {\bibfnamefont {N.}~\bibnamefont {Bobrovska}}, \bibinfo
  {author} {\bibfnamefont {M.~D.}\ \bibnamefont {Fraser}}, \bibinfo {author}
  {\bibfnamefont {M.}~\bibnamefont {Steger}}, \bibinfo {author} {\bibfnamefont
  {L.}~\bibnamefont {Pfeiffer}}, \bibinfo {author} {\bibfnamefont
  {K.}~\bibnamefont {West}}, \bibinfo {author} {\bibfnamefont {T.~C.~H.}\
  \bibnamefont {Liew}}, \bibinfo {author} {\bibfnamefont {M.}~\bibnamefont
  {Matuszewski}}, \bibinfo {author} {\bibfnamefont {D.~W.}\ \bibnamefont
  {Snoke}}, \bibinfo {author} {\bibfnamefont {A.~G.}\ \bibnamefont {Truscott}},
  \ and\ \bibinfo {author} {\bibfnamefont {E.~A.}\ \bibnamefont
  {Ostrovskaya}},\ }\bibfield  {title} {\bibinfo {title} {\emph {Single-shot
  condensation of exciton polaritons and the hole burning effect}},\ }\href
  {\doibase 10.1038/s41467-018-05349-4} {\bibfield  {journal} {\bibinfo
  {journal} {Nature Commun.}\ }\textbf {\bibinfo {volume} {9}},\ \bibinfo
  {pages} {2944} (\bibinfo {year} {2018})}\BibitemShut {NoStop}%
\bibitem [{\citenamefont {Baboux}\ \emph {et~al.}(2018)\citenamefont {Baboux},
  \citenamefont {De~Bernardis}, \citenamefont {Goblot}, \citenamefont
  {Gladilin}, \citenamefont {Gomez}, \citenamefont {Galopin}, \citenamefont
  {Le~Gratiet}, \citenamefont {Lemaître}, \citenamefont {Sagnes},
  \citenamefont {Carusotto}, \citenamefont {Wouters}, \citenamefont {Amo},\
  and\ \citenamefont {Bloch}}]{Baboux2018}%
  \BibitemOpen
  \bibfield  {author} {\bibinfo {author} {\bibfnamefont {F.}~\bibnamefont
  {Baboux}}, \bibinfo {author} {\bibfnamefont {D.}~\bibnamefont
  {De~Bernardis}}, \bibinfo {author} {\bibfnamefont {V.}~\bibnamefont
  {Goblot}}, \bibinfo {author} {\bibfnamefont {V.~N.}\ \bibnamefont
  {Gladilin}}, \bibinfo {author} {\bibfnamefont {C.}~\bibnamefont {Gomez}},
  \bibinfo {author} {\bibfnamefont {E.}~\bibnamefont {Galopin}}, \bibinfo
  {author} {\bibfnamefont {L.}~\bibnamefont {Le~Gratiet}}, \bibinfo {author}
  {\bibfnamefont {A.}~\bibnamefont {Lemaître}}, \bibinfo {author}
  {\bibfnamefont {I.}~\bibnamefont {Sagnes}}, \bibinfo {author} {\bibfnamefont
  {I.}~\bibnamefont {Carusotto}}, \bibinfo {author} {\bibfnamefont
  {M.}~\bibnamefont {Wouters}}, \bibinfo {author} {\bibfnamefont
  {A.}~\bibnamefont {Amo}}, \ and\ \bibinfo {author} {\bibfnamefont
  {J.}~\bibnamefont {Bloch}},\ }\bibfield  {title} {\bibinfo {title} {\emph
  {Unstable and stable regimes of polariton condensation}},\ }\href {\doibase
  10.1364/OPTICA.5.001163} {\bibfield  {journal} {\bibinfo  {journal} {Optica}\
  }\textbf {\bibinfo {volume} {5}},\ \bibinfo {pages} {1163} (\bibinfo {year}
  {2018})}\BibitemShut {NoStop}%
\bibitem [{\citenamefont {Altman}\ \emph {et~al.}(2015)\citenamefont {Altman},
  \citenamefont {Sieberer}, \citenamefont {Chen}, \citenamefont {Diehl},\ and\
  \citenamefont {Toner}}]{Altman2015}%
  \BibitemOpen
  \bibfield  {author} {\bibinfo {author} {\bibfnamefont {E.}~\bibnamefont
  {Altman}}, \bibinfo {author} {\bibfnamefont {L.~M.}\ \bibnamefont
  {Sieberer}}, \bibinfo {author} {\bibfnamefont {L.}~\bibnamefont {Chen}},
  \bibinfo {author} {\bibfnamefont {S.}~\bibnamefont {Diehl}}, \ and\ \bibinfo
  {author} {\bibfnamefont {J.}~\bibnamefont {Toner}},\ }\bibfield  {title}
  {\bibinfo {title} {\emph {Two-Dimensional Superfluidity of Exciton Polaritons
  Requires Strong Anisotropy}},\ }\href {\doibase 10.1103/PhysRevX.5.011017}
  {\bibfield  {journal} {\bibinfo  {journal} {Phys. Rev. X}\ }\textbf {\bibinfo
  {volume} {5}},\ \bibinfo {pages} {011017} (\bibinfo {year}
  {2015})}\BibitemShut {NoStop}%
\bibitem [{\citenamefont {Squizzato}\ \emph {et~al.}(2018)\citenamefont
  {Squizzato}, \citenamefont {Canet},\ and\ \citenamefont
  {Minguzzi}}]{Squizatto2018}%
  \BibitemOpen
  \bibfield  {author} {\bibinfo {author} {\bibfnamefont {D.}~\bibnamefont
  {Squizzato}}, \bibinfo {author} {\bibfnamefont {L.}~\bibnamefont {Canet}}, \
  and\ \bibinfo {author} {\bibfnamefont {A.}~\bibnamefont {Minguzzi}},\
  }\bibfield  {title} {\bibinfo {title} {\emph {Kardar-Parisi-Zhang
  universality in the phase distributions of one-dimensional
  exciton-polaritons}},\ }\href {\doibase 10.1103/PhysRevB.97.195453}
  {\bibfield  {journal} {\bibinfo  {journal} {Phys. Rev. B}\ }\textbf {\bibinfo
  {volume} {97}},\ \bibinfo {pages} {195453} (\bibinfo {year}
  {2018})}\BibitemShut {NoStop}%
\bibitem [{\citenamefont {Deligiannis}\ \emph {et~al.}(2021)\citenamefont
  {Deligiannis}, \citenamefont {Squizzato}, \citenamefont {Minguzzi},\ and\
  \citenamefont {Canet}}]{Deligiannis2021}%
  \BibitemOpen
  \bibfield  {author} {\bibinfo {author} {\bibfnamefont {K.}~\bibnamefont
  {Deligiannis}}, \bibinfo {author} {\bibfnamefont {D.}~\bibnamefont
  {Squizzato}}, \bibinfo {author} {\bibfnamefont {A.}~\bibnamefont {Minguzzi}},
  \ and\ \bibinfo {author} {\bibfnamefont {L.}~\bibnamefont {Canet}},\
  }\bibfield  {title} {\bibinfo {title} {\emph {Accessing Kardar-Parisi-Zhang
  universality sub-classes with exciton polaritons}},\ }\href {\doibase
  10.1209/0295-5075/132/67004} {\bibfield  {journal} {\bibinfo  {journal} {EPL
  (Europhysics Letters)}\ }\textbf {\bibinfo {volume} {132}},\ \bibinfo {pages}
  {67004} (\bibinfo {year} {2021})}\BibitemShut {NoStop}%
\bibitem [{\citenamefont {Gavrilov}\ \emph {et~al.}(2010)\citenamefont
  {Gavrilov}, \citenamefont {Brichkin}, \citenamefont {Dorodnyi}, \citenamefont
  {Tikhodeev}, \citenamefont {Gippius},\ and\ \citenamefont
  {Kulakovskii}}]{Gavrilov2010}%
  \BibitemOpen
  \bibfield  {author} {\bibinfo {author} {\bibfnamefont {S.~S.}\ \bibnamefont
  {Gavrilov}}, \bibinfo {author} {\bibfnamefont {A.~S.}\ \bibnamefont
  {Brichkin}}, \bibinfo {author} {\bibfnamefont {A.~A.}\ \bibnamefont
  {Dorodnyi}}, \bibinfo {author} {\bibfnamefont {S.~G.}\ \bibnamefont
  {Tikhodeev}}, \bibinfo {author} {\bibfnamefont {N.~A.}\ \bibnamefont
  {Gippius}}, \ and\ \bibinfo {author} {\bibfnamefont {V.~D.}\ \bibnamefont
  {Kulakovskii}},\ }\bibfield  {title} {\bibinfo {title} {\emph {Polarization
  instability in a polariton system in semiconductor microcavities}},\ }\href
  {\doibase 10.1134/S0021364010150105} {\bibfield  {journal} {\bibinfo
  {journal} {JETP Letters}\ }\textbf {\bibinfo {volume} {92}},\ \bibinfo
  {pages} {171} (\bibinfo {year} {2010})}\BibitemShut {NoStop}%
\bibitem [{\citenamefont {Klembt}\ \emph {et~al.}(2015)\citenamefont {Klembt},
  \citenamefont {Durupt}, \citenamefont {Datta}, \citenamefont {Klein},
  \citenamefont {Baas}, \citenamefont {L{\'{e}}ger}, \citenamefont {Kruse},
  \citenamefont {Hommel}, \citenamefont {Minguzzi},\ and\ \citenamefont
  {Richard}}]{Klembt2015}%
  \BibitemOpen
  \bibfield  {author} {\bibinfo {author} {\bibfnamefont {S.}~\bibnamefont
  {Klembt}}, \bibinfo {author} {\bibfnamefont {E.}~\bibnamefont {Durupt}},
  \bibinfo {author} {\bibfnamefont {S.}~\bibnamefont {Datta}}, \bibinfo
  {author} {\bibfnamefont {T.}~\bibnamefont {Klein}}, \bibinfo {author}
  {\bibfnamefont {A.}~\bibnamefont {Baas}}, \bibinfo {author} {\bibfnamefont
  {Y.}~\bibnamefont {L{\'{e}}ger}}, \bibinfo {author} {\bibfnamefont
  {C.}~\bibnamefont {Kruse}}, \bibinfo {author} {\bibfnamefont
  {D.}~\bibnamefont {Hommel}}, \bibinfo {author} {\bibfnamefont
  {A.}~\bibnamefont {Minguzzi}}, \ and\ \bibinfo {author} {\bibfnamefont
  {M.}~\bibnamefont {Richard}},\ }\bibfield  {title} {\bibinfo {title} {\emph
  {{Exciton-polariton gas as a nonequilibrium coolant}}},\ }\href {\doibase
  10.1103/PhysRevLett.114.186403} {\bibfield  {journal} {\bibinfo  {journal}
  {Phys. Rev, Lett.}\ }\textbf {\bibinfo {volume} {114}},\ \bibinfo {pages}
  {186403} (\bibinfo {year} {2015})}\BibitemShut {NoStop}%
\bibitem [{\citenamefont {Walker}\ \emph {et~al.}(2017)\citenamefont {Walker},
  \citenamefont {Tinkler}, \citenamefont {Royall}, \citenamefont {Skryabin},
  \citenamefont {Farrer}, \citenamefont {Ritchie}, \citenamefont {Skolnick},\
  and\ \citenamefont {Krizhanovskii}}]{Walker2017}%
  \BibitemOpen
  \bibfield  {author} {\bibinfo {author} {\bibfnamefont {P.~M.}\ \bibnamefont
  {Walker}}, \bibinfo {author} {\bibfnamefont {L.}~\bibnamefont {Tinkler}},
  \bibinfo {author} {\bibfnamefont {B.}~\bibnamefont {Royall}}, \bibinfo
  {author} {\bibfnamefont {D.~V.}\ \bibnamefont {Skryabin}}, \bibinfo {author}
  {\bibfnamefont {I.}~\bibnamefont {Farrer}}, \bibinfo {author} {\bibfnamefont
  {D.~A.}\ \bibnamefont {Ritchie}}, \bibinfo {author} {\bibfnamefont {M.~S.}\
  \bibnamefont {Skolnick}}, \ and\ \bibinfo {author} {\bibfnamefont {D.~N.}\
  \bibnamefont {Krizhanovskii}},\ }\bibfield  {title} {\bibinfo {title} {\emph
  {Dark Solitons in High Velocity Waveguide Polariton Fluids}},\ }\href
  {\doibase 10.1103/PhysRevLett.119.097403} {\bibfield  {journal} {\bibinfo
  {journal} {Phys. Rev. Lett.}\ }\textbf {\bibinfo {volume} {119}},\ \bibinfo
  {pages} {097403} (\bibinfo {year} {2017})}\BibitemShut {NoStop}%
\bibitem [{\citenamefont {Porras}\ \emph {et~al.}(2002)\citenamefont {Porras},
  \citenamefont {Ciuti}, \citenamefont {Baumberg},\ and\ \citenamefont
  {Tejedor}}]{Porras2002}%
  \BibitemOpen
  \bibfield  {author} {\bibinfo {author} {\bibfnamefont {D.}~\bibnamefont
  {Porras}}, \bibinfo {author} {\bibfnamefont {C.}~\bibnamefont {Ciuti}},
  \bibinfo {author} {\bibfnamefont {J.~J.}\ \bibnamefont {Baumberg}}, \ and\
  \bibinfo {author} {\bibfnamefont {C.}~\bibnamefont {Tejedor}},\ }\bibfield
  {title} {\bibinfo {title} {\emph {Polariton dynamics and Bose-Einstein
  condensation in semiconductor microcavities}},\ }\href {\doibase
  10.1103/PhysRevB.66.085304} {\bibfield  {journal} {\bibinfo  {journal} {Phys.
  Rev. B}\ }\textbf {\bibinfo {volume} {66}},\ \bibinfo {pages} {085304}
  (\bibinfo {year} {2002})}\BibitemShut {NoStop}%
\bibitem [{\citenamefont {Ouellet-Plamondon}\ \emph {et~al.}(2015)\citenamefont
  {Ouellet-Plamondon}, \citenamefont {Sallen}, \citenamefont {Jabeen},
  \citenamefont {Oberli},\ and\ \citenamefont {Deveaud}}]{Ouellet2015}%
  \BibitemOpen
  \bibfield  {author} {\bibinfo {author} {\bibfnamefont {C.}~\bibnamefont
  {Ouellet-Plamondon}}, \bibinfo {author} {\bibfnamefont {G.}~\bibnamefont
  {Sallen}}, \bibinfo {author} {\bibfnamefont {F.}~\bibnamefont {Jabeen}},
  \bibinfo {author} {\bibfnamefont {D.~Y.}\ \bibnamefont {Oberli}}, \ and\
  \bibinfo {author} {\bibfnamefont {B.}~\bibnamefont {Deveaud}},\ }\bibfield
  {title} {\bibinfo {title} {\emph {Multiple polariton modes originating from
  the coupling of quantum wells in planar microcavity}},\ }\href {\doibase
  10.1103/PhysRevB.92.075313} {\bibfield  {journal} {\bibinfo  {journal} {Phys.
  Rev. B}\ }\textbf {\bibinfo {volume} {92}},\ \bibinfo {pages} {075313}
  (\bibinfo {year} {2015})}\BibitemShut {NoStop}%
\end{thebibliography}%

\end{document}